\documentclass[10pt,american,3p]{elsarticle}
\usepackage[T1]{fontenc}
\usepackage[latin9]{inputenc}
\usepackage{color}
\usepackage{float}
\usepackage{amsmath}
\usepackage{amssymb}
\usepackage{graphicx}
\usepackage{esint}

\makeatletter

\newcommand{\noun}[1]{\textsc{#1}}
\providecommand{\tabularnewline}{\\}
\newcommand{\lyxdot}{.}

\floatstyle{ruled}
\newfloat{algorithm}{tbp}{loa}
\providecommand{\algorithmname}{Algorithm}
\floatname{algorithm}{\protect\algorithmname}

\usepackage{pslatex}
\usepackage{xcolor}
\journal{Computer Physics Communications}
\usepackage{mathrsfs}
\renewcommand{\boldsymbol}[1]{\pmb{#1}} 
\biboptions{sort&compress}

\usepackage{babel}

\@ifundefined{showcaptionsetup}{}{%
 \PassOptionsToPackage{caption=false}{subfig}}
\usepackage{subfig}
\makeatother

\usepackage{babel}
\begin{document}
\begin{frontmatter}

\title{Multiple-Relaxation-Time Lattice Boltzmann scheme for Fractional Advection-Diffusion Equation}

\author[CEA]{\noun{Alain Cartalade}\corref{cor1}}

\address[CEA]{Den -- DM2S, STMF, LMSF, CEA, Université de Paris-Saclay, F-91191, Gif-sur-Yvette, France.}

\ead{alain.cartalade@cea.fr}

\author[CEA,Framatome]{\noun{Amina Younsi}}

\address[Framatome]{Framatome ANP, Département Développement Codes \& Méthodes -- Tour AREVA, 1 Place Jean Millier, F-92400 Courbevoie, France.}

\ead{amina.younsi@framatome.com}

\author[Av]{\noun{Marie-Christine Néel}}

\address[Av]{Université d'Avignon et des Pays de Vaucluse, UMR 1114 EMMAH, 84018 Avignon Cedex, France.}

\ead{mcneel@avignon.inra.fr}

\cortext[cor1]{Corresponding author. Tel.:+33 (0)1 69 08 40 67}

\begin{abstract}

Partial differential equations (p.d.e) equipped with spatial derivatives
of fractional order capture anomalous transport behaviors observed
in diverse fields of Science. A number of numerical methods approximate
their solutions in dimension one. Focusing our effort on such p.d.e.
in higher dimension with Dirichlet boundary conditions, we present
an approximation based on Lattice Boltzmann Method with Bhatnagar-Gross-Krook
(BGK) or Multiple-Relaxation-Time (MRT) collision operators. First,
an equilibrium distribution function is defined for simulating space-fractional
diffusion equations in dimensions 2 and 3. Then, we check the accuracy
of the solutions by comparing with \emph{i)} random walks derived
from stable Lévy motion, and \emph{ii)} exact solutions. Because of
its additional freedom degrees, the MRT collision operator provides
accurate approximations to space-fractional advection-diffusion equations,
even in the cases which the BGK fails to represent because of anisotropic
diffusion tensor or of flow rate destabilizing the BGK LBM scheme.

\end{abstract}

\begin{keyword}

Fractional Advection-Diffusion Equation, Lattice Boltzmann method,
Multiple-Relaxation-Time, Random Walk, Stable Process.

\end{keyword}

\end{frontmatter}

\section{\label{sec:Introduction}Introduction}

Among diverse non-Fickian transport behaviors observed in all fields
of Science, heavy tailed spatial concentration profiles recorded on
chemical species, living cells or organisms, suggest displacements
more rapid than the classical Advection Diffusion Equation (ADE) predicts
\citep{Clarke_etal_GroundWat2005,Wheatcraft-Tyler_WRR1988,Metzler-Klafter_JPA2004,Clark_etal_Ecology1999,Fedotov-Iomin_PRL2007}.
Such super-dispersive phenomena include plumes that lack finite second
moment, or whose mean and peak do not coincide (see \citep{Benson_etal_TiPM2001}).
Possible explanations may be large scale heterogeneity or multiple
coupling between many simple sub-systems which separately would not
exhibit such abnormalities. Similar strange behaviors are observed
often enough to suggest exploring alternative models as fractional
partial differential equations. It turns out that \citep{Deng_etal_JHE2004,Benson_etal_TiPM2001,Benson_etal_WRR2000,Schumer_etal_WRR2001,Kelly_WRR019748}
many tracer tests in rivers and underground porous media are accurately
represented by the more general conservation equation 
\begin{equation}
\frac{\partial C}{\partial t}(\mathbf{x},\,t)+\boldsymbol{\nabla}\cdot(\mathbf{u}C)(\mathbf{x},\,t)=\boldsymbol{\nabla}\cdot\overline{\overline{\mathbf{D}}}(\mathbf{x})\boldsymbol{\mathcal{F}}^{\boldsymbol{\alpha}\boldsymbol{p}\boldsymbol{g}}(C)+S_{c}(\mathbf{x},\,t).\label{eq:FADE_moreGeneral}
\end{equation}
It models mass spreading for passive solute at concentration $C$
in incompressible fluid flowing at average flow rate $\mathbf{u}=\sum_{\mu=1}^{d}u_{\mu}\mathbf{b}_{\mu}$
super-imposed to small scale velocity field whose complexity causes
non-Fickian dispersive flux $-\overline{\overline{\mathbf{D}}}\boldsymbol{\mathcal{F}}^{\boldsymbol{\alpha}\boldsymbol{p}\boldsymbol{g}}(C)$.
The space variable $\mathbf{x}$ belongs to some domain $\Omega$
of $\mathbb{R}^{d}$, and is described in the orthonormal basis $\{\mathbf{b}_{\mu};\,\mbox{for }\mu=1,\,...,\,d\}$
of $\mathbb{R}^{d}$ by its coordinates noted $x_{\mu}$: greek subscripts
refer to spatial coordinates. Moreover $S_{c}$ is a source rate.
The coordinates of vector $\boldsymbol{\mathcal{F}}^{\boldsymbol{\alpha}\boldsymbol{p}\boldsymbol{g}}(C)$
are composed of partial derivatives of $C$ with respect to (w.r.t.)
the $x_{\mu}$ which reflect the variations of $C$ when all the other
coordinates of $\mathbf{x}$ are fixed. These derivatives are of order
one in the classical ADE, but in Eq. (\ref{eq:FADE_moreGeneral})
they may have fractional order $\alpha_{\mu}-1$ specified by the
entries of vector $\boldsymbol{\alpha}=(\alpha_{1},\,...,\,\alpha_{d})^{T}$
which belong to $]1,\,2]$. Yet, in general fractional derivatives
are not completely determined by their order. This is why vector $\boldsymbol{\mathcal{F}}^{\boldsymbol{\alpha}\boldsymbol{p}\boldsymbol{g}}(C)$
also depends on parameters $p_{\mu}\in[0,\,1]$ which we gather in
a vector $\boldsymbol{p}=(p_{1},\,...\,p_{d})^{T}$. The coordinates
$g_{\mu}\in\mathbb{R}_{+}$ of vector $\boldsymbol{g}=(g_{1},\,...\,g_{d})^{T}$
are just positive auxiliary factors, and $\overline{\overline{\mathbf{D}}}$
is a regular diffusivity tensor.

Actually, fractional derivatives w.r.t. $x_{\mu}$ can be viewed as
regular derivatives w.r.t. this variable compounded with fractional
integrals of the form $I_{x_{\mu}\pm}^{\gamma}$, often thought of
as integro-differential operators of negative order $-\gamma$. The
fractional integrals are convolutions whose kernel is a Dirac mass
at point $0$ if $\gamma=0$, or $(x_{\mu})_{\pm}^{\gamma-1}/\Gamma(\gamma)$
if $\gamma>0$: subscript $\pm$ designates positive or negative part
and $\Gamma$ is the Euler Gamma function defined by $\Gamma(x)=\int_{0}^{\infty}t^{x-1}e^{-t}dt$
\citep{Samko-Kilbas-Marichev_Book1993}. With these notations, the
vector $\boldsymbol{\mathcal{F}}^{\boldsymbol{\alpha}\boldsymbol{p}\boldsymbol{g}}(C)$
is defined by its components $\mathcal{F}_{\mu}^{\alpha_{\mu}p_{\mu}g_{\mu}}(C)$
in $\{\mathbf{b}_{\mu}\}$ basis 
\begin{equation}
\boldsymbol{\mathcal{F}}^{\boldsymbol{\alpha}\boldsymbol{p}\boldsymbol{g}}(C)=\sum_{\mu=1}^{d}\mathcal{F}_{\mu}^{\alpha_{\mu}p_{\mu}g_{\mu}}(C)\mathbf{b}_{\mu},\quad\mathcal{F}_{\mu}^{\alpha_{\mu}p_{\mu}g_{\mu}}(C)=\frac{\partial}{\partial x_{\mu}}\left[p_{\mu}I_{x_{\mu}+}^{2-\alpha_{\mu}}(g_{\mu}C)+(1-p_{\mu})I_{x_{\mu}-}^{2-\alpha_{\mu}}(g_{\mu}C)\right],\label{eq:Fg}
\end{equation}
in which $p_{\mu}$ and $1-p_{\mu}$ weight the two integrals $I_{x_{\mu}\pm}^{2-\alpha_{\mu}}$.
Hence, $\alpha_{\mu}$ sums up all integro-differential orders in
the contribution of $\mathcal{F}_{\mu}^{\alpha_{\mu}p_{\mu}g_{\mu}}(C)\mathbf{b}_{\mu}$
to $\boldsymbol{\nabla}\cdot\overline{\overline{\mathbf{D}}}(\mathbf{x})\boldsymbol{\mathcal{F}}^{\boldsymbol{\alpha}\boldsymbol{p}\boldsymbol{g}}(C)$,
which writes $\sum_{\nu=1}^{d}\partial(D_{\nu\mu}\mathcal{F}_{\mu}^{\alpha_{\mu}p_{\mu}g_{\mu}}(C))/\partial x_{\nu}$.
Since $I_{x_{\mu}+}^{0}$ and $I_{x_{\mu}-}^{0}$ coincide with operator
Identity ($\mathrm{Id}$), we immediately see that Eq. (\ref{eq:FADE_moreGeneral})
is the classical ADE for all values of $\boldsymbol{p}$ in the limit
case $\boldsymbol{\alpha}=\boldsymbol{2}=(2,\,...,\,2)$. Transport
phenomena deviating from this classical paradigm and reported in \citep{Clarke_etal_GroundWat2005,Wheatcraft-Tyler_WRR1988,Metzler-Klafter_JPA2004,Clark_etal_Ecology1999,Fedotov-Iomin_PRL2007,Deng_etal_JHE2004,Benson_etal_TiPM2001,Benson_etal_WRR2000,Schumer_etal_WRR2001,Kelly_WRR019748}
are better accounted by $\boldsymbol{\alpha}\neq\boldsymbol{2}$.
For $\gamma>0$ the convolution that defines $I_{x_{\mu}+}^{\gamma}C(\mathbf{x},\,t)$
results from integration over interval $\omega_{+}(\mathbf{x},\,\mu)=\{\mathbf{y}\in\Omega/y_{\nu}=x_{\nu}\,\,\mbox{for}\,\,\nu\neq\mu,\,y_{\mu}<x_{\mu}\}$
ending at point $\mathbf{x}$ and parallel to $\mathbf{b}_{\mu}$.
In other words, the elements of this interval have exactly the same
coordinates as $\mathbf{x}$ except for the coordinate of rank $\mu$
along which the integration is carried out. The other integral $I_{x_{\mu}-}^{\gamma}C$
corresponds to the opposite interval $\omega_{-}(\mathbf{x},\,\mu)=\{\mathbf{y}\in\Omega/y_{\nu}=x_{\nu}\,\,\mbox{for}\,\,\nu\neq\mu,\,y_{\mu}>x_{\mu}\}$,
and the complete definition of $I_{x_{\mu}\pm}^{\gamma}$ is 
\begin{equation}
(I_{x_{\mu}+}^{\gamma}C)(\mathbf{x})=\frac{1}{\Gamma(\gamma)}\int_{-\infty}^{x_{\mu}}\frac{1_{\omega_{+}(\mathbf{x},\,\mu)}(\mathbf{y})C(\mathbf{y})}{(x_{\mu}-y_{\mu})^{1-\gamma}}dy_{\mu},\qquad\mbox{and}\qquad(I_{x_{\mu}-}^{\gamma}C)(\mathbf{x})=\frac{1}{\Gamma(\gamma)}\int_{x_{\mu}}^{+\infty}\frac{1_{\omega_{-}(\mathbf{x},\,\mu)}(\mathbf{y})C(\mathbf{y})}{(y_{\mu}-x_{\mu})^{1-\gamma}}dy_{\mu}.\label{eq:Def_FracInt}
\end{equation}
In Eq. (\ref{eq:Def_FracInt}), $1_{E}$ represents the set function
of any subset $E$ of $\Omega$, i.e. $1_{E}(\mathbf{x})=1$ for $\mathbf{x}\in E$
and $1_{E}(\mathbf{x})=0$ for $\mathbf{x}\notin E$. Here we more
especially consider the domain $\Omega=\Pi_{\mu=1}^{d}]\ell_{\mu},\,L_{\mu}[$,
and the two above integrals correspond to intervals $]\ell_{\mu},\,x_{\mu}[$
and $]x_{\mu},\,L_{\mu}[$. For instance, if we assume $d=3$ and
$\mu=1$, Eqs (\ref{eq:Def_FracInt}) write 
\begin{equation}
(I_{x_{1}+}^{\gamma}C)(\mathbf{x})=\frac{1}{\Gamma(\gamma)}\int_{\ell_{1}}^{x_{1}}\frac{C(y_{1},\,x_{2},\,x_{3})}{(x_{1}-y_{1})^{1-\gamma}}dy_{1},\qquad\mbox{and}\qquad(I_{x_{1}-}^{\gamma}C)(\mathbf{x})=\frac{1}{\Gamma(\gamma)}\int_{x_{1}}^{L_{1}}\frac{C(y_{1},\,x_{2},\,x_{3})}{(y_{1}-x_{1})^{1-\gamma}}dy_{1}.\label{eq:Def_FracIntspe}
\end{equation}

The objective of this paper is to propose a LBM scheme applicable
to tensor $\overline{\overline{\mathbf{D}}}$ and vector $\boldsymbol{g}$
allowed to depend on $\mathbf{x}$ because this includes a variety
of configurations considered by \citep{Yong_etal_JSP2006,Meerschaert-Sikorskii_Book2012}
for $\boldsymbol{\alpha}\neq\boldsymbol{2}$ and by \citep{Delay_etal_VZJ2005,Ginzburg_AdWR2005a}
in the case of the ADE. Parameters $\boldsymbol{\alpha}$ and $\boldsymbol{p}$
are nevertheless assumed constant in time and space. The definition
(\ref{eq:Fg}) of $\boldsymbol{\mathcal{F}}^{\boldsymbol{\alpha}\boldsymbol{p}\boldsymbol{g}}(C)$
takes the same form when $\overline{\overline{\mathbf{D}}}$ and $\boldsymbol{g}$
depend on $\mathbf{x}$ or not. Moreover its structure is especially
well adapted to the design of LBM schemes approximating Eq. (\ref{eq:FADE_moreGeneral}),
and very comfortable for coding. This is why we prefer the formulation
Eq. (\ref{eq:Fg}) to equivalent expressions of $\boldsymbol{\mathcal{F}}^{\boldsymbol{\alpha}\boldsymbol{p}\boldsymbol{g}}(C)$
exhibiting fractional derivatives of Riemann-Liouville type \citep{Samko-Kilbas-Marichev_Book1993}
defined by 
\begin{equation}
\frac{\partial^{\alpha'}C}{\partial_{\pm}x_{\mu}^{\alpha'}}(\mathbf{x})=\pm\frac{\partial}{\partial x_{\mu}}(I_{x_{\mu}\pm}^{1-\alpha'}C)(\mathbf{x})\quad\mbox{for}\quad\alpha'\in]0,\,1[,\qquad\frac{\partial^{\alpha'}C}{\partial_{\pm}x_{\mu}^{\alpha'}}(\mathbf{x})=\frac{\partial^{2}}{\partial x_{\mu}^{2}}(I_{x_{\mu}\pm}^{2-\alpha'}C)(\mathbf{x})\quad\mbox{for}\quad\alpha'\in]1,\,2[.\label{eq:Def_FracDer_plus}
\end{equation}
For example, Eq. (\ref{eq:FADE_moreGeneral}) writes 
\begin{equation}
\frac{\partial C}{\partial t}(\mathbf{x},\,t)+\boldsymbol{\nabla}\cdot(\mathbf{u}C)(\mathbf{x},\,t)=\sum_{\mu=1}^{d}D_{\mu\mu}\left[p_{\mu}\frac{\partial^{\alpha_{\mu}}g_{\mu}C}{\partial_{+}x_{\mu}^{\alpha_{\mu}}}+(1-p_{\mu})\frac{\partial^{\alpha_{\mu}}g_{\mu}C}{\partial_{-}x_{\mu}^{\alpha_{\mu}}}\right]+S_{c}(\mathbf{x},\,t)\label{eq:FADE_moreGeneral3}
\end{equation}
when tensor $\overline{\overline{\mathbf{D}}}$ is diagonal and spatially
homogeneous, but this formulation becomes heavier than Eqs (\ref{eq:FADE_moreGeneral})-(\ref{eq:Fg})
in more general cases. Moreover, we assume non-dimensional variables
and parameters in Eqs (\ref{eq:FADE_moreGeneral}) and (\ref{eq:Fg}).

Eq. (\ref{eq:FADE_moreGeneral}) is more than just a model for solute
transport. It rules the evolution of the probability density function
(p.d.f) of a wide set of stochastic processes \citep{Yong_etal_JSP2006}
called stable. Stable processes include finite or infinite variance
and are more general than Brownian motion which corresponds to the
particular case $\boldsymbol{\alpha}=\boldsymbol{2}$. They are related
to stable probability laws which deserve the attention of physicists
because they are attractors (for limits of sums of independent identically
distributed random variables) \citep{Kyprianou_Book2006}. Moreover,
experimental techniques (not restricted to concentrations of particles)
document individual trajectories of animals \citep{Metzler-Klafter_JPA2004}
or characteristic functions of molecular displacements \citep{Guillon_etal_PRE2013}
measured by Pulse Field Gradient Nuclear Magnetic Resonance. They
reveal stable motion in different fields of Science such as biology
and fluid dynamics \citep{Neel_etal_PRE2014}. More specifically,
the latter reference points stable motion at small scale in water
flowing through packed grains exhibiting sharp edges, a material that
looks homogeneous: heterogeneity of flow inside pores might explain
anomalous transport in this case. Though multidimensional stable motions
are more diverse \citep{Meerschaert-Sikorskii_Book2012,Zoia_etal_PRE2007}
we just mention here those which have independent stable projections
on the $\mathbf{b}_{\mu}$: their density satisfies Eq. (\ref{eq:FADE_moreGeneral})
equipped with spatially homogeneous $\overline{\overline{\mathbf{D}}}$
tensor diagonal in this basis. Moreover, the Riemann-Liouville derivatives
included in the definition (\ref{eq:Fg}) of $\boldsymbol{\mathcal{F}}^{\boldsymbol{\alpha}\boldsymbol{p}\boldsymbol{g}}$
actually represent the particle fluxes generated by such stable process
\citep{Neel_etal_JPA2007}. This designates these derivatives as privileged
elements of this vector beside variants discussed in \citep{Baeumer_etal_JCOAM2018}
and in Subsection \ref{sub:Variant} of the present paper.

The many possible applications of Eq. (\ref{eq:FADE_moreGeneral})
motivate simulation efforts. Finite difference/volume/element schemes
are available \citep{Gorenflo_etal_ChemPhys2002,Meerschaert_etal_JCP2006,Fix-Roop_CAMWA2004},
and particle tracking is a powerful method \citep{Yong_etal_JSP2006}
based on the tight relationship to stable processes \citep{Meerschaert-Sikorskii_Book2012}.
Yet, the computing time requested by these methods (even the latter)
causes that increasing the space dimension $d$ enhances the attractivity
of the LBM. The latter method simulates evolution equations by considering
an ensemble of fictitious particles whose individual velocities belong
to a discrete set, namely the elementary grid of a lattice. More specifically,
the densities of populations experiencing each of these velocities
evolve according to Boltzmann Equation modeling displacements interspersed
with instantaneous collisions. Adapting the collision rules causes
that the total density satisfies the p.d.e which we want to simulate.
This method was applied to equations of the form of Eq. (\ref{eq:FADE_moreGeneral})
in the very particular case of the ADE (i.e. $\boldsymbol{\alpha}=\boldsymbol{2}$)
\citep{ServanCamas-Tsai_AdWR2008,Yoshida-Nagaoka_JCP2010,Ginzburg_AdWR2005a}.
It was also applied to the Cahn-Hilliard equation \citep{Huang_etal_IJNMF2009,Fakhari-Rahimian_PRE2010}
similar to Eq. (\ref{eq:FADE_moreGeneral}), with $\boldsymbol{\mathcal{F}}^{\boldsymbol{\alpha}\boldsymbol{p}\boldsymbol{g}}$
and $\overline{\overline{\mathbf{D}}}$ replaced respectively by the
gradient of the chemical potential and the mobility coefficient. LBM
schemes proposed in these references use equilibrium functions that
split into two items: the first one represents the convective part
$\mathbf{u}C$ of the flux, and is standard in ADE literature (see
\citep{Yoshida-Nagaoka_JCP2010}). The second item must be adapted
to the aim of the simulation, and indeed the reference \citep{Zhou_etal_PRE2016}
demonstrates for the one-dimensional Eq. (\ref{eq:FADE_moreGeneral})
a LBM scheme based on Bhatnagar-Gross-Krook (BGK) collision rule \citep{BGK_PR1954}.
Here we have in view (\emph{i}) higher dimension, (\emph{ii}) the
effect of advection which causes instabilities better damped by LBM
schemes equipped with Multiple-Relaxation-Times (MRT) collision rule
and (\emph{iii}) possible anisotropy that BGK collision does not account.

The aim of this paper is to present an accurate and efficient MRT
LBM that solves the two- and three-dimensional Eq. (\ref{eq:FADE_moreGeneral})
for greater flow velocity than BGK and for diffusion coefficient that
can be a tensor. The method should moreover adapt to anisotropic diffusion
tensor ($D_{\mu\nu}\neq0$), not necessarily uniform. We briefly present
in Section \ref{sec:LBM-FADE3D} the principle of LBM schemes with
BGK or MRT collision operators adapted to the multidimensional fractional
ADE Eq. (\ref{eq:FADE_moreGeneral}) and satisfying these requirements.
In Section \ref{sec:Validations}, the accuracy of these schemes is
discussed by comparing \emph{i)} with random walk approximations and
\emph{ii)} with exact solutions available for very specific $S_{c}$
and $\boldsymbol{p}$. Comparisons \emph{ii)} include a variant of
vector $\boldsymbol{\mathcal{F}}^{\boldsymbol{\alpha}\boldsymbol{p}\boldsymbol{g}}(C)$
defined by Caputo derivatives to which we easily adapt the LBM scheme.

\section{\label{sec:LBM-FADE3D}Lattice Boltzmann schemes for Eq. (\ref{eq:FADE_moreGeneral})}

The LBM simulates $d$-dimensional evolution equations by considering
the total density of an ensemble of fictitious particles. Their individual
velocities belong to the elementary grid $\{\mathbf{e}_{i}\Delta x\slash\Delta t,\,i=0,\,...,\,\mathcal{N}\}$
of a lattice of $\mathbb{R}^{d}$, $\Delta t$ and $\Delta x$ being
time- and space-steps. For $i=0,\,...,\,\mathcal{N}$, $f_{i}(\mathbf{x},\,t)$
is the distribution function of the population evolving at velocity
$\mathbf{e}_{i}\Delta x/\Delta t$ at time $t$. The Boltzmann Equation
evolves these partial densities: more specifically each $f_{i}(\mathbf{x},\,t)$
is subjected to the translation of amplitude $\Delta x\mathbf{e}_{i}$
during successive time intervals of duration $\Delta t$ separated
by instantaneous collisions which tend to let the vector $\left|\boldsymbol{f}\right\rangle =(f_{0},\,...,\,f_{\mathcal{N}})^{T}$
relax to the vector $\left|\boldsymbol{f}^{eq}\right\rangle =(f_{0}^{eq},\,...,\,f_{\mathcal{N}}^{eq})^{T}$
called equilibrium distribution function. The BGK collision rule assumes
one relaxation rate $\lambda$, in fact a non-dimensional parameter
determined by the generalized diffusivity of the p.d.e which we want
to simulate. With more freedom degrees, the MRT collision rule gives
us the opportunity of better controlling instabilities. BGK and MRT
LBMs differ in the form of their collision operator which depends
on relaxation rates and is applied to the deviation between distribution
function and equilibrium function.

Prescribing a Lattice Boltzmann scheme is tantamount to specify velocity
lattice, collision operator, initial and boundary conditions. We briefly
describe simple choices of such elements whose combination returns
approximations of Eq. (\ref{eq:FADE_moreGeneral}), here associated
with Dirichlet boundary conditions in $\Omega=\Pi_{\mu=1}^{d}]\ell_{\mu},\,L_{\mu}[$.

\subsection{\label{sub:Definition-of-lattices}Minimal lattices and auxiliary
vectors of $\mathbb{R}^{\mathcal{N}+1}$ used to solve Eq. (\ref{eq:FADE_moreGeneral})}

\begin{figure}
\begin{centering}
\subfloat[\label{fig:LatticeD3Q7}Three-dimensional lattice D3Q7.]{\protect

\protect\begin{centering}
\protect\includegraphics[scale=0.35]{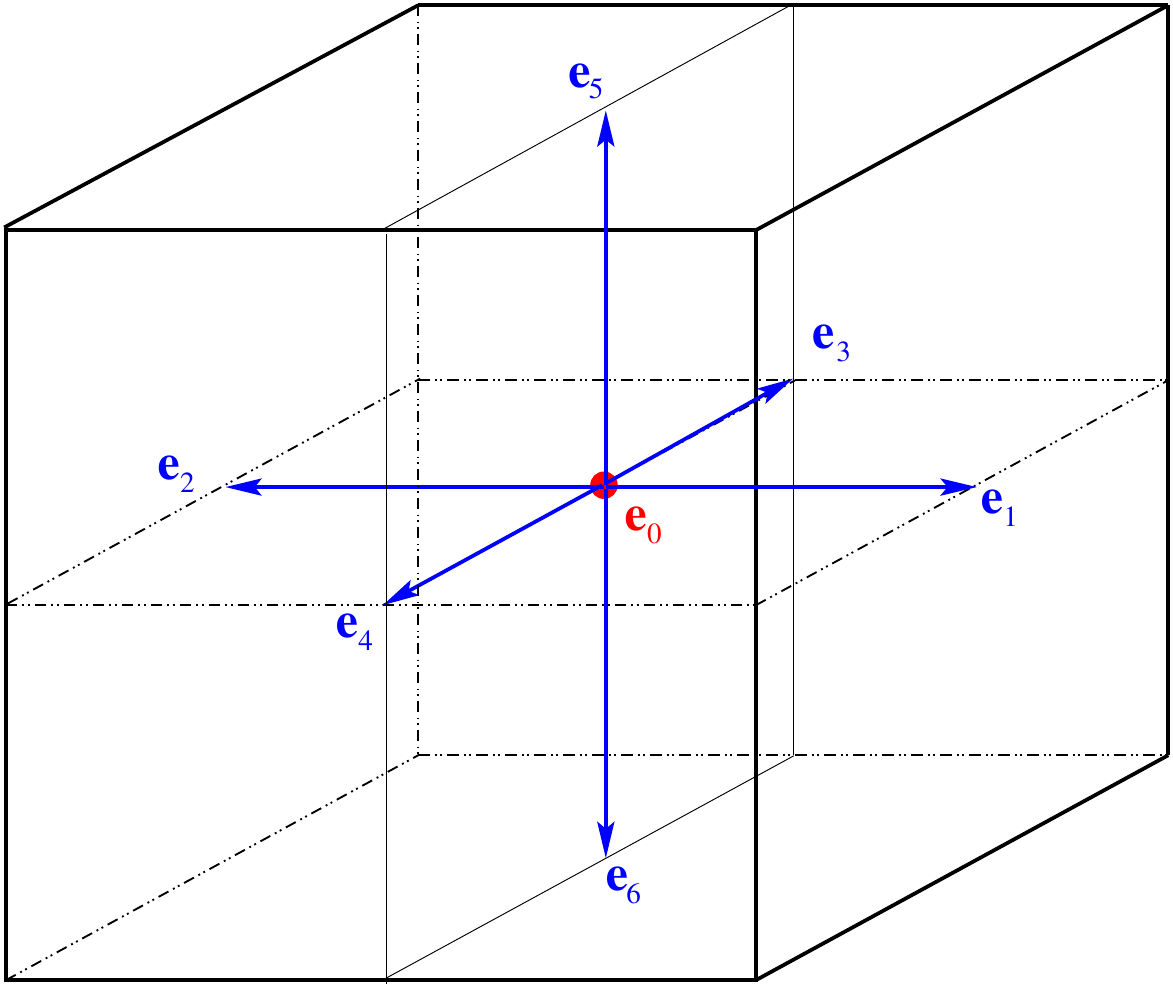}\protect
\par\end{centering}

}$\qquad\qquad$\subfloat[\label{fig:LatticeD2Q5}Two-dimensional lattice D2Q5.]{\protect\begin{centering}
\protect\includegraphics[scale=0.45]{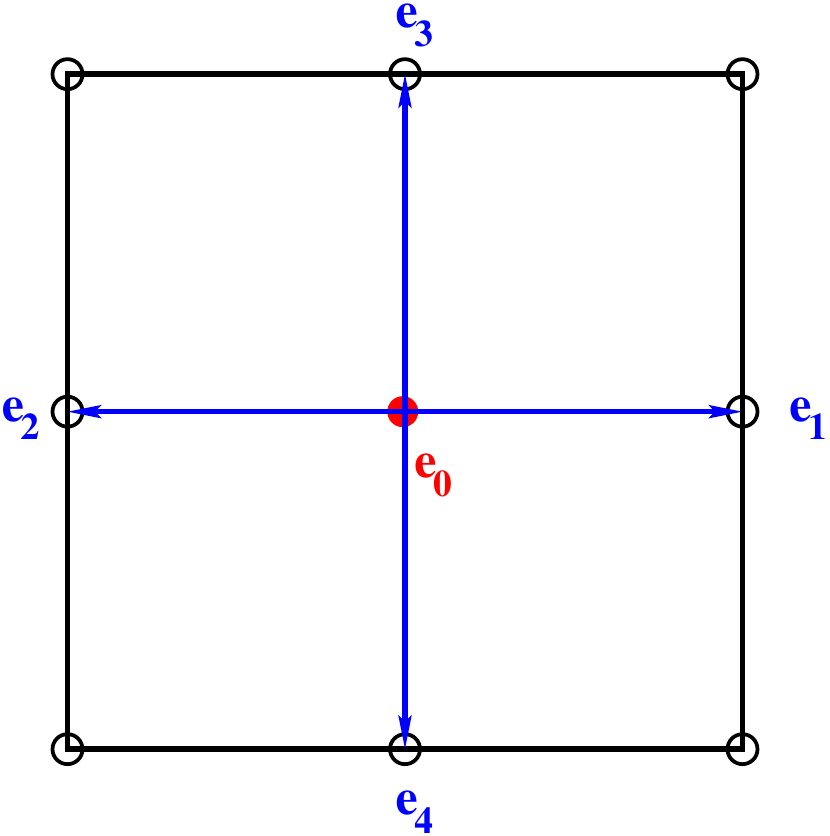}\protect
\par\end{centering}

}
\par\end{centering}

\protect\caption{Lattices considered in this work. On the left: D3Q7. On the right:
D2Q5. The weights of D3Q7 are are $w_{0}=1/4$ and $w_{1,...,6}=1/8$,
and its lattice coefficient is $e^{2}=1/4$.The weights of D2Q5 are
are $w_{0}=1/3$ and $w_{1,...,4}=1/6$, and its lattice coefficient
is $e^{2}=1/3$.}
\label{fig:Lattice} 
\end{figure}

The simplest choices of  velocity lattices in three/two dimensions
are the centered cubic/square D3Q7/D2Q5 represented on Fig. \ref{fig:Lattice}.
The three-dimensional lattice D3Q7 is composed of $\mathcal{N}+1=7$
vectors $\mathbf{e}_{i}$ (of $\mathbb{R}^{d}$), $i=0,\,...,\,\mathcal{N}$.
Their coordinates in $\{\mathbf{b}_{\mu}\}$ basis can be viewed as
the $\mathcal{N}+1$ columns of array $\mathbf{e}=[\mathbf{e}_{0},\,...,\,\mathbf{e}_{\mathcal{N}}]$,
in which we call $\bigl\langle e_{\mu}\bigr|$ the row of rank $\mu$,
with $\mu=1,\,2,\,3$. The elements of lattice D3Q7 are $\mathbf{e}_{0}=(0,\,0,\,0)^{T}$,
$\mathbf{e}_{1}=(1,\,0,\,0)^{T}$, $\mathbf{e}_{2}=(-1,\,0,\,0)^{T}$,
$\mathbf{e}_{3}=(0,\,1,\,0)^{T}$, $\mathbf{e}_{4}=(0,\,-1,\,0)^{T}$,
$\mathbf{e}_{5}=(0,\,0,\,1)^{T}$ and $\mathbf{e}_{\mathcal{N}=6}=(0,\,0,\,-1)^{T}$.
In two dimensions we obtain D2Q5 by just skipping $\mathbf{e}_{5}$
and $\mathbf{e}_{6}$, with of course $\mathcal{N}=4$. The definition
of the equilibrium function will need vector $\left|\mathbf{w}\right\rangle =(w_{0},\,...,\,w_{\mathcal{N}})^{T}$
whose entries are positive non-dimensional weights satisfying 
\begin{equation}
\langle\mathbf{1}|\mathbf{w}\rangle=1,\quad\langle e_{\mu}|\mathbf{w}\rangle=0,\quad\langle e_{\mu}e_{\nu}|\mathbf{w}\rangle=e^{2}\delta_{\mu\nu}\,\,\mbox{for }\mu,\nu\in\left\{ 1,\,...,\,d\right\} \label{eq:weights_c}
\end{equation}
where $\delta_{\mu\nu}$ is the Kronecker symbol, $e^{2}$ is a positive
(non-dimensional) coefficient attached to the lattice, and $\left\langle \mathbf{1}\right|=(1,\,...,\,1)$.
Note that $\left\langle \mathbf{a}\right|$ and $\left|\mathbf{a}\right\rangle $
respectively represent a row vector of $\mathbb{R}^{\mathcal{N}+1}$
and its transpose (a column), $\bigl\langle\bigr|\bigr\rangle$ standing
for the Euclidean scalar product of $\mathbb{R}^{\mathcal{N}+1}$.
The caption of Figs. \ref{fig:LatticeD3Q7}-\ref{fig:LatticeD2Q5}
documents weights and lattice coefficients attached to D3Q7 and D2Q5
and satisfying (\ref{eq:weights_c}) in which we see vector $\langle e_{\mu}e_{\nu}|$
whose entries are the products of those of $\langle e_{\mu}|$ and
$\langle e_{\nu}|$. Such vectors arise from the Taylor expansion
that introduces the derivatives of $\left|\boldsymbol{f}\right\rangle $.

\subsection{\label{sub:EquilibriumDistributionFunction}Equilibrium distribution
function adapted to Eq. (\ref{eq:FADE_moreGeneral}) and BGK collision}

For each $t=m\Delta t$ the Boltzmann Equation with BGK collision
is (for $i=0,\,...,\,\mathcal{N}$): 
\begin{align}
f_{i}(\mathbf{x}+\mathbf{e}_{i}\Delta x,\,t+\Delta t) & =f_{i}(\mathbf{x},\,t)-\frac{1}{\lambda}\left[f_{i}(\mathbf{x},\,t)-f_{i}^{eq}(\mathbf{x},\,t)\right]+w_{i}S_{c}(\mathbf{x},\,t)\Delta t,\label{eq:LBE_FADE}
\end{align}
where $S_{c}(\mathbf{x},\,t)$ is the source term of Eq. (\ref{eq:FADE_moreGeneral})
while the\textcolor{red}{{} }$f_{i}^{eq}$ are the entries of a vector
called equilibrium distribution function and noted $\bigl|\boldsymbol{f}^{eq}\bigr\rangle$.
The coefficient $\lambda$ is the relaxation rate which relaxes $f_{i}$
towards the equilibrium $f_{i}^{eq}$. The collision operator $\left|\boldsymbol{f}\right\rangle \mapsto\lambda^{-1}\left[\left|\boldsymbol{f}\right\rangle -\left|\boldsymbol{f}^{eq}\right\rangle \right]$
must preserve the total mass, a requirement equivalent to $\sum_{i}f_{i}=\sum_{i}f_{i}^{eq}$.
If we set $\boldsymbol{\Lambda}(\mathbf{x})=\lambda^{-1}(\mathbf{x})\mathbf{I}$
($\mathbf{I}$ being the $\mathcal{N}+1$-dimensional identity matrix),
the system of all equations Eq. (\ref{eq:LBE_FADE}) writes

\begin{subequations}
\begin{equation}
\mathbb{T}\bigl|\boldsymbol{f}(\mathbf{x},\,t+\Delta t)\bigr\rangle=\bigl|\boldsymbol{f}(\mathbf{x},\,t)\bigr\rangle-\boldsymbol{\Lambda}(\mathbf{x})\left[\bigl|\boldsymbol{f}(\mathbf{x},\,t)\bigr\rangle-\bigl|\boldsymbol{f}^{eq}(\mathbf{x},\,t)\bigr\rangle\right]+\bigl|\mathbf{w}\bigr\rangle S_{c}(\mathbf{x},\,t)\Delta t\label{eq:LBE_Ket}
\end{equation}
where operator $\mathbb{T}$ accounts for the translations experienced
by the fictitious particles of microscopic displacement vectors $\mathbf{e}_{0}$,~...,~$\mathbf{e}_{\mathcal{N}}$
between two successive collision steps, according to
\begin{equation}
\mathbb{T}\bigl|\boldsymbol{f}(\mathbf{x},\,t)\bigr\rangle=\left(f_{0}(\mathbf{x},\,t),\,f_{1}(\mathbf{x}+\mathbf{e}_{1}\Delta x,\,t),\,...,\,f_{\mathcal{N}}(\mathbf{x}+\mathbf{e}_{\mathcal{N}}\Delta x,\,t)\right)^{T}\label{eq:Def_Translation}
\end{equation}

\end{subequations}

Though partial derivatives are absent from Eqs (\ref{eq:LBE_FADE})
and (\ref{eq:LBE_Ket})-(\ref{eq:Def_Translation}), they arise when
we plug the Taylor expansion of $\mathbb{T}\left|\boldsymbol{f}(\mathbf{x},\,t+\Delta t)\right\rangle -\left|\boldsymbol{f}(\mathbf{x},\,t)\right\rangle $
into Eq. (\ref{eq:LBE_Ket}). A standard reasoning of LB literature
applies a multiple scale procedure which assumes that there exists
a small parameter $\varepsilon$ and a change of variables for $\mathbf{x}$
and $t$ defined by 
\begin{equation}
\frac{\partial}{\partial t}=\varepsilon\frac{\partial}{\partial t_{1}}+\varepsilon^{2}\frac{\partial}{\partial t_{2}}\quad\mathrm{and}\quad\frac{\partial}{\partial x_{\mu}}=\varepsilon\frac{\partial}{\partial x'_{\mu}}\quad\mathrm{for}\quad\mu=1,\,...,\,d.\label{eq:multiscale}
\end{equation}
This reasoning also assumes bounded derivatives w.r.t. the new variables
($t_{1}$, $t_{2}$, $x'_{\mu}$) for all items $\bigl|\boldsymbol{f}^{(k)}\bigr\rangle$
of the power expansion $\bigl|\boldsymbol{f}\bigr\rangle=\sum_{k=0}^{+\infty}\varepsilon^{k}\bigl|\boldsymbol{f}^{(k)}\bigr\rangle$.
Collecting items of the order of $\varepsilon^{k}$ for successive
integer $k$ returns a sequence of equations among which $C=\langle\mathbf{1}|\boldsymbol{f}^{(0)}\rangle$
approximately solves Eq. (\ref{eq:FADE_moreGeneral}) in several particular
cases (see e.g. \citep{Walsh-Saar_WRR2010} and \citep{Yoshida-Nagaoka_JCP2010}
for the ADE and \citep{Cartalade_etal_CAMWA2016,Younsi-Cartalade_JCP2016}
for the phase-field models of crystallization) provided the equilibrium
function $\left|\boldsymbol{f}^{eq}\right\rangle $ is appropriately
chosen.

\ref{sec:asp} proves that the LB equation approximates Eq. (\ref{eq:FADE_moreGeneral})
equipped with general $\boldsymbol{\alpha}$ and spherical tensor
$\overline{\overline{\mathbf{D}}}=D\overline{\overline{\mathbf{Id}}}$
($\overline{\overline{\mathrm{\mathbf{Id}}}}$ being the Identity
of $\mathbb{R}^{d}$) when the relaxation rate $\lambda$ is related
to the diffusion coefficient $D$ by 
\begin{equation}
D=e^{2}\left(\lambda-\frac{1}{2}\right)\frac{\Delta x^{2}}{\Delta t}\label{eq:Diffusion_Coefficient_3D}
\end{equation}
and the moments of zeroth-, first- and second-order of the equilibrium
distribution function $|\boldsymbol{f}^{eq}\rangle$ satisfy

\begin{subequations} 
\begin{align}
\langle\mathbf{1}|\boldsymbol{f}^{eq}\rangle & =C\label{eq:M0_FADE_3D}\\
\langle e_{\mu}|\boldsymbol{f}^{eq}\rangle & =\frac{\Delta t}{\Delta x}Cu_{\mu}\label{eq:M1_FADE_3D}\\
\langle e_{\mu}e_{\nu}|\boldsymbol{f}^{eq}\rangle & =e^{2}\left[p_{\mu}I_{x_{\mu}+}^{2-\alpha_{\mu}}(g_{\mu}C)+(1-p_{\mu})I_{x_{\mu}-}^{2-\alpha_{\mu}}(g_{\mu}C)\right]\delta_{\mu\nu}.\label{eq:M2_FADE_3D}
\end{align}

\end{subequations}

The equilibrium distribution function

\begin{subequations}\textcolor{red}{{} }
\begin{equation}
\mathbf{|}\boldsymbol{f}^{eq}(\mathbf{x},\,t)\rangle=\left|\boldsymbol{\mathcal{A}}(\mathbf{x},\,t)\right\rangle +\frac{C(\mathbf{x},\,t)}{e^{2}}\frac{\Delta t}{\Delta x}u_{\mu}(\mathbf{x},\,t)\bigl|e_{\mu}\mathbf{w}\bigr\rangle\label{eq:Feq}
\end{equation}
satisfies Eqs (\ref{eq:M0_FADE_3D})-(\ref{eq:M2_FADE_3D}) if each
component $\mathcal{A}_{i}$ of $\left|\boldsymbol{\mathcal{A}}(\mathbf{x},\,t)\right\rangle $
is a functional of $C$ given by (for $\mu=1,\,...,\,d$) 
\begin{equation}
\mathcal{A}_{i}(\mathbf{x},\,t)=\begin{cases}
C(\mathbf{x},\,t)-w_{0}\sum_{\mu=1}^{d}\left[p_{\mu}I_{x_{\mu}+}^{2-\alpha_{\mu}}(g_{\mu}C)+(1-p_{\mu})I_{x_{\mu}-}^{2-\alpha_{\mu}}(g_{\mu}C)\right] & \mbox{if}\,\,i=0\\
w_{i}\left[p_{\mu}I_{x_{\mu}+}^{2-\alpha_{\mu}}(g_{\mu}C)+(1-p_{\mu})I_{x_{\mu}-}^{2-\alpha_{\mu}}(g_{\mu}C)\right] & \mbox{if}\,\,i=2(\mu-1)+1,\,2(\mu-1)+2
\end{cases}\label{eq:CoeffA_FADE3D}
\end{equation}
The second item of Eq. (\ref{eq:Feq}) is a standard element of LBM
schemes applied to classical ADE: it accounts for the advective term
$\nabla\cdot(\mathbf{u}C)$ of Eq. (\ref{eq:FADE_moreGeneral}). The
first item $\left|\boldsymbol{\mathcal{A}}(\mathbf{x},\,t)\right\rangle $
defined by (\ref{eq:CoeffA_FADE3D}) generalizes to higher dimensions
the equilibrium function proposed by \citep{Zhou_etal_PRE2016} for
the 1D case.

\end{subequations}

In the particular case $\boldsymbol{\alpha}=\mathbf{2}$, Eq. (\ref{eq:M2_FADE_3D})
writes $\langle e_{\mu}e_{\nu}|\boldsymbol{f}^{eq\,(0)}\rangle=C\delta_{\mu\nu}$,
and $\left|\boldsymbol{\mathcal{A}}(\mathbf{x},\,t)\right\rangle =C\left|\mathbf{w}\right\rangle $
retrieves the equilibrium function commonly used for the classical
ADE \citep{Yoshida-Nagaoka_JCP2010}. Since only the $\mathcal{A}_{i}(\mathbf{x},\,t)$
depend on $\boldsymbol{\alpha}$, simulating the general version of
Eq. (\ref{eq:FADE_moreGeneral}) in any D2Q5 or D3Q7 Lattice Boltzmann
code just requires plugging discrete fractional approximations of
integrals $I_{x_{\mu}\pm}^{\alpha_{\mu}}g_{\mu}C$ into the equilibrium
distribution function according to Eq. (\ref{eq:CoeffA_FADE3D}).
This deserves an algorithm that is described in subsection \ref{sub:Discrete-integrals}.

Eq. (\ref{eq:Diffusion_Coefficient_3D}) intimately links the unique
relaxation rate of the BGK LBM to spherical tensor $\overline{\overline{\mathbf{D}}}=D\overline{\overline{\mathbf{Id}}}$
where $D$ may depend on $\mathbf{x}$ and $t$. One relaxation rate
still accounts for non-spherical but diagonal $\overline{\overline{\mathbf{D}}}$
if instead of each $g_{\mu}$ in $I_{x_{\mu}\pm}^{\alpha_{\mu}}g_{\mu}C$
we set $h_{\mu}g_{\mu}$ with $h_{\mu}$ such that $D_{\mu\mu}=e^{2}\left(\lambda-\frac{1}{2}\right)h_{\mu}\Delta x^{2}/\Delta t$
provided $\overline{\overline{\mathbf{D}}}$ does not depend on $\mathbf{x}$.
Nevertheless, even if the latter condition is satisfied non-trivial
off-diagonal entries of $\overline{\overline{\mathbf{D}}}$ remain
excluded. This fosters us considering in Section \ref{sub:MRT-Description}
more flexible MRT collision rules still based on the e.d.f. defined
by Eqs (\ref{eq:Feq})-(\ref{eq:CoeffA_FADE3D}), but involving more
freedom degrees. These collision rules moreover help us damping instabilities
caused by large velocity $\mathbf{u}$ even with diagonal diffusion
tensor.

\subsection{\label{sub:MRT-Description}MRT collision operator}

The MRT collision operator \citep{Lallemand-Luo_PRE2000,dHumieres_etal_PhilTranRoySoc2002}
$\mathbf{M}^{-1}\boldsymbol{\Lambda}(\mathbf{x})\mathbf{M}\left[\left|\boldsymbol{f}(\mathbf{x},\,t)\right\rangle -\left|\boldsymbol{f}^{eq}(\mathbf{x},\,t)\right\rangle \right]$
includes invertible $(\mathcal{N}+1)\times(\mathcal{N}+1)$ matrices
$\mathbf{M}$ and $\boldsymbol{\Lambda}$, and here the above defined
equilibrium function is still used. The LB equation (\ref{eq:LBE_Ket})
is replaced by 
\begin{equation}
\mathbb{T}\bigl|\boldsymbol{f}(\mathbf{x},\,t+\Delta t)\bigr\rangle=\bigl|\boldsymbol{f}(\mathbf{x},\,t)\bigr\rangle-\mathbf{M}^{-1}\boldsymbol{\Lambda}(\mathbf{x})\mathbf{M}\left[\bigl|\boldsymbol{f}(\mathbf{x},\,t)\bigr\rangle-\bigl|\boldsymbol{f}^{eq}(\mathbf{x},\,t)\bigr\rangle\right]+\bigl|\mathbf{w}\bigr\rangle S_{c}(\mathbf{x},\,t)\Delta t,\label{eq:MRT-Coll}
\end{equation}
in which matrix $\mathbf{M}$ represents a change of basis. In MRT
collision rule the latter associates to each $\left|\boldsymbol{f}\right\rangle $
a set of $\mathcal{N}+1$ independent moments equivalent to scalar
products of $\left|\boldsymbol{f}\right\rangle $ by independent elements
of $\mathbb{R}^{\mathcal{N}+1}$ including $\left|\mathbf{1}\right\rangle $,
the $\bigl|e_{\mu}\bigr\rangle$ and combinations of products of these
vectors. In the new basis, the $d+1$ first new coordinates of $\left|\boldsymbol{f}\right\rangle $
are the total mass $\left\langle \mathbf{1}|\boldsymbol{f}\right\rangle $
(a conserved quantity) and the $\bigl\langle\boldsymbol{f}\bigr|e_{\mu}\bigr\rangle$,
proportional to fictive particle fluxes in the $d$ physical directions.
These $d+1$ first rows turn out to be mutually orthogonal. Reference
\citep{Yoshida-Nagaoka_JCP2010} suggests complementing them with
$\mathcal{N}-d$ rows associated to second order moments related to
kinetic energy $\left|\mathbf{e}^{2}\right\rangle =\sum_{\mu=1}^{d}\bigl|e_{\mu}e_{\mu}\bigr\rangle$
and to other linear combinations of the $\bigl|e_{\mu}e_{\mu}\bigr\rangle$
with $\mu<d$, mutually orthogonal and orthogonal to the $d+1$ first
rows. Any linear combination of these vectors is suitable if all rows
are mutually orthogonal. This requirement, justified by accumulated
experience, seems necessary to achieve numerical stability when $\mathbf{u}$
is different from zero. It is satisfied by several configurations
among which we choose the ones suggested by \citep{Yoshida-Nagaoka_JCP2010}:
\begin{equation}
\mathbf{M}=\left(\begin{array}{ccccccc}
1 & 1 & 1 & 1 & 1 & 1 & 1\\
0 & 1 & -1 & 0 & 0 & 0 & 0\\
0 & 0 & 0 & 1 & -1 & 0 & 0\\
0 & 0 & 0 & 0 & 0 & 1 & -1\\
6 & -1 & -1 & -1 & -1 & -1 & -1\\
0 & 2 & 2 & -1 & -1 & -1 & -1\\
0 & 0 & 0 & 1 & 1 & -1 & -1
\end{array}\right)\;\mathrm{or}\;\;\mathbf{M}=\left(\begin{array}{ccccc}
1 & 1 & 1 & 1 & 1\\
0 & 1 & -1 & 0 & 0\\
0 & 0 & 0 & 1 & -1\\
4 & -1 & -1 & -1 & -1\\
0 & 1 & 1 & -1 & -1
\end{array}\right),\label{eq:MatrixD2}
\end{equation}
for D3Q7 or D2Q5 respectively. The $\boldsymbol{\Lambda}(\mathbf{x})$
matrix associated to the D3Q7 lattice is defined by its inverse 
\begin{equation}
\boldsymbol{\Lambda}^{-1}(\mathbf{x})=\left(\begin{array}{ccccccc}
\lambda_{0}(\mathbf{x}) & 0 & 0 & 0 & 0 & 0 & 0\\
0 & \lambda_{11}(\mathbf{x}) & \lambda_{12}(\mathbf{x}) & \lambda_{13}(\mathbf{x}) & 0 & 0 & 0\\
0 & \lambda_{21}(\mathbf{x}) & \lambda_{22}(\mathbf{x}) & \lambda_{23}(\mathbf{x}) & 0 & 0 & 0\\
0 & \lambda_{31}(\mathbf{x}) & \lambda_{32}(\mathbf{x}) & \lambda_{33}(\mathbf{x}) & 0 & 0 & 0\\
0 & 0 & 0 & 0 & \lambda_{4}(\mathbf{x}) & 0 & 0\\
0 & 0 & 0 & 0 & 0 & \lambda_{5}(\mathbf{x}) & 0\\
0 & 0 & 0 & 0 & 0 & 0 & \lambda_{6}(\mathbf{x})
\end{array}\right)\label{eq:MatrixS}
\end{equation}
whose elements are relaxation rates. In dimension two with D2Q5, we
just skip elements $\lambda_{3\mu}$ and $\lambda_{\mu3}$ and replace
$\lambda_{5}$ and $\lambda_{6}$ by $\lambda_{3}$. In both cases,
we call $\overline{\overline{\boldsymbol{\lambda}}}$ the $d\times d$
matrix of elements $\lambda_{\mu\nu}$.

Assuming $\Delta t\sim\Delta x^{2}$, the reference \citep{Yoshida-Nagaoka_JCP2010}
proves that the moment of zeroth-order $C=\langle\mathbf{1}|\boldsymbol{f}\rangle$
deduced from Eqs (\ref{eq:MRT-Coll}), (\ref{eq:MatrixS}), or (\ref{eq:MatrixD2})
solves the anisotropic ADE (i.e. Eq. (\ref{eq:FADE_moreGeneral})
with $\boldsymbol{\alpha}=\mathbf{2}$) within an error of the order
of $\varepsilon^{2}$ provided diffusion tensor and relaxation parameters
satisfy 
\begin{equation}
\overline{\mathbf{\overline{D}}}(\mathbf{x})=e^{2}\left(\overline{\overline{\boldsymbol{\lambda}}}(\mathbf{x})-\frac{1}{2}\overline{\overline{\mathrm{\mathbf{Id}}}}\right)\frac{\Delta x^{2}}{\Delta t}\label{eq:D_ab}
\end{equation}
which generalizes Eq. (\ref{eq:Diffusion_Coefficient_3D}). \ref{sec:aspMRT}
extends this statement to general $\boldsymbol{\alpha}$. \textcolor{red}{{} }Eq.
(\ref{eq:D_ab}) relates the elements $\lambda_{\mu\nu}(\mathbf{x})$
of matrix $\boldsymbol{\Lambda}^{-1}(\mathbf{x})$ ($\mu,\,\nu=1,\,...,\,d$)
to the generalized diffusion tensor, and $\lambda_{0}(\mathbf{x})$
(applied on the conserved quantity) has no effect.\textcolor{red}{{}
}Section \ref{sub:Validation_Advection} demonstrates that the diagonal
elements $\lambda_{\mu}(\mathbf{x})$ with $\mu=d+1,\,...,\,\mathcal{N}$
can be viewed as additional freedom degrees influencing the stability
and the accuracy of the algorithm.

\subsection{\label{sub:Discrete-integrals}Computation of discrete integrals}

Updating the discrete equilibrium function at each time step requires
discrete fractional integrals, in BGK as well as in MRT setting. Accurately
discretizing the fractional integrals $I_{x_{\mu}\pm}^{2-\alpha_{\mu}}$
involved in the $\mathcal{A}_{i}(\mathbf{x},\,t)$ significantly improves
the efficiency of LBM schemes applied to Eq. (\ref{eq:FADE_moreGeneral}).
Discrete schemes are available for one-dimensional fractional integrals
of any continuous function $f(x)$ of a real variable $x\in[0,\,N\Delta x]$.
\textcolor{red}{{} }Approximations of the order of $\Delta x$ may be
sufficient in the one-dimensional case as in Ref. \citep{Zhou_etal_PRE2016}.
Since higher $d$ dimension fosters us avoiding too small mesh, we
disregard discrete algorithms using step functions to interpolate
$f(x)$, and prefer those of \citep{Diethelm_etal_CMAME2005} which
stem from the trapezoidal rule and return errors of the order of $\Delta x^{2}$: 

\begin{equation}
I_{n\Delta x+}^{\gamma}f\approx\frac{\Delta x^{\gamma}}{\Gamma(2+\gamma)}\sum_{l=0}^{n}f(l\Delta x)c_{ln}^{+}(\gamma)\quad\quad I_{n\Delta x-}^{\gamma}f\approx\frac{\Delta x^{\gamma}}{\Gamma(2+\gamma)}\sum_{l=n}^{N}f(l\Delta x)c_{ln}^{-}(\gamma,\,N).\label{eq:1Dminus_discret}
\end{equation}
These equations need the Gamma function for which an intrinsic Fortran
2008 function exists, and coefficients $c_{ln}^{+}(\gamma)$ and $c_{ln}^{-}(\gamma,\,N)$
are defined by:

\begin{subequations} 
\begin{equation}
c_{ln}^{+}(\gamma)=\begin{cases}
(1+\gamma)n^{\gamma}-n^{\gamma+1}+(n-1)^{\gamma+1} & \textrm{for}\,\,l=0,\\
(n-l+1)^{\gamma+1}-2(n-l)^{\gamma+1}+(n-l-1)^{\gamma+1} & \textrm{for}\,\,0<l<n,\\
1 & \textrm{for}\,\,l=n
\end{cases}\label{eq:c+}
\end{equation}
and: 
\begin{equation}
c_{ln}^{-}(\gamma,\,N)=\begin{cases}
(1+\gamma)(N-n)^{\gamma}-(N-n)^{\gamma+1}+(N-n-1)^{\gamma+1} & \textrm{for}\,\,l=N,\\
(l-n+1)^{\gamma+1}-2(l-n)^{\gamma+1}+(l-n-1)^{\gamma+1} & \textrm{for}\,\,n<l<N\\
1 & \textrm{for}\,\,l=n
\end{cases}\label{eq:c-}
\end{equation}

\end{subequations}

For $d>1$ and each $\mu$, we use Eq. (\ref{eq:1Dminus_discret})
in $(I_{x_{\mu}\pm}^{\gamma_{\mu}}C)(\mathbf{x})$ which is an integral
of the form $(I_{x_{\mu}\pm}^{\gamma_{\mu}}f)(x_{\mu})$ if we set
$f(x_{\mu})\equiv C(\mathbf{x})$ in which we fix all coordinates
of $\mathbf{x}$ of rank different from $\mu$. For each $\mathbf{x}=(x_{1},\,...,\,x_{d})$
belonging to the domain $\Omega=\Pi_{\mu=1}^{d}[\ell_{\mu},\,L_{\mu}=\ell_{\mu}+N_{\mu}\Delta x]$,
we set $x_{\mu}=\ell_{\mu}+n\Delta x$ and yield 

\begin{equation}
(I_{x_{\mu}+}^{\gamma_{\mu}}C)(\mathbf{x})=\frac{\Delta x^{\gamma_{\mu}}}{\Gamma(2+\gamma_{\mu})}\sum_{l=0}^{n}C(\mathbf{y}_{l})c_{ln}^{+}(\gamma_{\mu}),\quad\quad(I_{x_{\mu}-}^{\gamma_{\mu}}C)(\mathbf{x})=\frac{\Delta x^{\gamma_{\mu}}}{\Gamma(2+\gamma_{\mu})}\sum_{l=n}^{N_{\mu}}C(\mathbf{y}_{l})c_{ln}^{-}(\gamma_{\mu},\,N_{\mu})\label{eq:XXminus_discret}
\end{equation}
where $\mathbf{y}_{l}$ has the same meaning as $\mathbf{y}$ in Eq.
(\ref{eq:Def_FracInt}). For instance, in the particular case $d=3$,
$\mathbf{y}_{l}=(\ell_{1}+l\Delta x,\,x_{2},\,x_{3})$ if $\mu=1$,
$\mathbf{y}_{l}=(x_{1},\,\ell_{2}+l\Delta x,\,x_{3})$ if $\mu=2$
and $\mathbf{y}_{l}=(x_{1},\,x_{2},\,\ell_{3}+l\Delta x)$ if $\mu=3$.

Because fractional integrals are non-local, updating the equilibrium
function at each computing step needs the complete description of
the concentration field in $\Omega$. Parallel computing would need
specific programming effort in this case.

\subsection{\label{sub:BC-and-Algo}Boundary conditions and algorithm}

Here we consider homogeneous and non-homogeneous Dirichlet boundary
conditions at the boundary of domain $\Omega$ limited by hyperplanes
of inward normal unit vectors $\mathbf{e}_{b}$ with $b=1,\,...,\,\mathcal{N}$.
At each node $\mathbf{x}$ of such boundary, after the collision stage,
each displacement stage derives from Eq. (\ref{eq:Def_Translation})
all the components of the distribution function $\bigl|\boldsymbol{f}\bigr\rangle$
except $f_{b}$. This accounts for the possible asymmetry of the equilibrium
function since the collision stage has been performed. Imposing concentration
$C_{0}$ is equivalent to update the unknown function 
\begin{equation}
f_{b}=C_{0}-\sum_{i\neq b}^{\mathcal{N}}f_{i}.\label{eq:Dirichlet_BC}
\end{equation}
For instance, assuming $b=1$ with a three-dimensional D3Q7 lattice
at point $\mathbf{x}=(\ell_{1},\,x_{2},\,x_{3})^{T}$, the unknown
distribution function\textcolor{red}{{} }$f_{1}$ is given by $f_{1}=C_{0}-\sum_{i\neq1}^{\mathcal{N}}f_{i}=C_{0}-f_{0}-f_{2}-f_{3}-f_{4}-f_{5}-f_{6}$.
This method is also applied when the concentration $C_{0}$ varies
with position and time (see Section \ref{sub:Validation_AS}).

The main stages of the above described LBM scheme are summarized in
Algorithm \ref{alg:Steps-for-LBM-FADE3D}. Only the first two stages
of the time loop differ from standard LBM applied to classical ADE
because updating the equilibrium distribution function $\bigl|\boldsymbol{f}^{eq}\bigr\rangle$
requires fractional integrals $I_{x_{\mu}\pm}^{2-\alpha_{\mu}}(g_{\mu}C)$
discretized according to Section \ref{sub:Discrete-integrals}. Regarding
initializations, the third item is applicable to MRT LBM only. Parameters
$\lambda_{\mu}$ with $\mu=d+1,\,...,\,\mathcal{N}$ achieving stability
and accuracy of MRT LBM schemes approximating the fractional ADE may
belong to a subset of those adapted to the classical case $\boldsymbol{\alpha}=\boldsymbol{2}$.
Choosing these parameters was found necessary at large Péclet numbers
in the numerical experiments described in Section \ref{sec:Validations}.

\begin{algorithm}
Initializations 
\begin{enumerate}
\item Define time step $\Delta t$, space step $\Delta x$, moving vectors
$\mathbf{e}_{i}$, weights $w_{i}$ and lattice coefficient $e^{2}$. 
\item Being given $D$ (or $D_{\mu\nu}$), calculate the relaxation rates
$\lambda$ (or $\lambda_{\mu\nu}$) with Eq. (\ref{eq:Diffusion_Coefficient_3D})
(or Eq. (\ref{eq:D_ab})). 
\item Choose parameters $\lambda_{\mu}$ ($\mu=d+1,\,...,\,\mathcal{N}$)
in MRT case.
\item Read or define the initial conditions $C(\mathbf{x},\,0)$, and $\mathbf{u}(\mathbf{x},\,0)$.
\end{enumerate}
Start of time loop 
\begin{enumerate}
\item For each $\mu=1,\,...,\,d$, calculate the fractional term $\left[p_{\mu}I_{x_{\mu}+}^{2-\alpha_{\mu}}(g_{\mu}C)+(1-p_{\mu})I_{x_{\mu}-}^{2-\alpha_{\mu}}(g_{\mu}C)\right]$.
\item Calculate the e.d.f. $\bigl|\boldsymbol{f}^{eq}\bigr\rangle$ defined
by Eqs. (\ref{eq:Feq})-(\ref{eq:CoeffA_FADE3D}). This gives us the
right-hand side of Eqs (\ref{eq:LBE_Ket}) or (\ref{eq:MRT-Coll}) 
\item Apply operator $\mathbb{T}^{-1}$ (displacement) to this right-hand
side and get $\left|\boldsymbol{f}(\mathbf{x},\,t+\Delta t)\right\rangle $
\item Update the boundary conditions with Eq. (\ref{eq:Dirichlet_BC}).
\item Calculate the new concentration $C(\mathbf{x},\,t)=\bigl\langle\mathbf{1}\bigr|\boldsymbol{f}\bigr\rangle$. 
\end{enumerate}
End of time loop

\protect\caption{\label{alg:Steps-for-LBM-FADE3D}LBM algorithm for fractional ADE.}
\end{algorithm}

\section{\label{sec:Validations}LBM-FADE validations}

Numerical algorithms proposed to solve partial differential equations
can be checked by comparing with exact solutions, or with numerical
solutions issued of other approaches. Analytical solutions are available
for Eq. (\ref{eq:FADE_moreGeneral}) involving general $\boldsymbol{\alpha}$
provided the skewness parameter $\boldsymbol{p}$ takes the\textcolor{green}{{}
}value $\mathbf{1}$ or $\mathbf{0}$ with, moreover, specific source
term $S_{c}$ and initial data. Though Eq. (\ref{eq:FADE_moreGeneral})
with $S_{c}=0$ constitutes an important simple case, it does not
have exact solutions in bounded domains. However, in this case Eq.
(\ref{eq:FADE_moreGeneral}) equipped with spatially homogeneous coefficients
rules the evolution of the probability density function (p.d.f) of
a wide class of stochastic processes described in subsection \ref{subRW2D}.
Sampling sufficiently many trajectories of such process yields random
walk approximation of Eq. (\ref{eq:FADE_moreGeneral}). Subsections
\ref{sub:Validations_LBM-BGK} and \ref{sub:Validation_Advection}
compare issues of the LBM scheme described in Section \ref{sec:LBM-FADE3D}
with such random walks when $\mathbf{u}=\mathbf{0}$ and $\mathbf{u}\neq\mathbf{0}$
respectively. Subsection \ref{sub:Validation_AS} considers anisotropic
and space-dependent diffusion tensor such that Eq. (\ref{eq:FADE_moreGeneral})
has an exact solution which is compared with LBM simulation. Subsection
\ref{sub:Variant} shows that the simulation adapts to slightly different
(still fractional) $\boldsymbol{\mathcal{F}}^{\boldsymbol{\alpha}\boldsymbol{p}\boldsymbol{g}}(C)$
in Eq. (\ref{eq:FADE_moreGeneral}). Hence, exact solutions and random
walks help us to validate LBM schemes in complementary situations.

\subsection{\label{subRW2D}Fractional random walks }

Eq. (\ref{eq:FADE_moreGeneral}) rules the evolution of stochastic
processes in $\mathbb{R}^{d}$, and in bounded domains under some
conditions. Though comparisons with numerical LBM simulations correspond
to the latter case, we start with unconstrained random walks  and
briefly comment the role of the several parameters of Eq. (\ref{eq:FADE_moreGeneral}).

\subsubsection*{\label{subRW2Dinfi}Equation (\ref{eq:FADE_moreGeneral}) and random
walks in $\mathbb{R}^{d}$ }

Eq. (\ref{eq:FADE_moreGeneral}) with $\boldsymbol{\alpha}=\mathbf{2}$
is the ADE, and random walks are commonly used to simulate the solutions
of this equation. To recall the principle of this method we assume
for simplicity a diagonal tensor $\overline{\overline{\mathbf{D}}}$
and a vector $\boldsymbol{g}$ independent of $\mathbf{x}$, here
equal to $\boldsymbol{1}$. We consider a $d$-dimensional random
variable $\mathbf{X}(0)$ of probability density function $\Phi$,
and call $\mathbf{B}(t)$ the standard $d$-dimensional Brownian motion,
and $\overline{\overline{\mathbf{D'}}}$ the tensor of entries $D'_{\mu\nu}=(D_{\mu\mu}g_{\mu})^{1/2}\delta_{\mu\nu}$.
Moreover, for each stochastic process $\mathbf{Y}(t)$ we call $\delta\mathbf{Y}(t,\,t+\delta t)$
its increment $\mathbf{Y}(t+\delta t)-\mathbf{Y}(t)$. With these
notations, the ADE with the initial condition $\Phi$ rules the p.d.f.
of the stochastic process $\mathbf{X}(t)$ that starts from $\mathbf{X}(0)$
and has increments satisfying 
\begin{equation}
\delta\mathbf{X}(t,\,t+\delta t)=\mathbf{u}(\mathbf{X}(t))\delta t+\overline{\overline{\mathbf{D'}}}\delta\mathbf{B}(t,\,t+\delta t)\label{eq:RW+_STD}
\end{equation}
for $\delta t>0$ \citep{Risken_Book1984}. For each $t\geq0$ and
each $\delta t>0$, it turns out that $\delta\mathbf{B}(t,\,t+\delta t)$
is distributed as $\sqrt{2\delta t}\mathbf{G}$, where $\mathbf{G}$
is a random variable of $\mathbb{R}^{d}$ with mutually independent
standard Gaussian entries, also independent of what happened before
$t$. Sampling $\mathcal{N}_{RW}$ independent trajectories of $\mathbf{X}(t)$
by applying Eq. (\ref{eq:RW+_STD}) to successive intervals of fixed
step $\delta t$ \citep{Delay_etal_VZJ2005} yields histograms that
approximate the p.d.f. of $\mathbf{X}(t)$, i.e. the solution of Eq.
(\ref{eq:FADE_moreGeneral}) started from $\Phi$, provided $\boldsymbol{\alpha}=\mathbf{2}$.
Increasing $\mathcal{N}_{RW}$ improves the accuracy of such random
walk approximation, still valid when $\boldsymbol{g}$ depends on
$\mathbf{x}$ if we plug $g_{\mu}(X(t))$ instead of $g_{\mu}$ in
the entries of $\overline{\overline{\mathbf{D'}}}$ that appear in
Eq. (\ref{eq:RW+_STD}). Note that this is not true if $\overline{\overline{\mathbf{D}}}$
is allowed to depend on $\mathbf{x}$ \citep{Delay_etal_VZJ2005}:
this fact motivates separating $\overline{\overline{\mathbf{D}}}$
and $\boldsymbol{g}$ in Eq. (\ref{eq:FADE_moreGeneral}). Decreasing
$\delta t$ is useful only when $\mathbf{u}$ or $\boldsymbol{g}$
do depend on space, in unbounded domain.

Actually, Eq. (\ref{eq:RW+_STD}) only describes a particular case
of more general $\mathbf{X}(t)$ defined for $t\geq0$ and $\delta t>0$
by 
\begin{equation}
\delta\mathbf{X}(t,\,t+\delta t)=\mathbf{u}(\mathbf{X}(t))\delta t+\overline{\overline{\mathbf{D'}}}\delta\mathbf{L}_{\boldsymbol{\alpha},\boldsymbol{\beta}}(t,\,t+\delta t)\label{eq:RW_STABLE}
\end{equation}
which is very similar to Eq. (\ref{eq:RW+_STD}), except that $D'_{\mu\nu}=(D_{\mu\mu}g_{\mu})^{1/\alpha_{\mu}}\delta_{\mu\nu}$
and the Brownian motion $\mathbf{B}$ is replaced by the stable $d$-dimensional
random process $\mathbf{L}_{\boldsymbol{\alpha},\boldsymbol{\beta}}$.
The latter is completely determined by two sets of parameters encapsulated
in two vectors, namely $\boldsymbol{\alpha}$ defined in Section \ref{sec:Introduction},
and another vector $\boldsymbol{\beta}$ of $\mathbb{R}^{d}$ whose
entries $\beta_{\mu}$ belong to $[-1,\,1]$. As $\delta\mathbf{B}(t,\,t+\delta t)$
above, $\delta\mathbf{L}_{\boldsymbol{\alpha},\boldsymbol{\beta}}(t,\,t+\delta t)$
has $d$ mutually independent components $\mathbf{b}_{\mu}\cdot\delta\mathbf{L}_{\boldsymbol{\alpha},\boldsymbol{\beta}}$
that are independent of what happened before instant $t$. For each
$\delta t>0$ they satisfy 
\begin{equation}
\mathbf{b}_{\mu}\cdot\delta\mathbf{L}_{\boldsymbol{\alpha},\boldsymbol{\beta}}(t,\,t+\delta t)\overset{d}{=}\left(-\cos\frac{\pi\alpha_{\mu}}{2}\delta t\right)^{1/\alpha_{\mu}}S(\alpha_{\mu},\,\beta_{\mu})\label{eq:distri}
\end{equation}
where symbol $\overset{d}{=}$ links equally distributed random variables.
Moreover, the one-dimensional stable random variable $S(\alpha_{\mu},\,\beta_{\mu})$
described in \ref{sec:ApA} is entirely determined by its stability
exponent $\alpha_{\mu}$ and its skewness parameter $\beta_{\mu}$.
Since $S(\alpha_{\mu},\,\beta_{\mu})$ is Gaussian in the limit case
$\alpha_{\mu}=2$ (whatever the value of $\beta_{\mu}$), we retrieve
$\mathbf{B}$ in $\mathbf{L}_{\boldsymbol{2},\boldsymbol{\beta}}$.
With these notations, the p.d.f of $\mathbf{X}(t)$ defined by Eq.
(\ref{eq:RW_STABLE}) satisfies Eq. (\ref{eq:FADE_moreGeneral}) in
$\mathbb{R}^{d}$ provided $\boldsymbol{\beta}=2\boldsymbol{p}-\mathbf{1}$
according to \citep{Yong_etal_JSP2006,Meerschaert-Sikorskii_Book2012}. 

We sample $\mathbf{X}(t)$ by applying Eqs (\ref{eq:RW_STABLE})-(\ref{eq:distri})
with prescribed $\delta t$, and the density of the sample approximately
solves Eq. (\ref{eq:FADE_moreGeneral}) as in the diffusive case.
This needs sampling many values of the $S(\alpha_{\mu},\,\beta_{\mu})$:
each one is given by applying algebraic formulas of \citep{Weron_SPL1996}
to a pair of independent random numbers which are drawn from two uniform
distributions. This procedure works for general initial data equivalent
to the p.d.f. $\Phi$ of $\mathbf{X}(0)$. When we assume spatially
homogeneous parameters, replacing $t$ by $0$ and $\delta t$ by
$t$ in Eqs. (\ref{eq:RW_STABLE})-(\ref{eq:distri}) returns the
distribution of $\mathbf{X}(t)$ in one shot. However, spatially inhomogeneous
parameters or bounded domain require small $\delta t$ in Eqs. (\ref{eq:RW_STABLE})-(\ref{eq:distri}).

\subsubsection*{\label{effect}Influence of parameters $\boldsymbol{\alpha}$ and
$\boldsymbol{\beta}$ }

The stable process $\mathbf{L}_{\boldsymbol{\alpha},\boldsymbol{\beta}}$
defined by Eq. (\ref{eq:distri}) exhibits super-diffusion in each
direction $x_{\mu}$ such that $\alpha_{\mu}<2$. This means infinite
second moment for its projection on $\mathbf{b}_{\mu}$, equivalent
to large displacements significantly\textcolor{red}{{} }more probable
than for $\alpha_{\mu}=2$. More specifically, the density of $S(\alpha_{\mu}<2,\,\beta_{\mu})$
is illustrated by Fig. \ref{fig:densial15} and falls off as $x^{-\alpha_{\mu}-1}$
while that of $S(\alpha_{\mu}=2,\,\beta_{\mu})$ decays exponentially.

The second parameter $\beta_{\mu}$ in $S(\alpha_{\mu},\,\beta_{\mu})$
quantifies the skewness degree of the distribution of this random
variable (see Fig. \ref{fig:densial15}), which also rules the\textcolor{red}{{}
}displacements of $\mathbf{L}_{\boldsymbol{\alpha},\boldsymbol{\beta}}$
in $x_{\mu}$ direction. Yet, stability exponent $\alpha_{\mu}$ approaching
$2$ decreases this influence which becomes evanescent in the limit
case $\alpha_{\mu}=2$. While Brownian motion has symmetrically distributed
displacements in each direction, for $\alpha_{\mu}<2$ the displacements
of $\mathbf{L}_{\boldsymbol{\alpha},\boldsymbol{\beta}}$ and $\mathbf{X}(t)$
in $\mathbf{b}_{\mu}$ direction are symmetric only if $\beta_{\mu}=0$.
Moreover, larger positive $\beta_{\mu}$ magnify large positive jumps
and decrease large negative jumps. Nevertheless the average remains
equal to $0$. This causes most probable jumps to be negative for
$\beta_{\mu}>0$ (see Fig. \ref{fig:densial15}), and in the limit
case $\beta_{\mu}=1$, large negative jumps occur even more scarcely
than in Brownian motion.

\subsubsection*{\label{subRW2bd}Random walks in bounded domain and equation associated
with Dirichlet boundary conditions }

The above described link between $\mathbf{X}(t)$ and Eq. (\ref{eq:FADE_moreGeneral})
in $\Omega=\mathbb{R}^{d}$ persists in bounded domain provided we
consider boundary conditions equivalent to restrictions imposed to
the sample paths of $\mathbf{X}(t)$ \citep{Baeumer_etal_JCOAM2018,Patie_etal_Pot2012}.
However, most boundary problems associated with space-fractional p.d.es
remain still open. It is only in specific cases that we actually know
slight modifications that transform the sample paths of $\mathbf{X}(t)$
into the ones of a closely related random walk whose p.d.f satisfies
Eq. (\ref{eq:FADE_moreGeneral}) and prescribed boundary conditions.
This occurs in the simple case of homogeneous Dirichlet conditions
\citep{Chen-Meerschaert-Nane_JMAA2012,Burch-Lehoucq_PRE2011,Du_etal_DCDSB2014,Baeumer_etal_submitted2018}
which we use for checks. Associating Eq. (\ref{eq:FADE_moreGeneral})
with non-homogeneous Dirichlet conditions returns problems whose well-posedness
is not assessed: solutions do exist but uniqueness apparently depends
on how we interpret these boundary conditions in the case of space-fractional
equations \citep{Baeumer_etal_submitted2018}. We disregard here Neumann
conditions because it is only in too miscellaneous cases that they
have been proved to correspond to sample paths transformations compatible
with density satisfying Eq. (\ref{eq:FADE_moreGeneral}). Just note
that Neumann conditions for space-fractional p.d.es as Eq. (\ref{eq:FADE_moreGeneral})
involve fractional derivatives of order $\alpha_{\mu}-1$ \citep{Baeumer_etal_JCOAM2018,Baeumer_etal_TAMS2016},
as the dispersive flux driven by stable process \citep{Neel_etal_JPA2007}.

Assuming space-independent parameters, we detail in \ref{sto} why
Eq. (\ref{eq:FADE_moreGeneral}) associated to homogeneous Dirichlet
conditions at the boundary of a rectangle rules the evolution of the
p.d.f of $\mathbf{X}^{\Omega}(t)$, a process derived from $\mathbf{X}(t)$
by killing each sample path at its first exit time $\tau_{\Omega}$
\citep{Chen-Meerschaert-Nane_JMAA2012}. From a practical point of
view, each sample path of $\mathbf{X}$ determines one value of the
random variable $\tau_{\Omega}$ and one sample path of $\mathbf{X}^{\Omega}$
composed of all its positions before time $\tau_{\Omega}$. The remainder
of the sample path of $\mathbf{X}$ does not contribute to that of
$\mathbf{X}^{\Omega}$, even if it returns to $\Omega$ after time
$\tau_{\Omega}$. Consequently, good approximations to sample path
of $\mathbf{X}^{\Omega}$ need small $\delta t$ in Eq. (\ref{eq:RW_STABLE})
which would not be necessary if no boundary condition were imposed
with spatially uniform parameters. Here we use random walks to approximate
the p.d.f. of $\mathbf{X}^{\Omega}(t)$ which we deduce from an histogram
of a sample. Therefore, we consider $\delta t$ small enough when
decreasing $\delta t$ does not modify the histogram. This criterion
is satisfied by $\delta t\leq0.001$ in all presented comparisons.

\subsection{\label{sub:Validations_LBM-BGK}Comparisons between LBM and RW for
$\mathbf{u}=\mathbf{0}$}

Without advection ($\mathbf{u}=\mathbf{0}$) and when the diffusion
tensor is isotropic ($\overline{\overline{\mathbf{D}}}=D\overline{\overline{\mathbf{Id}}}$),
BGK and MRT collision rules return quite comparable approximations
to Eq. (\ref{eq:FADE_moreGeneral}). This is what we check in this
subsection by comparing LBM with random walk simulations in dimensions
$d=2$ and $d=3$ with spatially uniform coefficients. Validations
are presented for three sets of parameters that exemplify the strange
behaviors included in Eq. (\ref{eq:FADE_moreGeneral}), and associated
to symmetric or skewed super-diffusion. Moreover, all these random
walk simulations are started from a sample of the $d$-dimensional
Gaussian random variable $\mathbf{X}(0)$ of standard deviation $\sigma_{0}$,
and centered at point $\mathbf{x}^{s}$ of coordinates $x_{\mu}^{s}$.
The p.d.f. of $\mathbf{X}(0)$ is the standard Gaussian hill 
\begin{equation}
C(\mathbf{x},\,0)=\frac{C_{0}}{(2\pi\sigma_{0}^{2})^{d/2}}\exp\left[-\frac{1}{2\sigma_{0}^{2}}\sum_{\mu=1}^{d}(x_{\mu}-x_{\mu}^{s})^{2}\right],\label{eq:Gaussian_IC_3D}
\end{equation}
a smooth initial condition suited for LBM, and approximated by the
distribution of the sample when $\mathcal{N}_{RW}$ is large enough.
Choosing $\sigma_{0}$ small concentrates this initial condition near
point $\mathbf{x}^{s}$. For all simulations of this paper, we set
$\sigma_{0}=0.002$ and $C_{0}=5$.

The two-dimensional validations 1 and 2 assume square domain $\Omega$
($\ell_{1}=\ell_{2}=0$, $L_{1}=L_{2}=2$) and the initial condition
is located at domain center: $\mathbf{x}^{s}=(1,\,1)^{T}$. Then,
the p.d.f. of $\mathbf{X}^{\Omega}(t)$ satisfies Eq. (\ref{eq:FADE_moreGeneral}).
It is of the form of $C(x_{1},\,x_{2},\,t)=P_{1}(x_{1},\,t)P_{2}(x_{2},\,t)$
where each $P_{\mu}$ solves the one-dimensional version of Eq. (\ref{eq:FADE_moreGeneral})
in $]\ell_{\mu},\,L_{\mu}[$, according to \ref{sto}. We moreover
assume $\overline{\overline{\mathbf{D}}}=0.5\overline{\overline{\mathbf{Id}}}$
and $\boldsymbol{g}=\mathbf{1}$.

\subsubsection*{Validation 1: the effect of $\alpha_{\mu}$ when $\beta_{\mu}=0$}

\begin{figure}
\begin{centering}
\subfloat[\label{fig:Comparisons1_Profiles}Narrow peak caused by super-diffusion
at early times. Left: profiles $C(x_{1},\,x_{2}^{s},\,t)$ in $x_{1}$-direction
along which super-diffusion occurs at times $t_{1}=0.01$ and $t_{2}=0.045$.
Right: $C(x_{1}^{s},\,x_{2},\,t)$ profiles at the same times in $x_{2}$-direction
along which diffusion is almost normal.]{\protect\begin{centering}
\begin{tabular}{ccc}
\protect\includegraphics[angle=-90,scale=0.31]{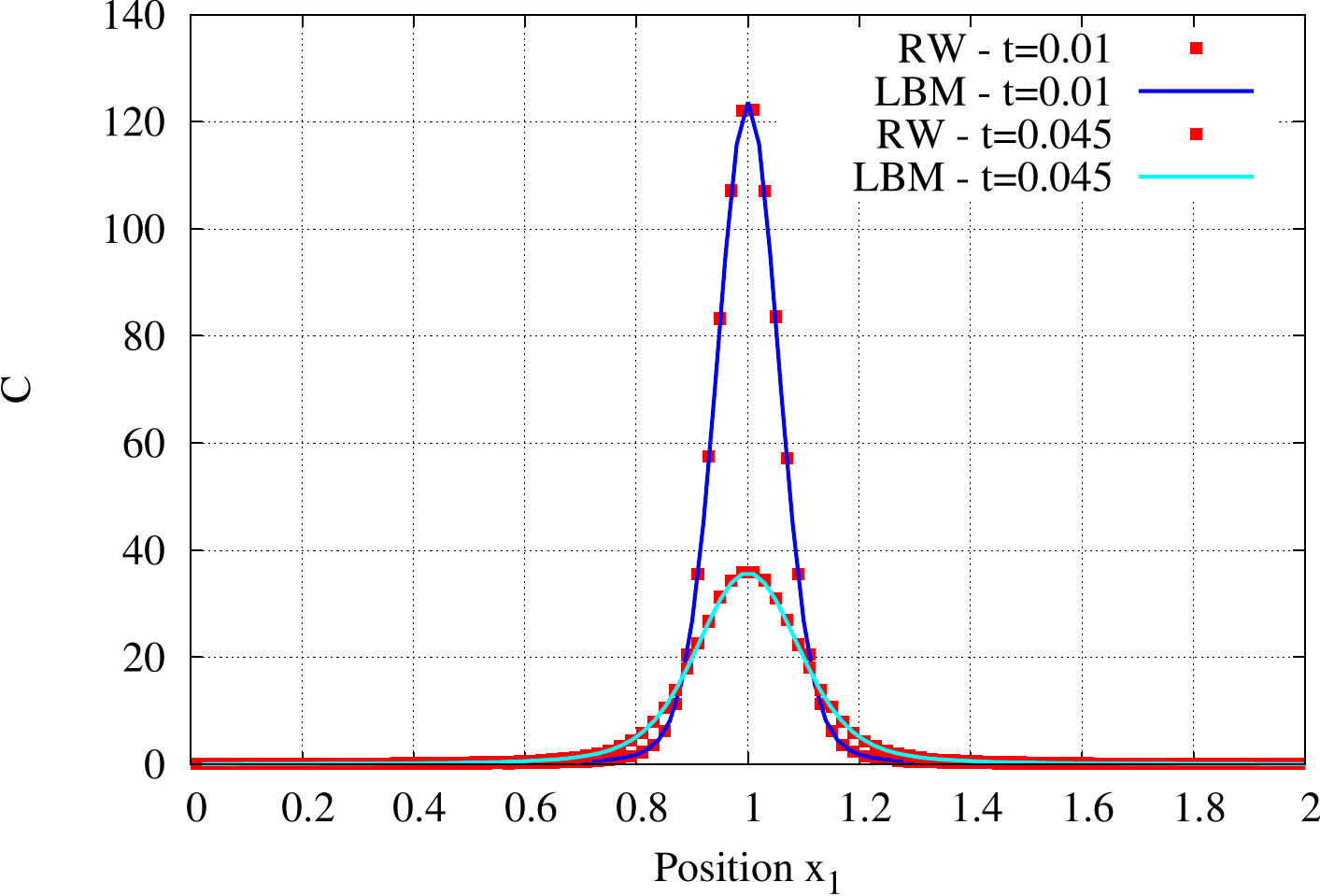}  &  & \protect\includegraphics[angle=-90,scale=0.31]{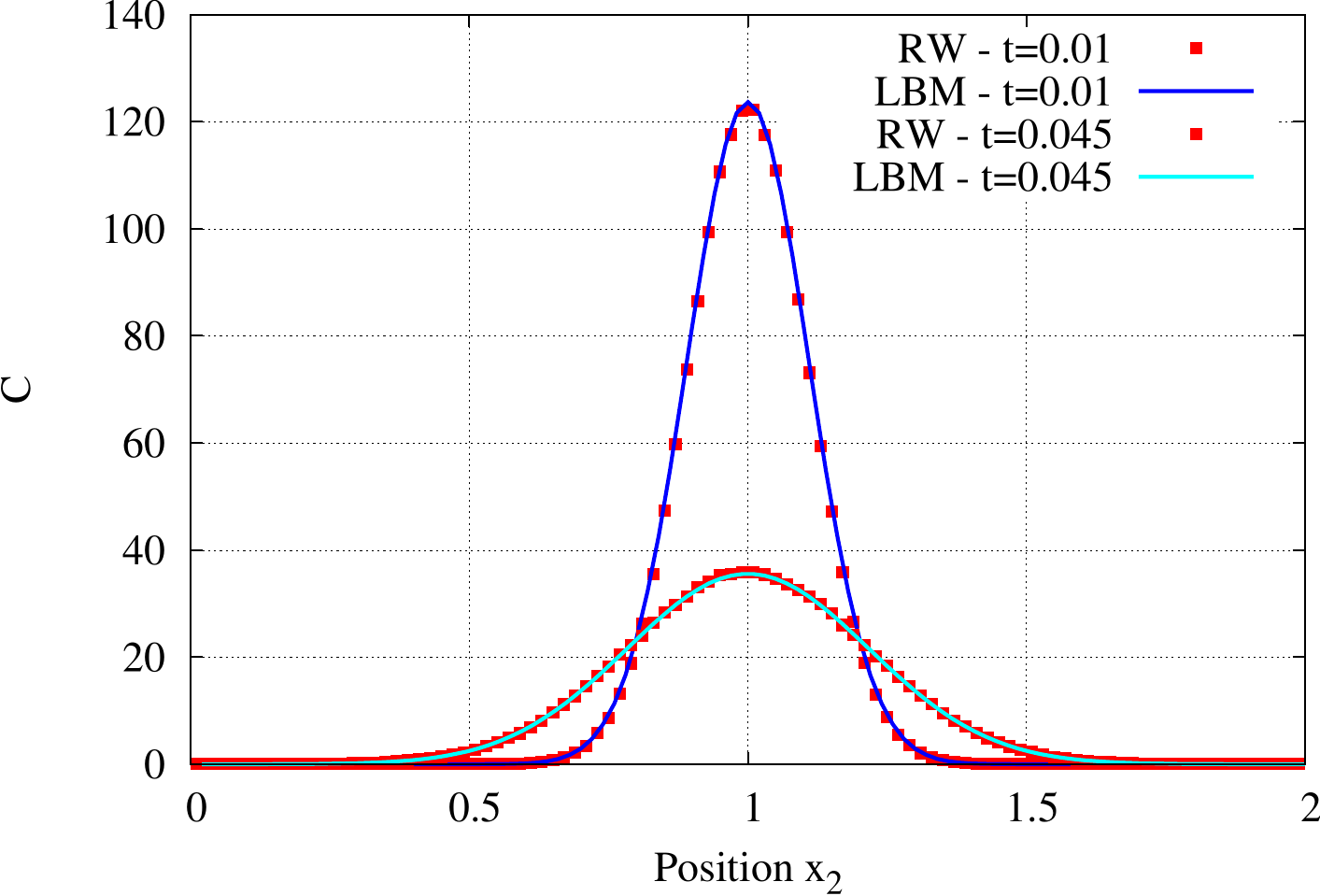}\tabularnewline
\end{tabular}\protect
\par\end{centering}

}
\par\end{centering}

~

\begin{centering}
\subfloat[\label{fig:Comparison1_Zoom}Left: global view of $x_{1}$- and $x_{2}$-profiles
at $t_{3}=0.1$, showing the region magnified at the right. Right:
magnification near right boundary. At time $t_{3}=0.1$, profiles
in $x_{1}$-direction exhibit heavier tails than earlier, and the
peak near the maximum is still narrow compared with $C(x_{1}^{s},\,x_{2},\,t_{3})$.]{\protect\begin{centering}
\begin{tabular}{ccc}
\protect\includegraphics[angle=-90,scale=0.31]{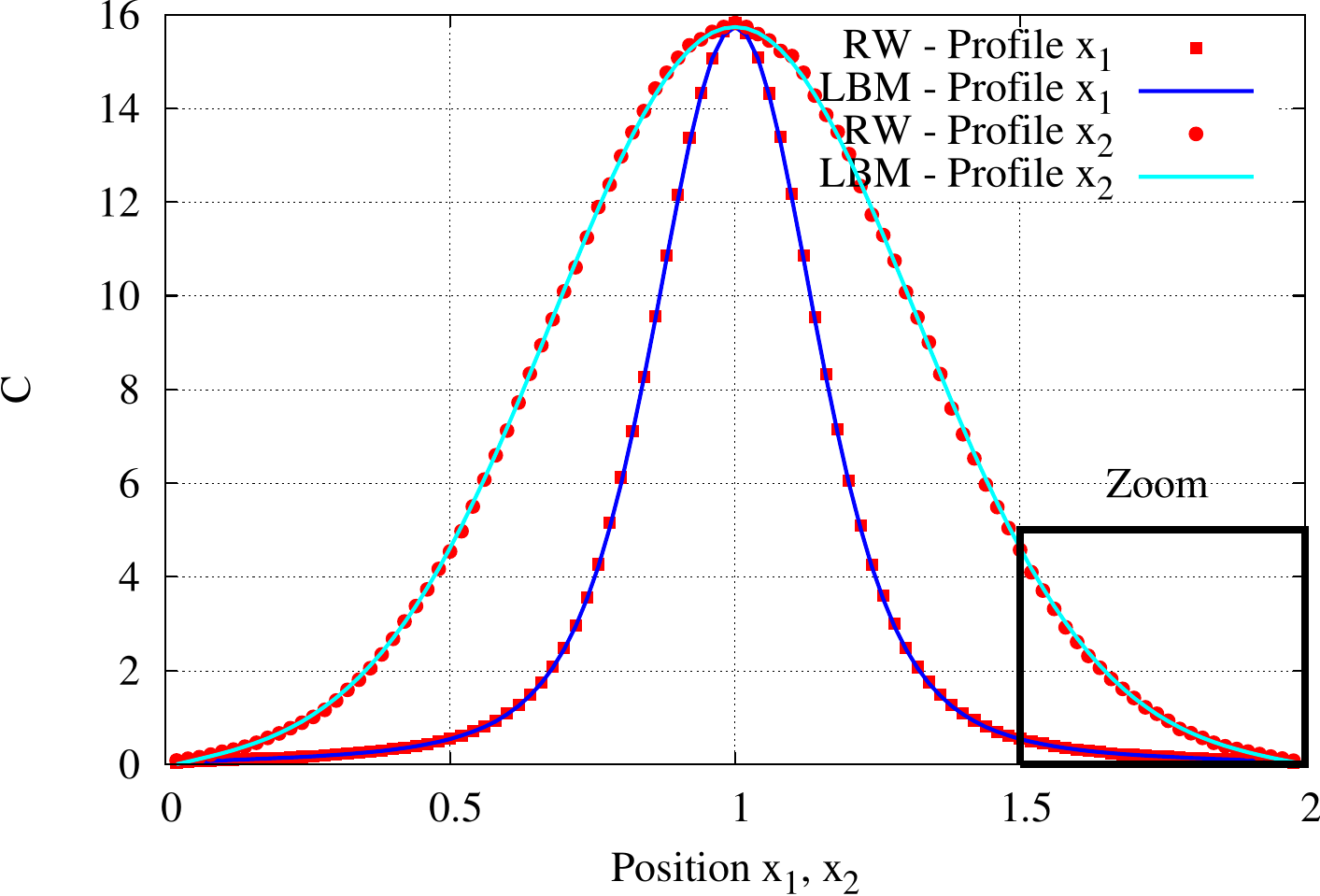}  &  & \protect\includegraphics[angle=-90,scale=0.31]{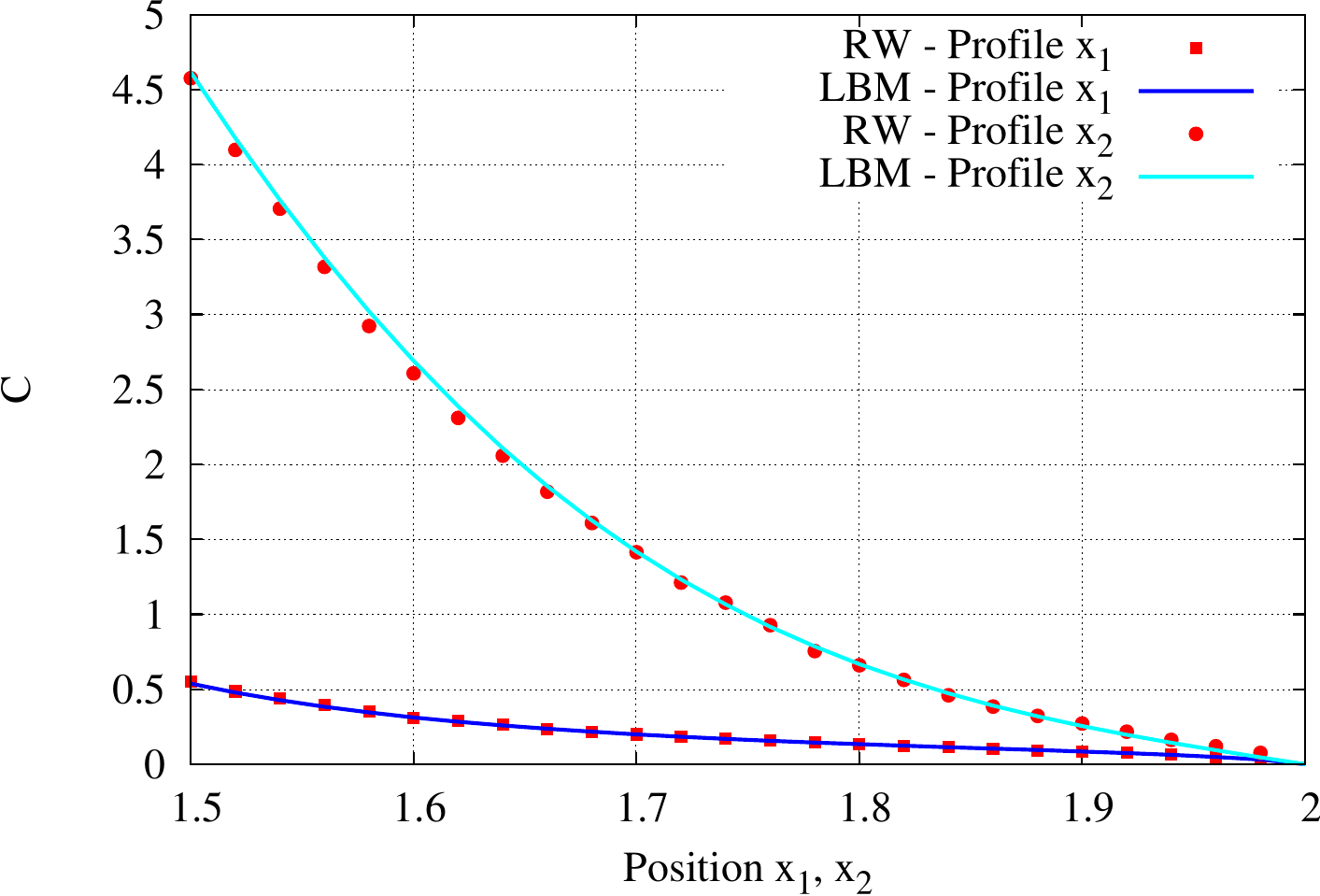}\tabularnewline
\end{tabular}\protect
\par\end{centering}

}
\par\end{centering}

\protect\caption{\label{fig:Bench1}Solutions of the 2D Eq. (\ref{eq:FADE_moreGeneral})
with symmetric integrals ($\boldsymbol{p}=\boldsymbol{1/2}$) w.r.t.
each coordinate. The stability parameter is equivalent to super-diffusion
only in $x_{1}$-direction: $\boldsymbol{\alpha}=(1.5,\,199)^{T}$.
LBM and RW return indistinguishable approximations, even near the
boundaries.}
\end{figure}

With parameters $\boldsymbol{\alpha}=(1.5,\,1.99)^{T}$ and $\boldsymbol{\beta}=\mathbf{0}$,
the process $\mathbf{X}^{\Omega}$ defined in Section \ref{subRW2bd}
accumulates symmetric displacements. This is equivalent to integrals
$I_{x_{\mu}\pm}^{2-\alpha_{\mu}}$ of equal weights in Eq. (\ref{eq:Fg}).
Here with $\mathbf{u}=\mathbf{0}$, we see on Fig. \ref{fig:Bench1}
that the maximum of the solution to Eq. (\ref{eq:FADE_moreGeneral})
stays immobile. Counting the sample paths that leave $\Omega$ reveals
that most of them exit through boundaries $x_{1}=\ell_{1}$ and $x_{1}=L_{1}$,
due to large displacements more probable in $x_{1}$-direction because
of $\alpha_{1}<\alpha_{2}$. At times $t$ satisfying $D_{\mu\mu}g_{\mu}t<1$
the left panel of Fig. \ref{fig:Comparisons1_Profiles} exhibits profiles
$C(x_{1},\,x_{2}^{s},\,t)$ (in $x_{1}$-direction where super-diffusion
occurs). They are thinner than profiles $C(x_{1}^{s},\,x_{2},\,t)$
represented on the right and describing the variations of $C$ in
$x_{2}$-direction where diffusion is almost normal. Actually, this
agrees with Eq. (\ref{eq:distri}) though this equation is only exact
in unbounded domain. At times satisfying $D_{\mu\mu}g_{\mu}t\geq1$
(not represented here) the information included in this equation is
no longer valid for the two profiles which become similar to each
other, but small because almost all the sample paths have left $\Omega$.
LBM and random walk simulation agree fairly well even in the neighborhood
of the boundaries magnified on Fig.\ref{fig:Comparison1_Zoom}, with
space and time steps $\Delta x=0.02$ and $\Delta t=10^{-4}$. This
necessitates $100^{2}$ nodes in the considered domain, and $10^{3}$
time steps completed in $9$ minutes on simple core desktop workstation.

\subsubsection*{Validation 2: non-symmetric super-diffusion }

\begin{figure}
\begin{centering}
\subfloat[\label{fig:Valid2_Profiles}$C(x_{1},\,x_{2}^{s},\,t)$ profiles at
$t_{1}=0.6$ (blue), $t_{2}=0.8$ (magenta) and $t_{3}=1$ (cyan).]{\protect\begin{centering}
\protect\includegraphics[angle=-90,scale=0.31]{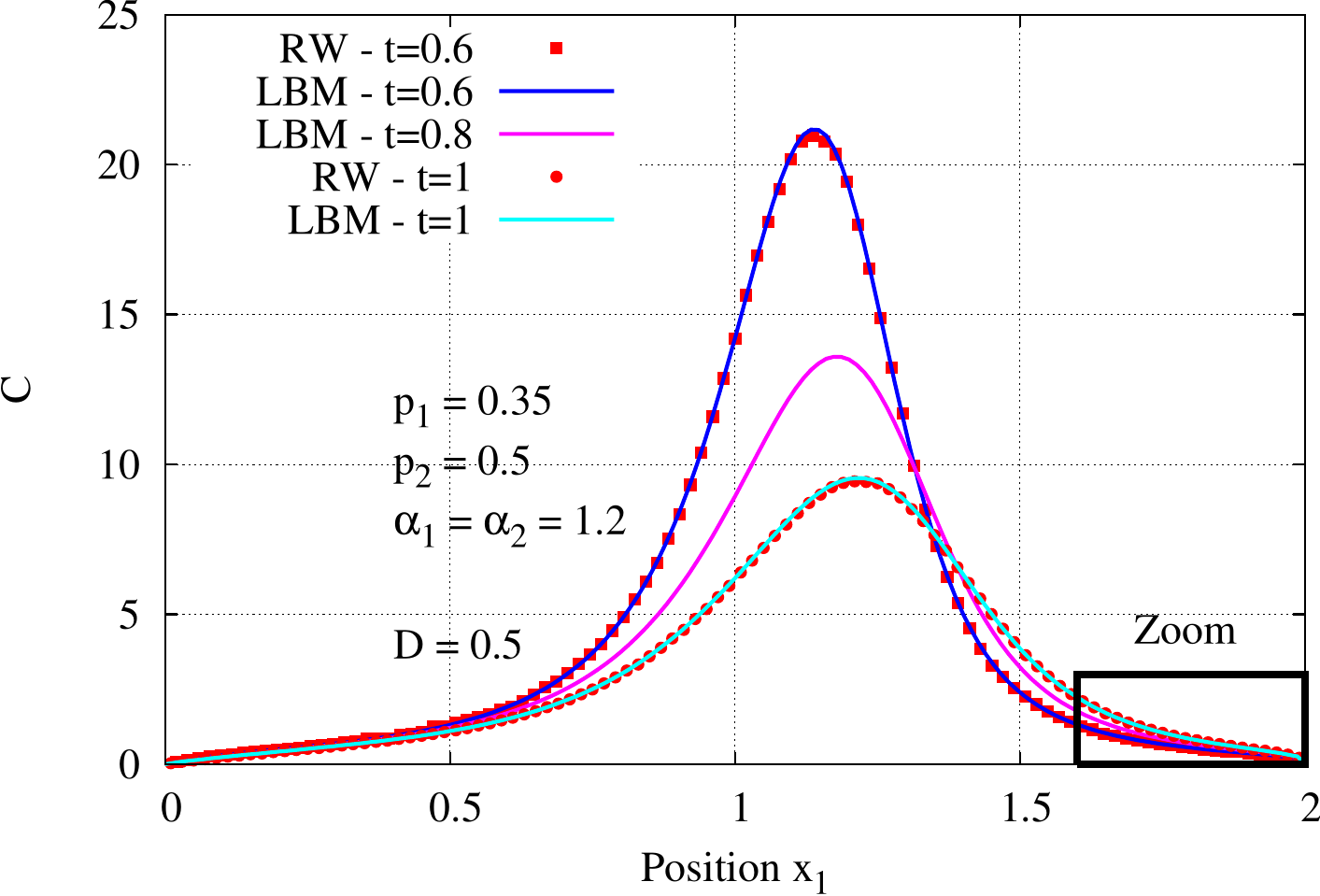}\protect
\par\end{centering}

}$\qquad$\subfloat[\label{fig:Valid2_Zoom}Magnification of the neighborhood of the right
boundary.]{\protect\begin{centering}
\protect\includegraphics[angle=-90,scale=0.31]{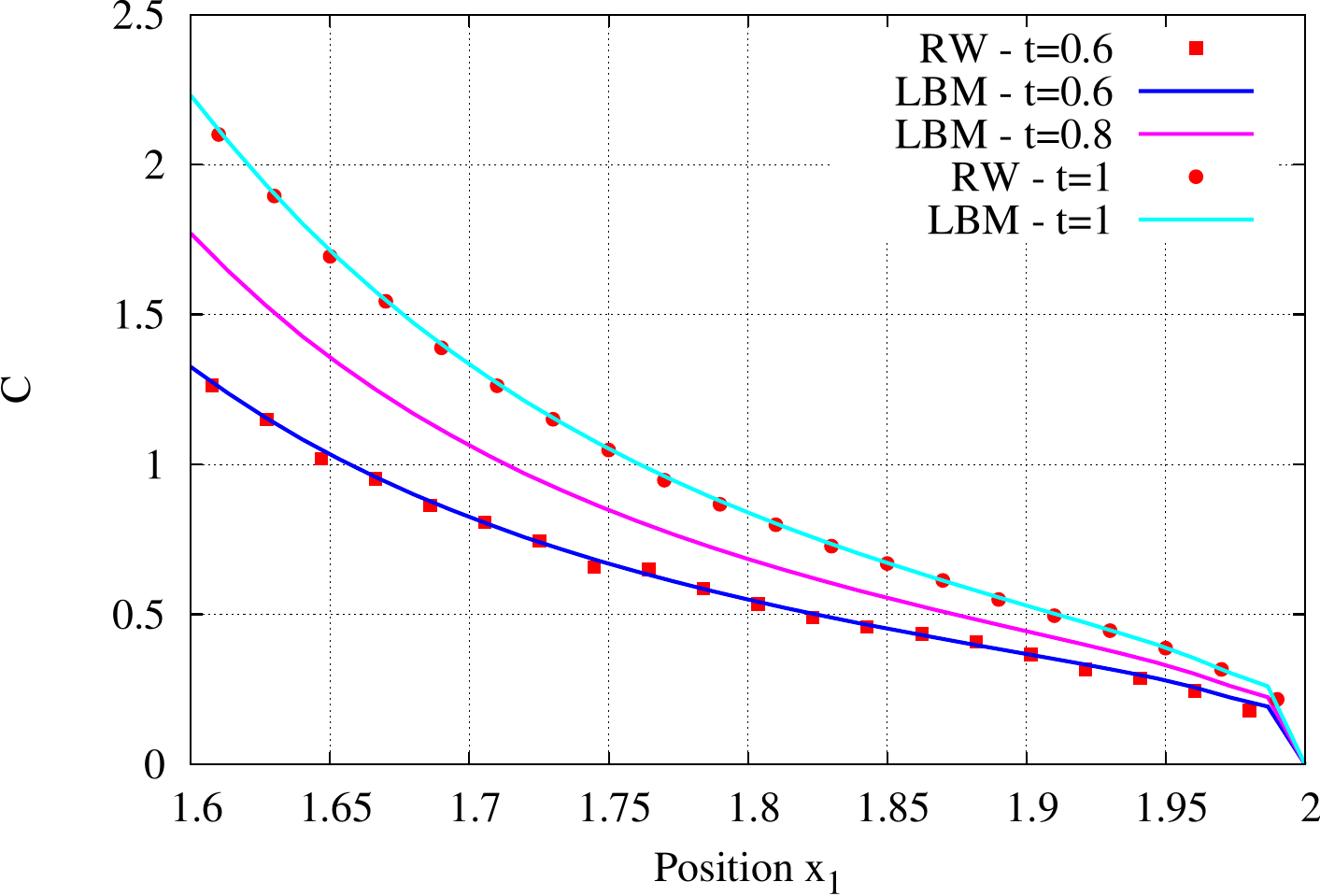}\protect
\par\end{centering}

}
\par\end{centering}

\protect\caption{\label{fig:Valid2}$C(x_{1},\,1,\,t)$ Profiles of solutions of the
2D Eq. (\ref{eq:FADE_moreGeneral}) equipped with the same stability
parameter in all directions and with non-symmetric integrals in $x_{1}$-direction
only: $\boldsymbol{p}=(0.35,\,0.5)^{T}$ and $\boldsymbol{\alpha}=\boldsymbol{1.2}$.
LBM and RW return indistinguishable approximations, even near the
right boundary where the solution decreases especially rapidly due
to $\alpha_{1}$ and to skewness. }
\end{figure}

\begin{figure}
\subfloat[\label{fig:Evolution-of-concentration}Evolution of non-symmetric
super-diffusive concentration field due to $p_{1}=0.35<0.5=p_{2}$
and$\boldsymbol{\alpha}=\boldsymbol{1.2}$ .]{\protect\begin{centering}
\begin{tabular}{cccc}
{\small{}$t=0.05$}  & {\small{}$t=0.2$}  & {\small{}$t=0.6$}  & {\small{}$t=1$}\tabularnewline
\protect\includegraphics[scale=0.12]{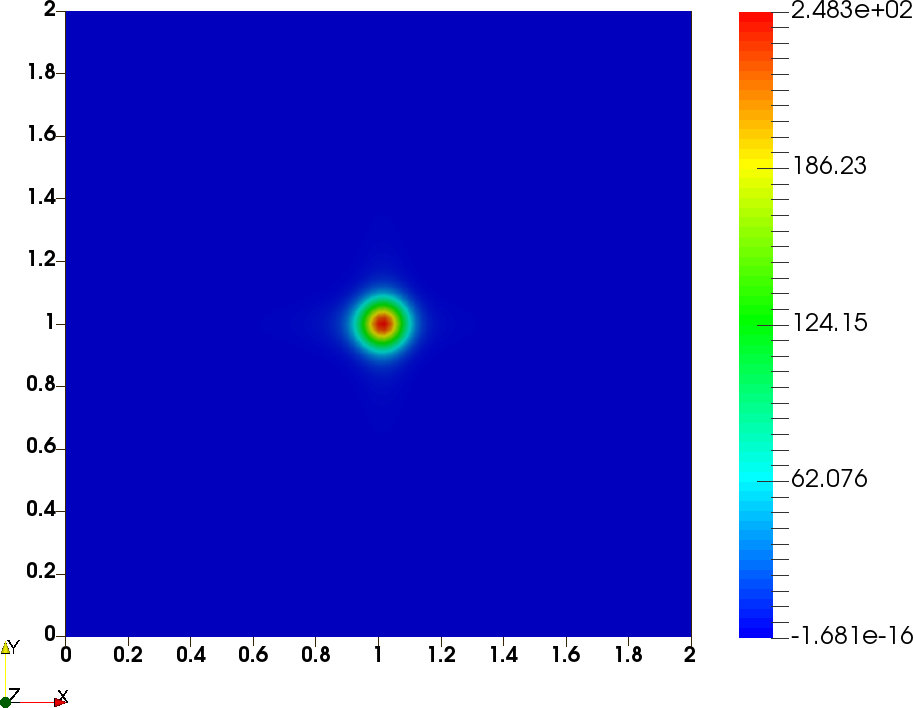} & \protect\includegraphics[scale=0.12]{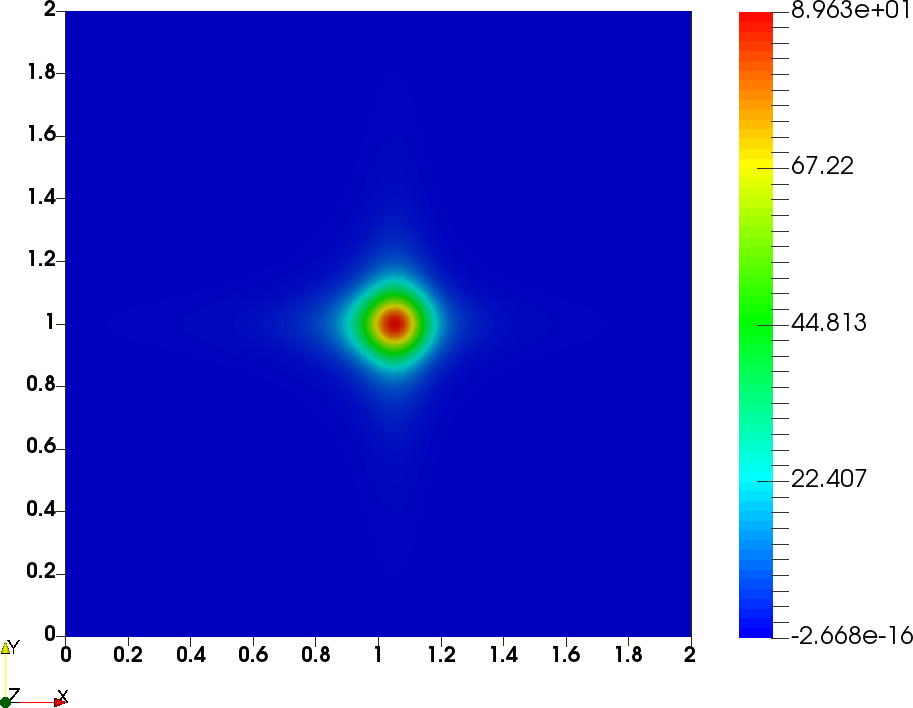} & \protect\includegraphics[scale=0.12]{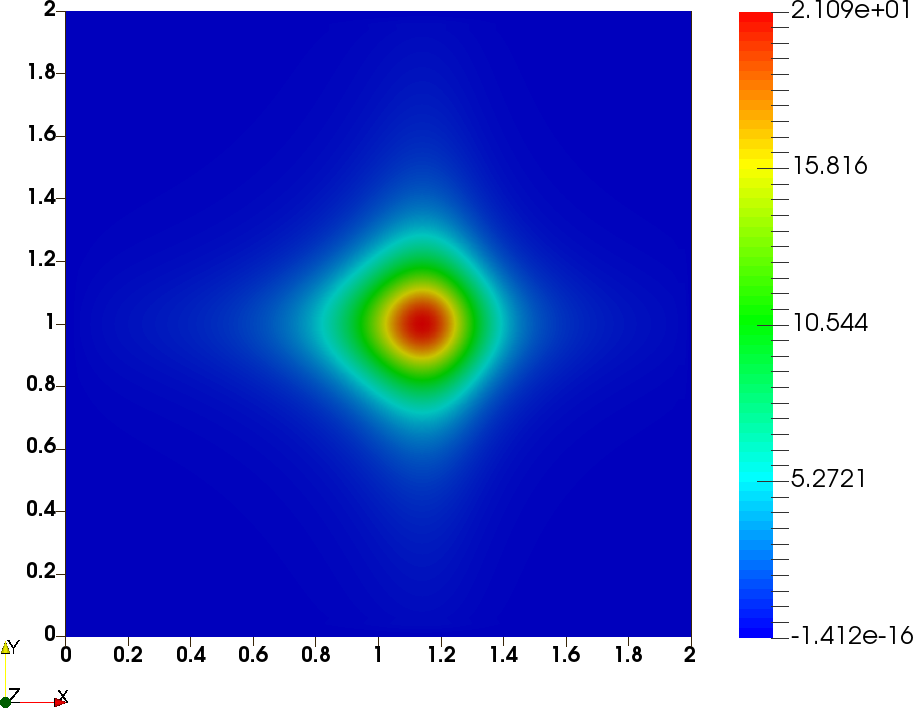} & \protect\includegraphics[scale=0.12]{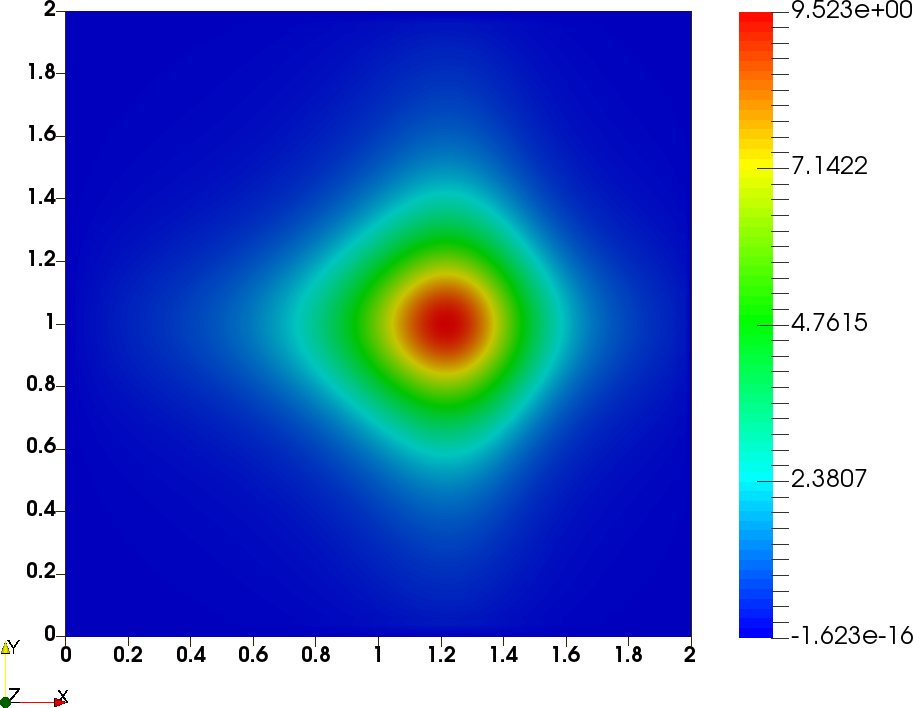}\tabularnewline
\end{tabular}\protect
\par\end{centering}

}

~

~

\begin{centering}
\subfloat[\label{fig:Valid2b}Iso-levels $C=0.5,\,5,\,10,\,15$ at $t=0.6$
for RW (red) and LBM (blue).]{\protect\begin{centering}
\protect\includegraphics[angle=-90,scale=0.4]{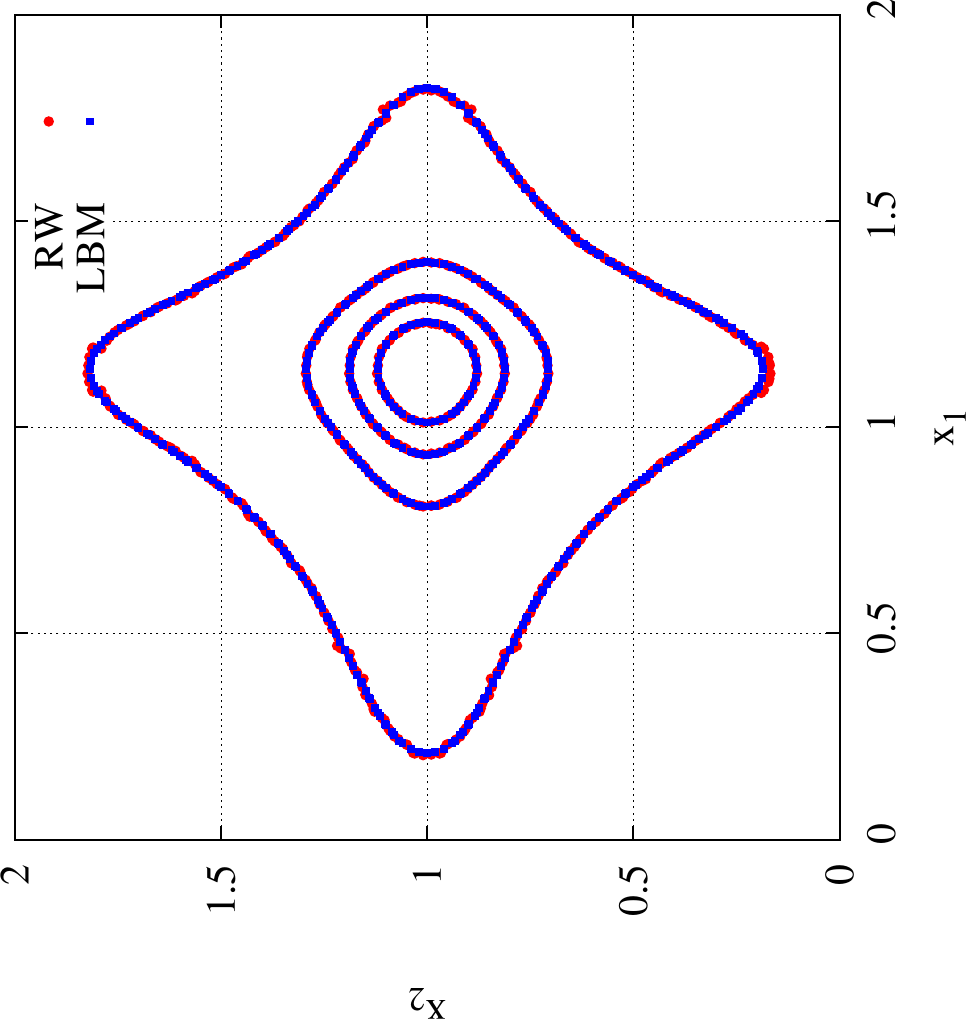}\protect
\par\end{centering}

}
\par\end{centering}

\protect\caption{Global views of the solutions to Eq. (\ref{eq:FADE_moreGeneral})
equipped with the same parameters as for Fig.\ref{fig:Valid2}, and
showing (a) time evolution and (b) iso-levels issued of LBM and RW.}
\end{figure}

For $\beta_{\mu}\neq0$ and $\alpha_{\mu}$ strictly between $1$
and $2$, $S(\alpha_{\mu},\,\beta_{\mu})$ is not symmetric and the
sign of its most probable value is that of $-\beta_{\mu}$, though
the average is zero. The same holds for $\mathbf{b}_{\mu}\cdot\mathbf{L}_{\boldsymbol{\alpha},\boldsymbol{\beta}}$
whose most probable value has a modulus that increases with time due
to Eq. (\ref{eq:distri}). Here with $\mathbf{u}=\mathbf{0}$, the
most probable value of $\mathbf{X}^{\Omega}$ (equivalent to the maximum
of the solution $C$ to Eq. (\ref{eq:FADE_moreGeneral})) exhibits
the same behavior if the initial data $\Phi$ is localized far from
the boundaries, and at times such that a small amount of tracer has
left $\Omega$. Fig. \ref{fig:Valid2_Profiles} (obtained with $\boldsymbol{\alpha}=\mathbf{1.2}$
and $\boldsymbol{p}=(0.35,\,0.5)^{T}$) illustrates the shift of this
maximum  and exhibits left tail slightly thicker than right tail.
This is due to $p_{1}<0.5$ that makes large negative displacements
in $x_{1}$-direction more probable. Fig. \ref{fig:Valid2_Zoom} reveals
that BGK collision rule achieves perfect agreement with random walk
even near the boundary, here with $\Delta x=0.01$ and $\Delta t=5\times10^{-5}$
necessitating $200^{2}$ nodes and $2\times10^{4}$ times steps (to
reach $t=1$) completed in $16.89$ hours by the above mentioned workstation.
Fig. \ref{fig:Evolution-of-concentration} represents the evolution
of  the solute plume that corresponds to Fig. \ref{fig:Valid2b}:
though $\mathbf{u}=\mathbf{0}$, the plume center shifts to the right
in $x_{1}$-direction. Nevertheless the iso-levels of $C$ extend
farther to the left than to the right in the direction of $x_{1}$.
They show larger curvature than if $p_{1}$ and $p_{2}$ were equal
(see \citep{Meerschaert-Sikorskii_Book2012}), and are reminiscent
of anisotropic diffusion though here $\overline{\overline{\mathbf{D}}}$
is spherical with $\alpha_{1}=\alpha_{2}$.

\subsubsection*{Validation 3: Three-dimensional simulations with LBM and RW}

\begin{figure}
\begin{centering}
\subfloat[\label{fig:3D_Profile-Iso}Comparisons between solutions random walk
(red) and LBM (blue) approximations to the 3D Eq. (\ref{eq:FADE_moreGeneral})
for $\mathbf{u}=\mathbf{0}$: parameter $p_{1}<0.5$ (equivalent to
$\beta_{1}<0$) shifts the maximum toward larger $x_{1}$. Left: $C(x_{1},\,x_{2}^{s},\,x_{3}^{s},\,t)$-profile
at $t=0.21$. Right: iso-levels $C=15,\,100,\,200,\,300$ for $x_{2}=x_{2}^{s}$.]{\protect\begin{centering}
\begin{tabular}{ccc}
\protect\includegraphics[angle=-90,scale=0.35]{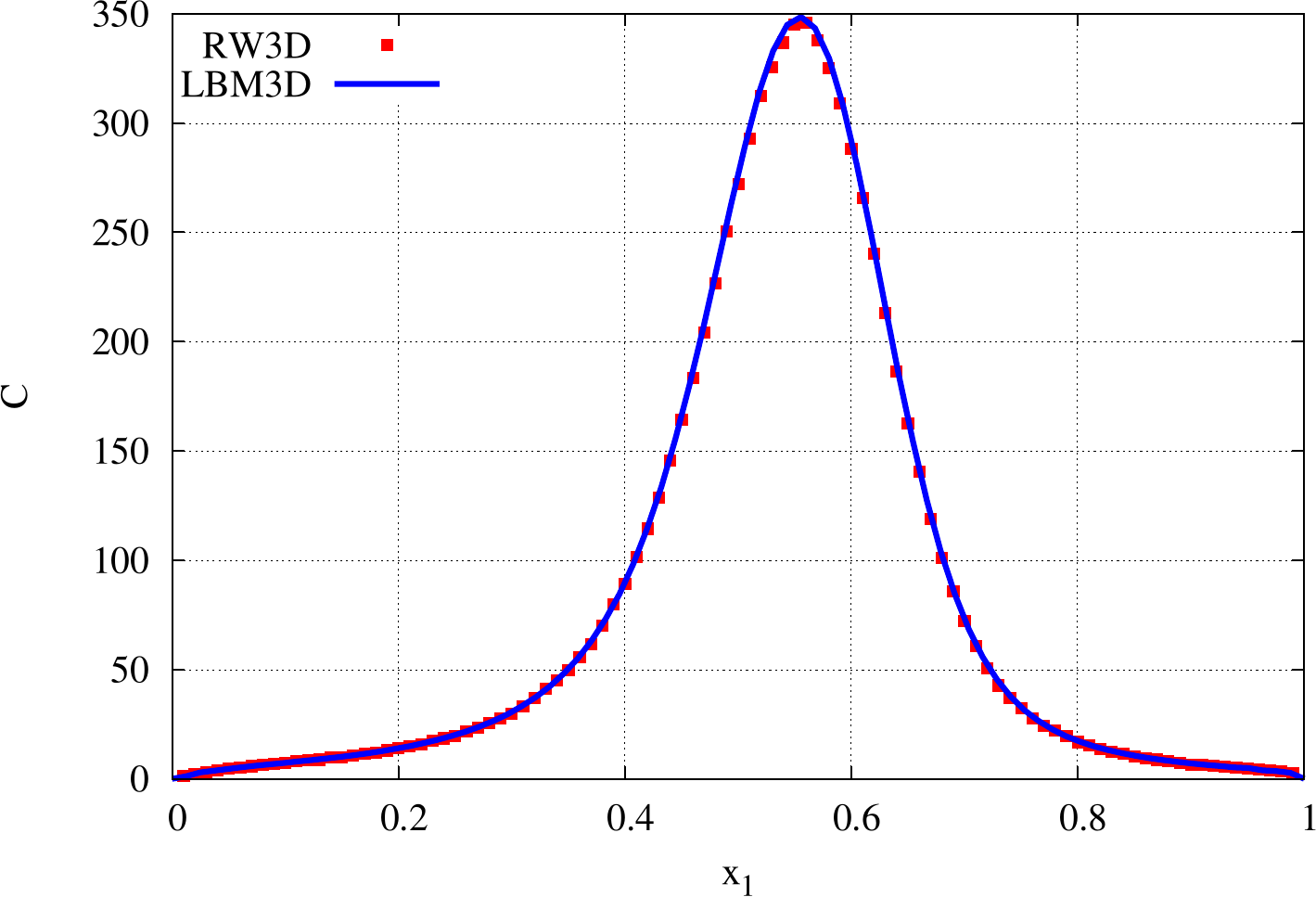}  & $\quad$  & \protect\includegraphics[angle=-90,scale=0.35]{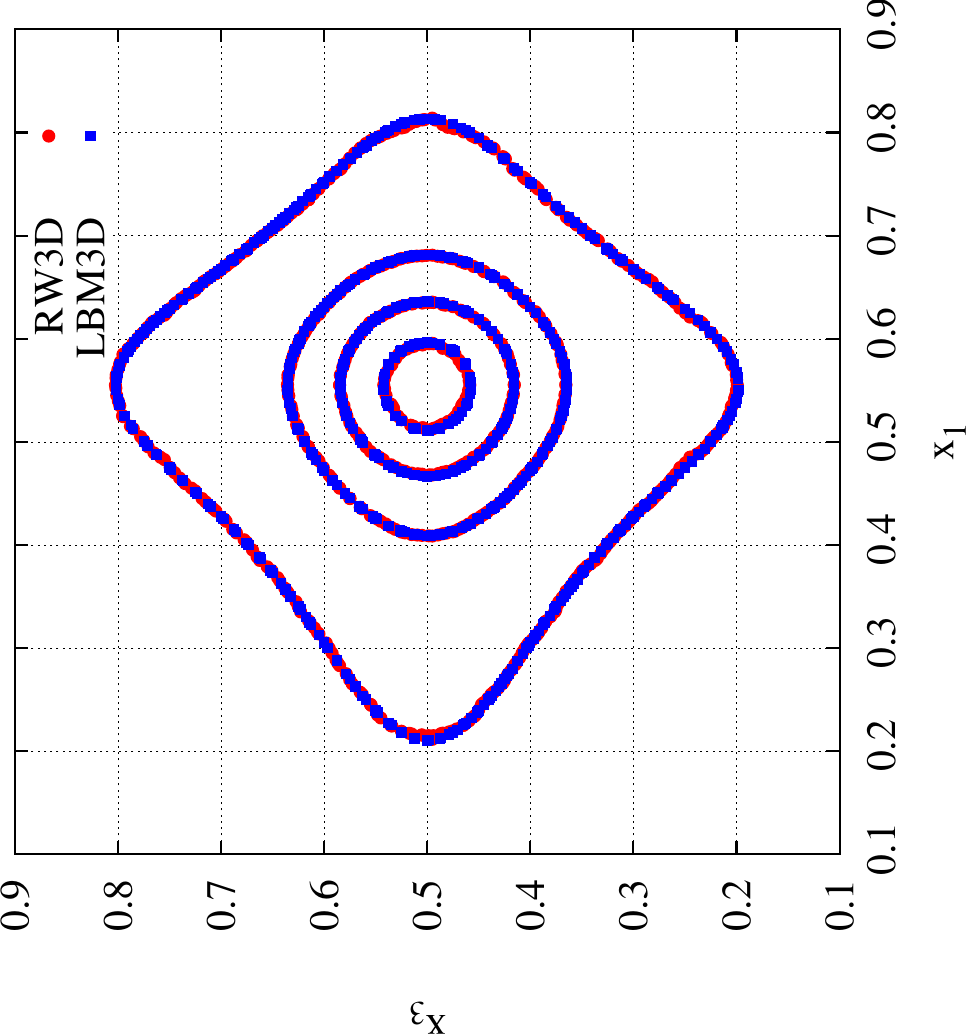}\tabularnewline
\end{tabular}\protect
\par\end{centering}

}
\par\end{centering}

~

~

\begin{centering}
\subfloat[\label{fig:3D_C-Iso}Global concentration field at $t=0.21$ illustrated
by color levels on section views of the three planes $x_{\mu}=0.5$
(left), and by perspective view of iso-surface $C=15$ (right).]{\protect\begin{centering}
\begin{tabular}{ccc}
\protect\includegraphics[scale=0.23]{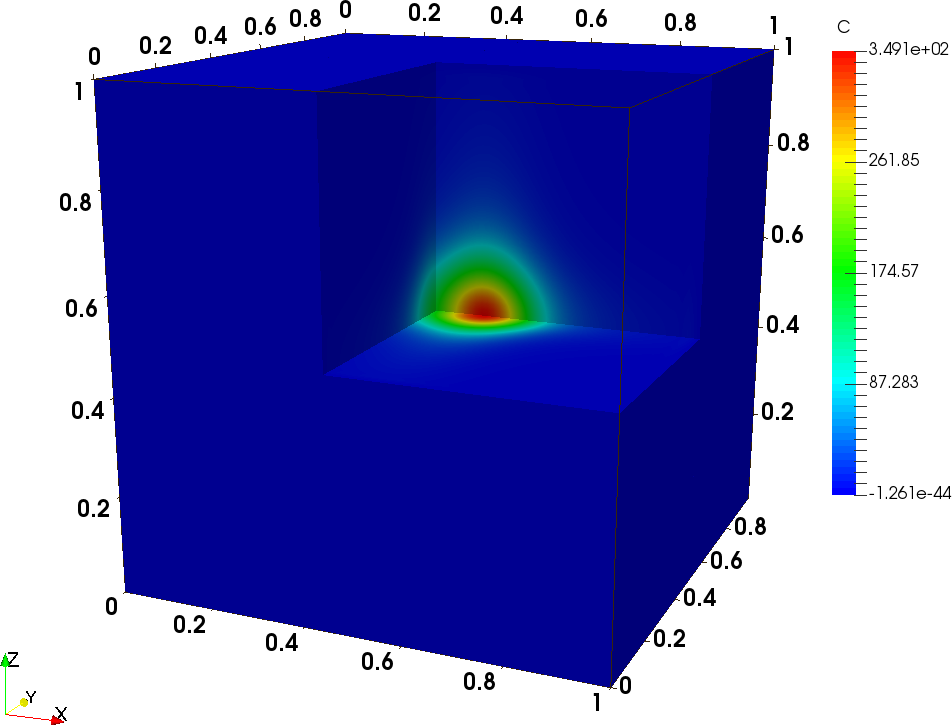}  & $\qquad$  & \protect\includegraphics[scale=0.23]{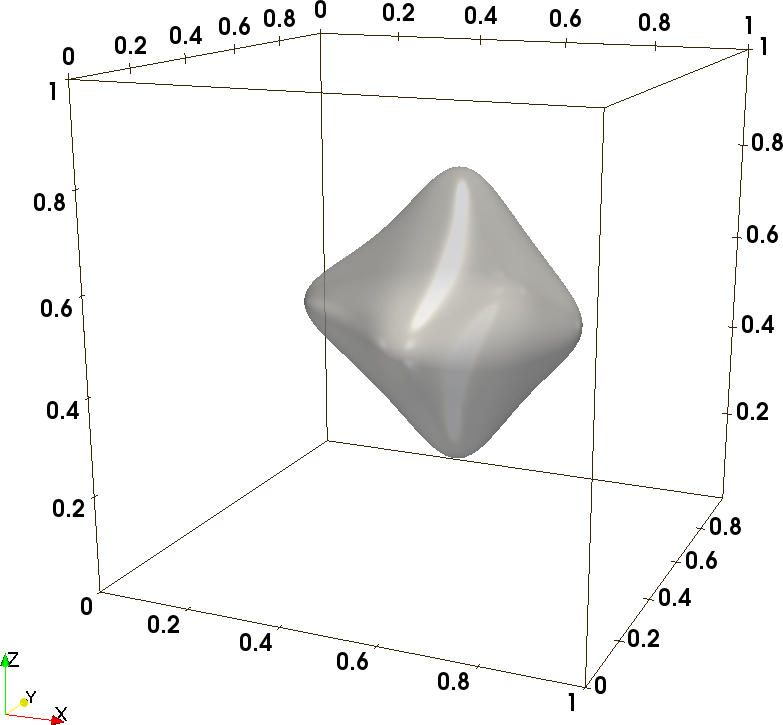}\tabularnewline
\end{tabular}\protect
\par\end{centering}

}
\par\end{centering}

~

~

\protect\caption{\label{fig:FieldsC_Isolevel}Solutions to three-dimensional Eq. (\ref{eq:FADE_moreGeneral})
in $\Omega=[0,\,1]^{3}$ with parameters $\boldsymbol{\alpha}=\boldsymbol{1.2}$,
$\boldsymbol{g}=\boldsymbol{1}$ and $\boldsymbol{p}=(0.35,\,0.5,\,0.5)^{T}$
supplemented by $\mathbf{u}=\mathbf{0}$ and spherical $\overline{\overline{\mathbf{D}}}$.
Above: LBM and random walk return indistinguishable $C(x_{1},\,x_{2}^{s},\,x_{3}^{s},\,t)$
profile (left) and iso-levels $C(x_{1},\,x_{2}^{s},\,x_{3},\,t)$
in $x_{1}x_{3}$-plane (right). Below: three-dimensional iso-surfaces. }
\end{figure}

BGK LBM captures the anisotropic contours in perfect agreement with
random walks in dimension three also. Fig. \ref{fig:FieldsC_Isolevel}
documents the solutions of Eq. (\ref{eq:FADE_moreGeneral}) in $\Omega=[0,\,1]^{3}$
for $\boldsymbol{\alpha}=\mathbf{1.2}$, $\boldsymbol{p}=(0.35,\,0.5,\,0.5)^{T}$
and $\overline{\overline{\mathbf{D}}}=0.5\overline{\overline{\mathbf{Id}}}$.
The initial Gaussian hill is located near the center $\mathbf{x}^{s}=(0.5,\,0.5,\,0.5)^{T}$
of the cube. Similar trend is observed as in dimension two, except
that the more confined geometry results into iso-contours (displayed
on Fig. \ref{fig:3D_Profile-Iso}) and iso-surfaces (displayed on
Fig. \ref{fig:3D_C-Iso}) showing smaller curvature. Global views
and especially non-spherical iso-surface (see Fig. \ref{fig:3D_C-Iso})
illustrate the anisotropy due to our choice of vector $\boldsymbol{p}$,
already visible on the right panel of Fig. \ref{fig:3D_Profile-Iso}.
The profiles recorded on several lines in $\Omega$ and the contour
levels recorded in several planes validate BGK LBM and random walks
simulations. Here we present on Fig. \ref{fig:3D_Profile-Iso} only
one particular $x_{1}$-profile and contour levels in the particular
$x_{1}x_{3}$-plane at $t=0.21$. These figures are issued from a
computation using space and time steps $\Delta x=0.0125$ and $\Delta t=3\times10^{-5}$,
with $80^{3}$ nodes. It took 13.12h to perform $7\times10^{3}$ time
steps on a single core.

These three tests validate the algorithm and the equilibrium distribution
function described in Section \ref{sec:LBM-FADE3D} based on fractional
integrals approximated at order two according to Section \ref{sub:Discrete-integrals}.
Lower order approximations do not return so good agreement between
LBM and RW.

\subsection{\label{sub:Validation_Advection}Instabilities due to $\mathbf{u}\protect\neq\mathbf{0}$
in the LBM}

Numerical approximations to the classical ADE are often subjected
to instabilities when Péclet numbers are increased. The additional
freedom degrees included in MRT collision operator help us fixing
such shortcoming better than BGK \citep{Yoshida-Nagaoka_JCP2010}.
Fractional equations \citep{Meerschaert_etal_JCP2006,Krepysheva_etal_PhysA2006}
also are subjected to such instabilities.

Finite difference schemes approximating the one-dimensional ADE become
unstable at large Péclet number, even in implicit versions. We also
experience it on the LBM equipped with BGK collision operator adapted
to the ADE with parameters with $D_{11}=D_{22}=2\times10^{-3}$, $\boldsymbol{p}=\boldsymbol{1}/\mathbf{2}$
and $\mathbf{u}=(5,\,0)^{T}$ at times $t>0.128$. The domain is $\Omega=[0,\,2]\times[0,\,1]$,
the initial condition described in \ref{sub:Validations_LBM-BGK}
is centered at $(0.5,\,0.5)$, and the space- and time-steps are $\Delta x=10^{-2}$
and $\Delta t=8\times10^{-4}$. Appropriately choosing $\lambda_{3}$
and $\lambda_{4}$ (e.g. $\lambda_{3}=\lambda_{4}>0.833$) in MRT
setting re-stabilizes the LBM without sacrificing accuracy, as in
\citep{Yoshida-Nagaoka_JCP2010}.

\begin{figure}
\begin{centering}
\subfloat[\label{fig:Conv-Profiles} $C(x_{1},\,0.5,\,t)$-profiles issued of
BGK LBM (cyan), random walk (red) and MRT LBM (blue squares) at times
$t=0.08$ and $t=1$: BGK is unstable for $t>0.08$, and not represented.]{\protect\begin{centering}
\protect\includegraphics[angle=-90,scale=0.36]{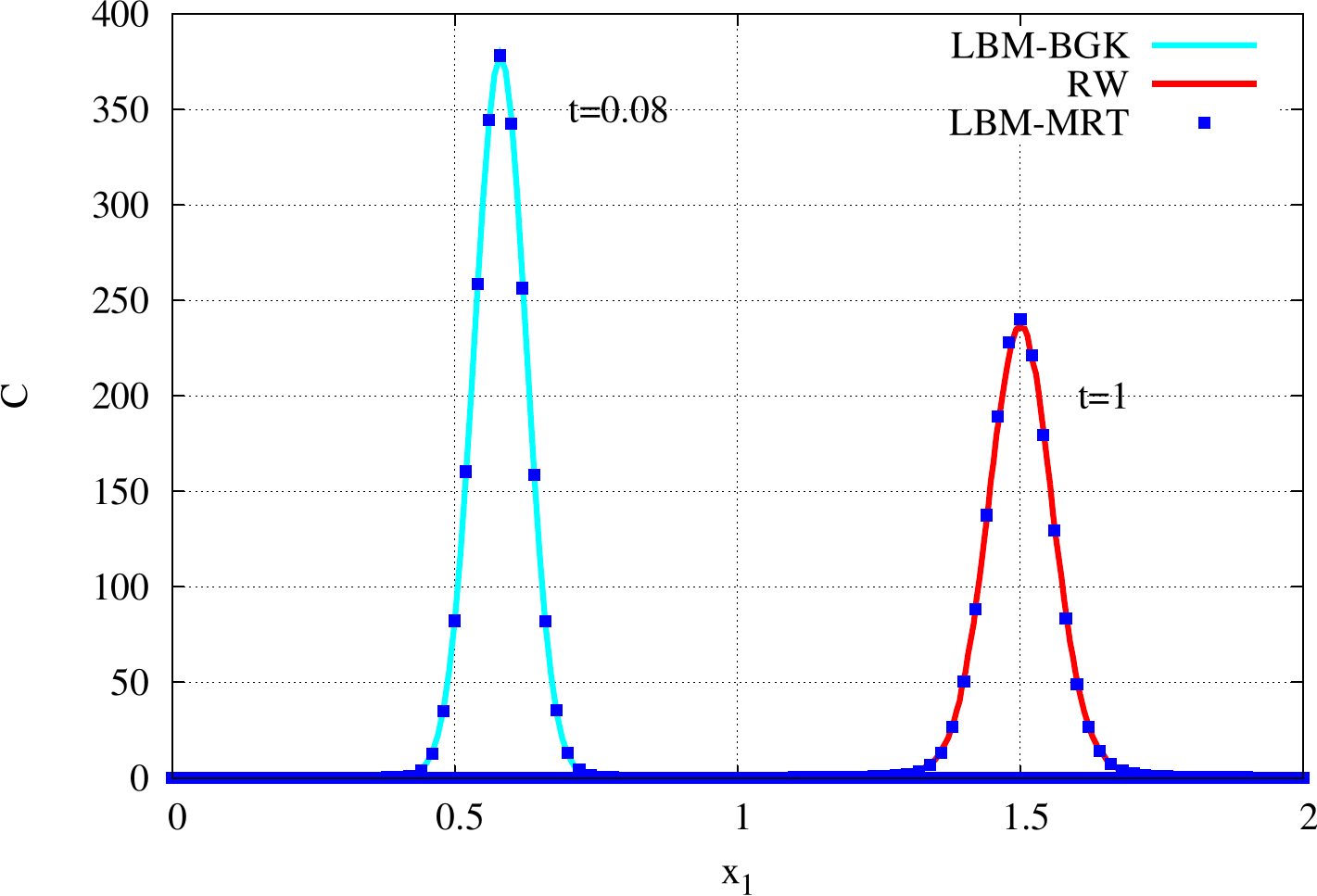}\protect
\par\end{centering}

}$\qquad$\subfloat[\label{fig:Conv-Iso}Iso-levels $C=0.5,\,2.5,\,15,\,50,\,150$ issued
of random walk (red dots) and MRT-LBM (blue squares) at $t=1$.]{\protect\begin{centering}
\protect\includegraphics[angle=-90,scale=0.36]{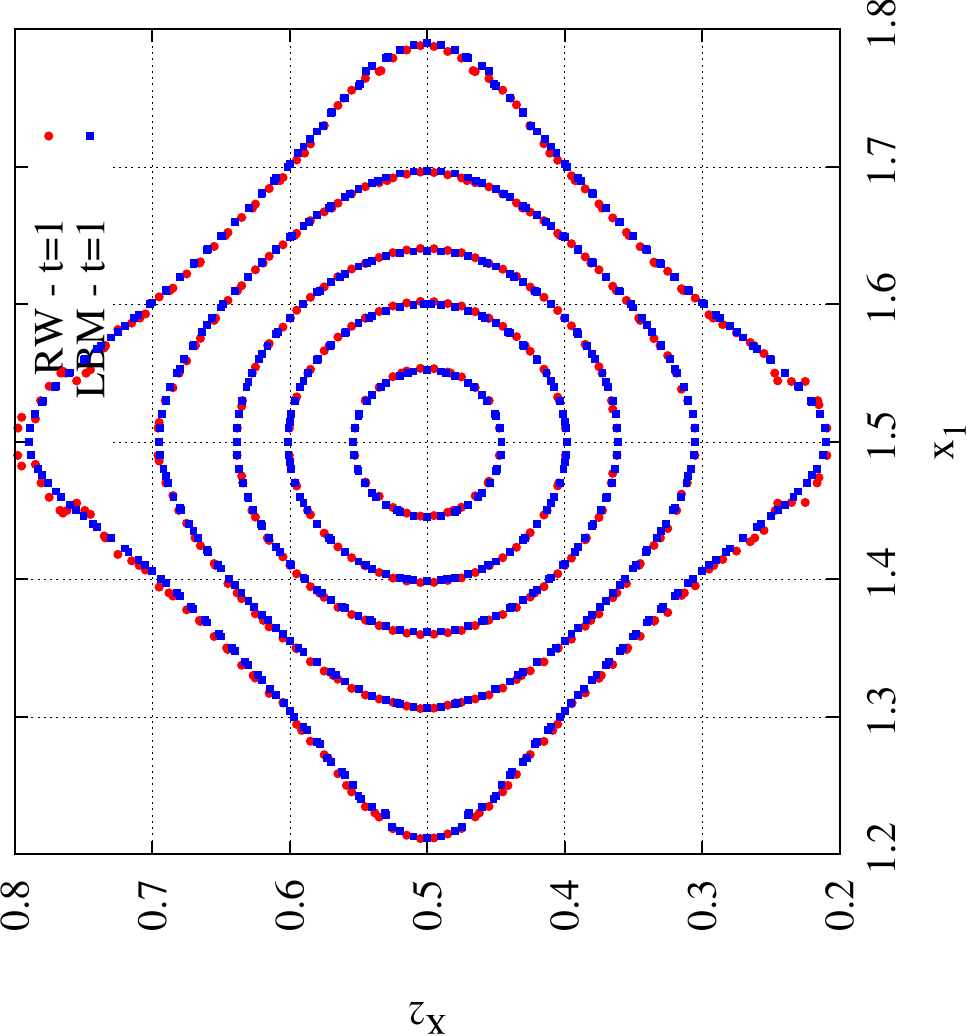}\protect
\par\end{centering}

}
\par\end{centering}

\protect\caption{\label{fig:Res_MRT-BGK-RW}LBM and random walk approximations to Eq.
(\ref{eq:FADE_moreGeneral}) accounting for super-diffusion super-imposed
to advection. The domain is $\Omega=[0,\,2]\times[0,\,1]$, the flow
rate is $\mathbf{u}=u_{1}\mathbf{b}_{1}$ with $u_{1}=1$ and diffusion
tensor is $\overline{\overline{\mathbf{D}}}=D\overline{\overline{\mathbf{Id}}}$
with $D=0.002$. Fractional parameters are $\boldsymbol{\alpha}=\mathbf{1.7}$,
$\boldsymbol{g}=\mathbf{1}$ and $\boldsymbol{p}=\mathbf{0.5}$.}
\end{figure}

The BGK LBM applied to Eq. (\ref{eq:FADE_moreGeneral}) equipped with
parameters $\boldsymbol{\alpha}=\mathbf{1.7}$, $\boldsymbol{p}=\mathbf{0.5}$,
$\boldsymbol{g}=\mathbf{1}$ and with the above spherical tensor $\overline{\overline{\mathbf{D}}}$
needs $\lambda=0.548$ if the space-time mesh is as above. The BGK
scheme is stable at $t=0.08$ (see Fig. \ref{fig:Conv-Profiles},
cyan curve), but definitely unstable at $t=1$, at smaller velocity
($u_{1}=1$) than when we apply it to the ADE. Profiles and iso-levels
represented on Figs \ref{fig:Conv-Profiles} and \ref{fig:Conv-Iso}
demonstrate that MRT LBM equipped with parameters $\lambda_{3}=\lambda_{4}=2.1277$
is stable at $t=1$, and in perfect agreement with random walk approximation.
Stability and accuracy are still preserved by MRT LBM at velocity
$u_{1}=1.3$ if we increase $\lambda_{3}=\lambda_{4}$ to $2.6316$.
At larger velocity, several values of $\lambda_{3}$ and $\lambda_{4}$
are found to ensure the stability (even for $u_{1}=1.5$), but the
LBM solution becomes less accurate. Nevertheless, the MRT collision
achieves stability and accuracy (together) at velocities two times
larger than the BGK which becomes poorly accurate at $u_{1}=0.7$
and does not preserve the symmetry of the concentration field.

\subsection{\label{sub:Validation_AS}Validation of the LBM with exact solution}

For certain boundary conditions and source term $S_{c}$, Eq. (\ref{eq:FADE_moreGeneral})
admits exact solutions \citep{Meerschaert_etal_JCP2006} that contribute
to validate our numerical method if we allow $\overline{\overline{\mathbf{D}}}$
to depend on $\mathbf{x}$. Analytical solutions are derived from
the relationship\textcolor{red}{{} }
\begin{equation}
\frac{\partial}{\partial x}I_{x+}^{1-\alpha'}(x^{a})=\frac{\partial^{\alpha'}x^{a}}{\partial_{+}x^{\alpha'}}=x^{a-\alpha'}\frac{\Gamma(a+1)}{\Gamma(a+1-\alpha')}\quad\mbox{for}\quad a\geq\alpha'\quad\mbox{and}\quad1\geq\alpha'\label{eq:derfracpuissce}
\end{equation}
which is used with $\boldsymbol{p}=\mathbf{1}$ in simulations. In
this case Eqs (\ref{eq:FADE_moreGeneral})-(\ref{eq:Fg}) become simpler
because the second fractional integrals $I_{x_{\mu}-}^{2-\alpha_{\mu}}$
disappear. Assuming non-dimensional Eq. (\ref{eq:FADE_moreGeneral})
equipped with $\mathbf{u}=\mathbf{0}$ and $\boldsymbol{g}=\boldsymbol{1}$
supplemented by a diagonal diffusion tensor $\overline{\overline{\mathbf{D}}}$
and a source term satisfying respectively

\begin{equation}
D_{\mu\mu}(\mathbf{x})=x_{\mu}^{\alpha_{\mu}}\frac{\Gamma(a_{\mu}+2-\alpha_{\mu})}{\Gamma(a_{\mu}+2)}\qquad\mbox{and}\qquad S_{c}(\mathbf{x},\,t)=-(d+1)e^{-t}\Pi_{\mu=1}^{d}x_{\mu}^{a_{\mu}},\label{eq:Def_D_AS}
\end{equation}
we easily check that

\begin{equation}
C(\mathbf{x},\,t)=e^{-t}\Pi_{\mu=1}^{d}x_{\mu}^{a_{\mu}}\label{exsolFADE_3D}
\end{equation}
solves Eq. (\ref{eq:FADE_moreGeneral}) in $\Omega=\Pi_{\mu=1}^{d}]0,\,L_{\mu}[$.
Of course, we use initial data and boundary conditions dictated by
Eq. (\ref{exsolFADE_3D}).

\begin{figure}
\begin{centering}
\begin{tabular}{cc}
(a)  & (b)\tabularnewline
\includegraphics[angle=-90,scale=0.45]{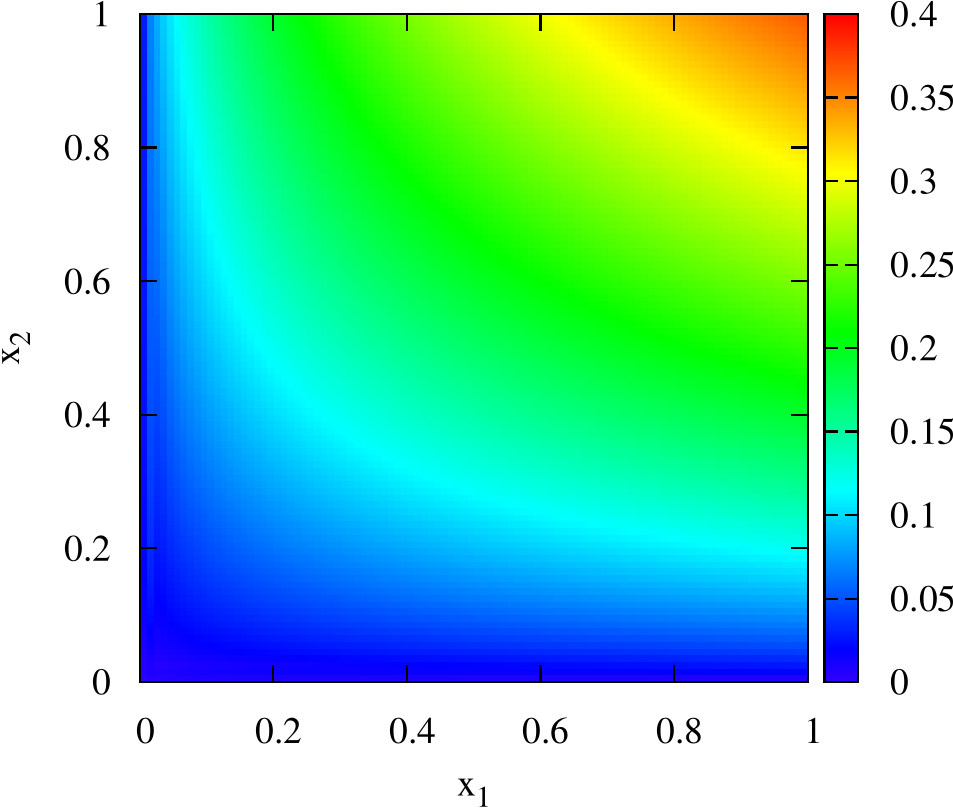}  & \includegraphics[angle=-90,scale=0.35]{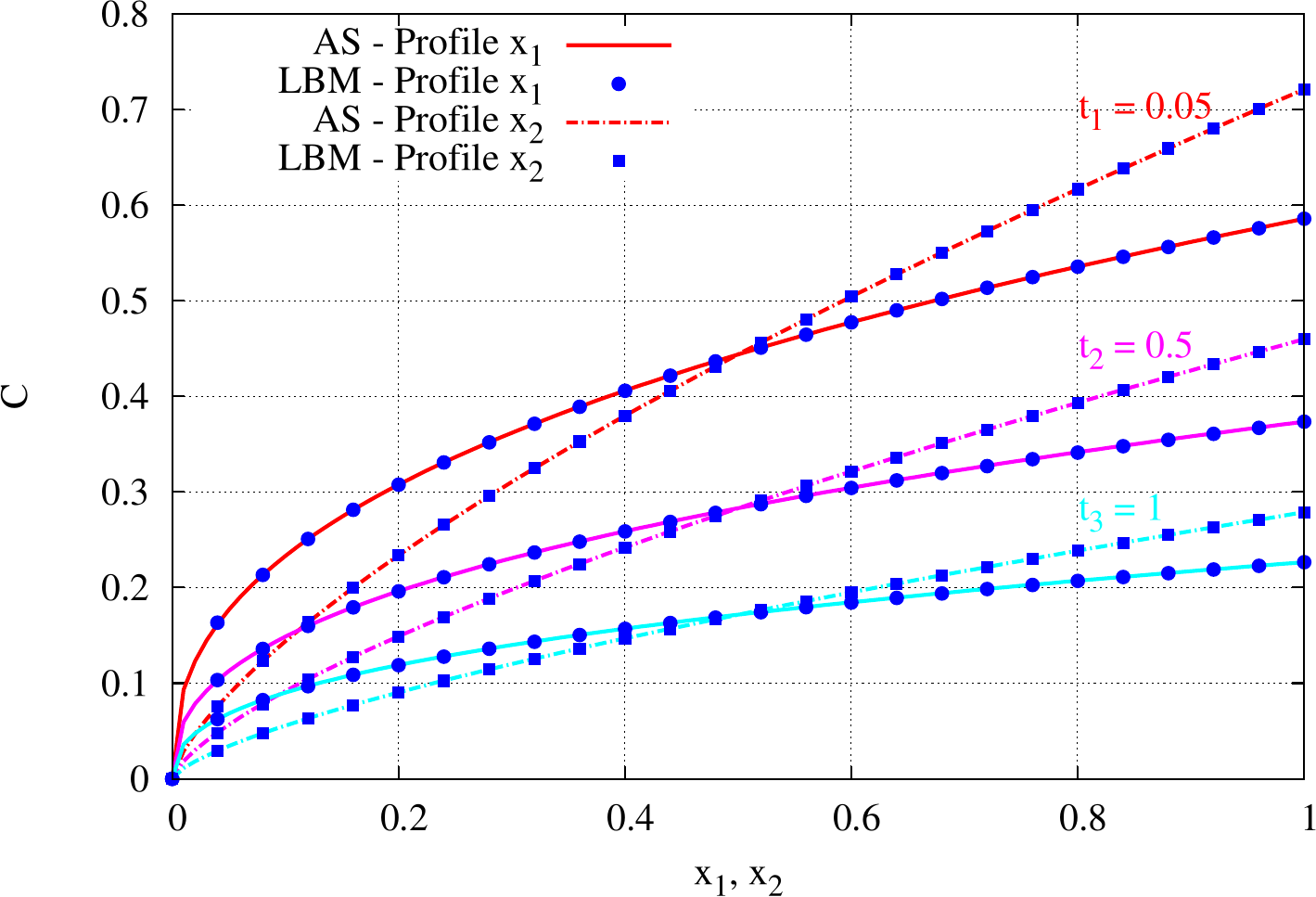}\tabularnewline
\end{tabular}
\par\end{centering}

\protect\caption{\label{fig:Bench_MRT-AnalySol}Solutions of the 2D Eq. (\ref{eq:FADE_moreGeneral})
in the conditions described by Eqs (\ref{eq:Def_D_AS}). (a) Concentration
field at $t=1$ from MRT LBM with $a_{1}=0.4$, $a_{2}=0.7$, $p_{1}=p_{2}=1$.
(b) Comparisons of profiles between LBM and Analytical Solution (AS)
$C(x_{1},\,x_{2},\,t)=e^{-t}x_{1}^{a_{1}}x_{2}^{a_{2}}$ for three
times: $t_{1}=0.05$ (red), $t_{2}=0.5$ (magenta) and $t_{3}=1$
(cyan).}
\end{figure}

The LBM scheme described by Algorithm \ref{alg:Steps-for-LBM-FADE3D}
with the MRT collision operator retrieves these solutions of the two-dimensional
Eq. (\ref{eq:FADE_moreGeneral}), according to Fig. \ref{fig:Bench_MRT-AnalySol}
where the stability parameters are $\alpha_{1}=1.5$, $\alpha_{2}=1.7$,
with $a_{1}=0.4$ and $a_{2}=0.7$. In this case, time-dependent boundary
condition and source term need being updated at each time step, and
the time- and space-steps are $\Delta t=5\times10^{-5}$ and $\Delta x=0.01$.
Fig. \ref{fig:Bench_MRT-AnalySol} displays $C(x_{1},\,x_{2}^{s},\,t)$
and $C(x_{1}^{s},\,x_{2},\,t)$ profiles that demonstrate the good
agreement between exact solution (solid and dash lines) and LBM simulation
(symbols) at $t_{1}=0.05$, $t_{2}=0.5$ and $t_{3}=1$.

\subsection{\textcolor{blue}{\label{sub:Variant}}A variant of Eq. (\ref{eq:Fg})}

Eq. (\ref{eq:Fg}) specifies the entries of vector $\boldsymbol{\mathcal{F}}^{\boldsymbol{\alpha}\boldsymbol{p}\boldsymbol{g}}(C)$
that determines the dispersive flux $\overline{\overline{\mathbf{D}}}(\mathbf{x})\boldsymbol{\mathcal{F}}^{\boldsymbol{\alpha}\boldsymbol{p}\boldsymbol{g}}(C)$
in Eq. (\ref{eq:FADE_moreGeneral}). When all its coefficients are
independent of $\mathbf{x}$, Eq. (\ref{eq:Fg}) is equivalent to
dispersive flux based on Riemann-Liouville derivatives as particle
fluxes in random walks based on stable process \citep{Neel_etal_JPA2007}.
This motivated our choice of the ordering of derivative and integral
in Eq. (\ref{eq:Fg}), used in most applications of space-fractional
diffusion equation. This choice returns an evolution equation (Eq.
(\ref{eq:FADE_moreGeneral3})) that preserves the positivity at least
in the particular cases studied in \citep{Baeumer_etal_JCOAM2018}.
However, it is sometimes questioned because it implies that Eq. (\ref{eq:FADE_moreGeneral})
does not admit independent of $\mathbf{x}$ equilibria for $\boldsymbol{\alpha}\neq\boldsymbol{2}$.
Replacing the Riemann-Liouville derivatives of the order of $\alpha_{\mu}$
in (\ref{eq:FADE_moreGeneral3}) by Caputo type derivatives of the
form of $I_{x_{\mu}\pm}^{2-\alpha}\frac{\partial^{2}}{\partial x_{\mu}^{2}}$
would return an evolution equation admitting such equilibria but badly
suited for mass transport studies because it does not preserve positivity
\citep{Baeumer_etal_JCOAM2018}. However, the same reference points
that replacing $\boldsymbol{\mathcal{F}}^{\boldsymbol{\alpha}\boldsymbol{p}\boldsymbol{g}}(C)$
by $\tilde{\boldsymbol{\mathcal{F}}}^{\boldsymbol{\alpha}\boldsymbol{p}\boldsymbol{g}}(C)$
defined by \citep{Patie_etal_Pot2012} 
\begin{equation}
\tilde{\boldsymbol{\mathcal{F}}}^{\boldsymbol{\alpha}\boldsymbol{p}\boldsymbol{g}}(C)=\sum_{\mu=1}^{d}\tilde{\mathcal{F}}_{\mu}^{\alpha_{\mu}p_{\mu}g_{\mu}}(C)\mathbf{b}_{\mu},\quad\tilde{\mathcal{F}}_{\mu}^{\alpha_{\mu}p_{\mu}g_{\mu}}(C)=p_{\mu}I_{x_{\mu}+}^{2-\alpha_{\mu}}\frac{\partial}{\partial x_{\mu}}(g_{\mu}C)+(1-p_{\mu})I_{x_{\mu}-}^{2-\alpha_{\mu}}\frac{\partial}{\partial x_{\mu}}(g_{\mu}C)\label{eq:Fg2}
\end{equation}
inserts Caputo type derivatives into the dispersive flux \citep{Patie_etal_Pot2012,Cushman_etal_WRR2000}.
This results into a variant of Eq. (\ref{eq:FADE_moreGeneral3}) that
preserves positivity (in the particular cases considered in \citep{Baeumer_etal_JCOAM2018}),
and of course admits uniform equilibria. Though we do not know experimental
data to which the variant was applied, adapting to it the LBM scheme
presented in Section \ref{sec:LBM-FADE3D} just needs slightly modified
equilibrium function.

Indeed, $\boldsymbol{\mathcal{F}}^{\boldsymbol{\alpha}\boldsymbol{p}\boldsymbol{g}}(C)$
and $\tilde{\boldsymbol{\mathcal{F}}}^{\boldsymbol{\alpha}\boldsymbol{p}\boldsymbol{g}}(C)$
only differ by a quantity related to the values of $C$ on the boundary
$\partial\Omega$ of $\Omega$, because in the dimension $1$ with
$\Omega=]\ell,\,L[$ integration by parts yields 
\begin{equation}
(I_{x+}^{2-\alpha}\frac{\partial}{\partial x}f)(x)=\frac{\partial}{\partial x}\left[I_{x+}^{2-\alpha}f(x)-f(\ell)\frac{(x-\ell)^{2-\alpha}}{\Gamma(3-\alpha)}\right],\qquad(I_{x-}^{2-\alpha}\frac{\partial}{\partial x}f)(x)=\frac{\partial}{\partial x}\left[I_{x-}^{2-\alpha}f(x)-f(L)\frac{(L-x)^{2-\alpha}}{\Gamma(3-\alpha)}\right].\label{eq:PSRL1D}
\end{equation}
In higher dimension $d$ we apply this formula to the left/right fractional
integrals w.r.t. the $x_{\mu}$ in Eqs. (\ref{eq:Def_FracInt}). The
integration intervals are $\omega_{\pm}(\mathbf{x},\,\mu)$, and the
link between $\boldsymbol{\mathcal{F}}^{\boldsymbol{\alpha}\boldsymbol{p}\boldsymbol{g}}(C)(\mathbf{x})$
and $\tilde{\boldsymbol{\mathcal{F}}}^{\boldsymbol{\alpha}\boldsymbol{p}\boldsymbol{g}}(C)(\mathbf{x})$
depends on the values taken by $C$ at the bounds of the $\omega_{\pm}(\mathbf{x},\,\mu)$
that stay on $\partial\Omega$. They correspond to $\ell$ or $L$
in (\ref{eq:PSRL1D}) depending on whether we have $+$ or $-$ in
$\pm$. For general domain $\Omega$ we call these points $\partial\omega_{\pm}(\mathbf{x},\,\mu)$,
and their coordinates $z_{\pm\nu}(\mathbf{x},\,\mu)$: the latter
coincide with the $x_{\nu}$ for $\nu\neq\mu$. For $\nu=\mu$ they
are the lower and the upper bound of the projection on $\mathbf{b}_{\mu}$
of $\omega_{+}(\mathbf{x},\,\mu)$ and $\omega_{-}(\mathbf{x},\,\mu)$
(the other bound of the interval is $x_{\mu}$). In the domain $\Omega=\Pi_{\mu=1}^{d}]\ell_{\mu},\,L_{\mu}[$
which we use for our comparisons $z_{+\nu}(\mathbf{x},\,\mu)=\ell_{\mu}$
and $z_{-\nu}(\mathbf{x},\,\mu)=L_{\mu}$.  With these notations,
for general $\Omega$ we have

\begin{subequations}

\begin{align}
(I_{x_{\mu}+}^{2-\alpha_{\mu}}\frac{\partial}{\partial x_{\mu}}C)(\mathbf{x},t) & =\frac{\partial}{\partial x_{\mu}}\left[I_{x_{\mu}+}^{2-\alpha_{\mu}}C(\mathbf{x},t)-C(\partial\omega_{+}(\mathbf{x},\,\mu),\,t)\frac{(x_{\mu}-z_{+\mu})^{2-\alpha_{\mu}}}{\Gamma(3-\alpha_{\mu})}\right],\label{eq:PSRLdD-a}\\
(I_{x_{\mu}-}^{2-\alpha_{\mu}}\frac{\partial}{\partial x_{\mu}}C)(\mathbf{x},t) & =\frac{\partial}{\partial x_{\mu}}\left[I_{x_{\mu}-}^{2-\alpha_{\mu}}C(\mathbf{x},t)-C(\partial\omega_{-}(\mathbf{x},\,\mu),\,t)\frac{(z_{-\mu}-x_{\mu})^{2-\alpha_{\mu}}}{\Gamma(3-\alpha_{\mu})}\right].\label{eq:PSRLdD-b}
\end{align}

\end{subequations}

\begin{figure}
\begin{centering}
\begin{tabular}{ccc}
(a) &  & (b)\tabularnewline
\includegraphics[angle=-90,scale=0.27]{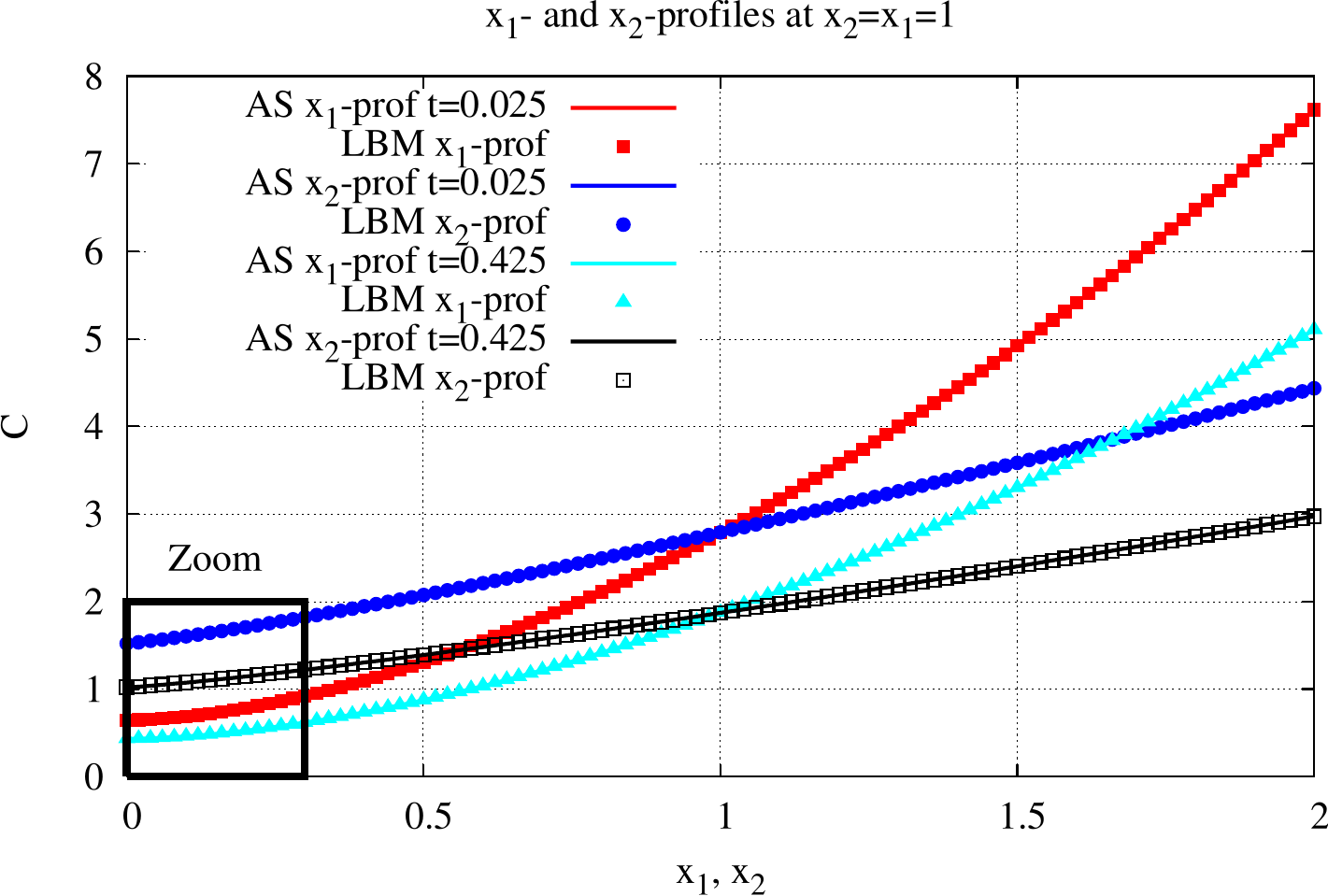} &  & \includegraphics[angle=-90,scale=0.36]{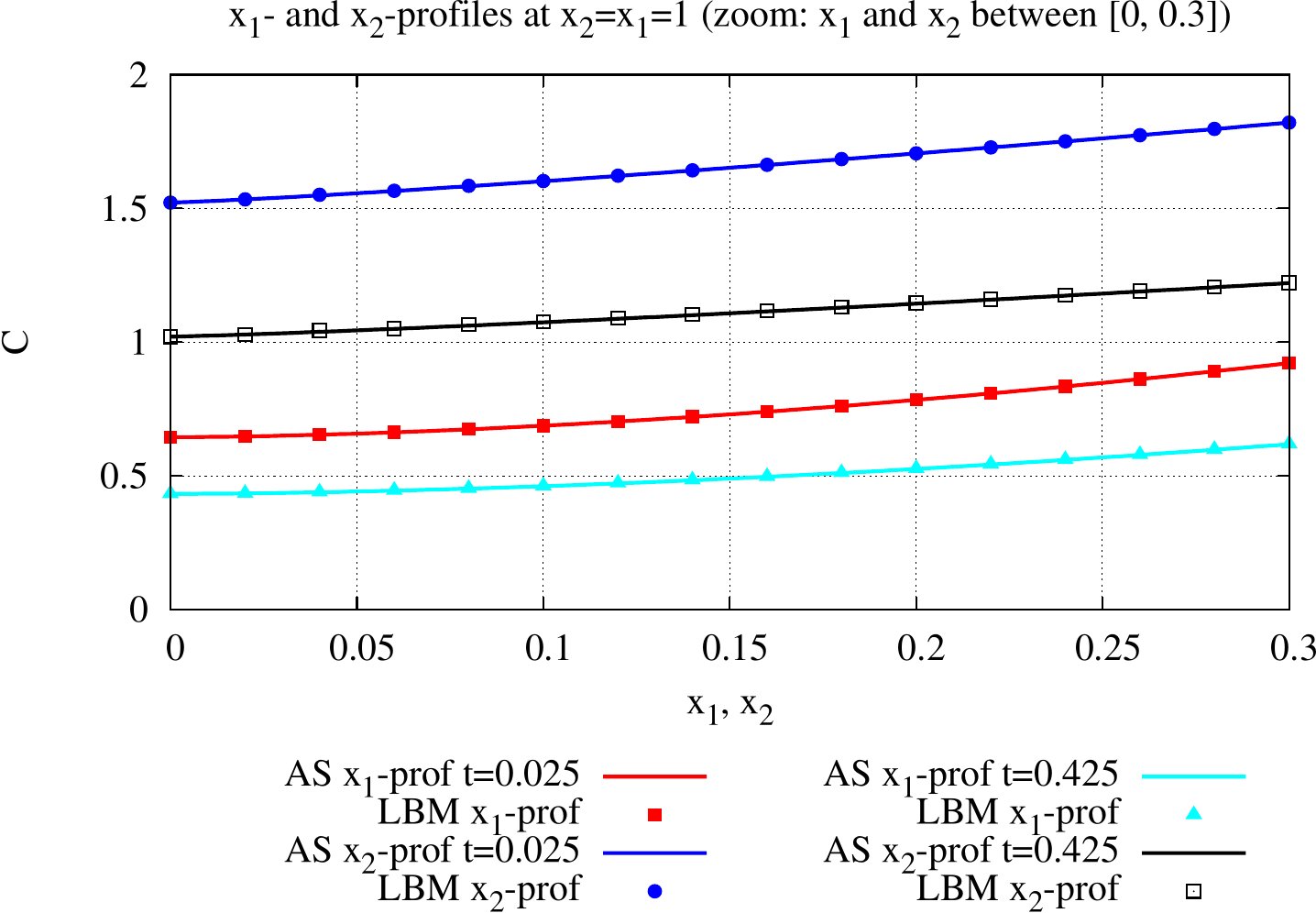}\tabularnewline
\end{tabular}
\par\end{centering}

\protect\caption{\textcolor{blue}{\label{fig:Comparison-LBM-AS_Patie-Simon}}Comparison
between LBM and Analytical Solution (AS) for Eq. (\ref{eq:FADE_moreGeneral})
equipped with $\tilde{\boldsymbol{\mathcal{F}}}^{\boldsymbol{\alpha}\boldsymbol{p}\boldsymbol{g}}(C)$
defined by Eq. (\ref{eq:Fg22}) instead of $\mathcal{F}_{\mu}^{\alpha_{\mu}p_{\mu}g_{\mu}}C(\mathbf{x},\,t)$.}
\end{figure}

This implies

\begin{align}
\tilde{\mathcal{F}}_{\mu}^{\alpha_{\mu}p_{\mu}g_{\mu}}C(\mathbf{x},t) & =\frac{\partial}{\partial x_{\mu}}\left[p_{\mu}\left\{ I_{x_{\mu}+}^{2-\alpha_{\mu}}(g_{\mu}C)(\mathbf{x},t)-g_{\mu}C(\partial\omega_{+}(\mathbf{x},\,\mu),\,t)\frac{(x_{\mu}-z_{+\mu})^{2-\alpha_{\mu}}}{\Gamma(3-\alpha_{\mu})}\right\} +\right.\nonumber \\
 & \qquad\qquad\qquad\qquad\left.(1-p_{\mu})\left\{ I_{x_{\mu}-}^{2-\alpha_{\mu}}(g_{\mu}C)(\mathbf{x},t)-g_{\mu}C(\partial\omega_{-}(\mathbf{x},\,\mu),\,t)\frac{(z_{-\mu}-x_{\mu})^{2-\alpha_{\mu}}}{\Gamma(3-\alpha_{\mu})}\right\} \right],\label{eq:Fg22}
\end{align}
which suggests an equilibrium function defined by Eq. (\ref{eq:Feq})
with 
\begin{equation}
\mathcal{A}_{i}(\mathbf{x},\,t)=\begin{cases}
C(\mathbf{x},\,t)-w_{0}\sum_{\mu=1}^{d}\left[p_{\mu}\left\{ I_{x_{\mu}+}^{2-\alpha_{\mu}}g_{\mu}C(\mathbf{x},t)-g_{\mu}C\left(\partial\omega_{+}(\mathbf{x},\,\mu),\,t\right)\frac{(x_{\mu}-z_{+\mu})^{2-\alpha_{\mu}}}{\Gamma(3-\alpha_{\mu})}\right\} +\right.\\
\qquad\qquad\qquad\qquad\qquad\left.(1-p_{\mu})\left\{ I_{x_{\mu}-}^{2-\alpha_{\mu}}g_{\mu}C(\mathbf{x},t)-g_{\mu}C\left(\partial\omega_{-}(\mathbf{x},\,\mu),\,t\right)\frac{(z_{-\mu}-x_{\mu})^{2-\alpha_{\mu}}}{\Gamma(3-\alpha_{\mu})}\right\} \right] & \mbox{if}\,\,i=0\\
\\
w_{i}\left[p_{\mu}\left\{ I_{x_{\mu}+}^{2-\alpha_{\mu}}g_{\mu}C(\mathbf{x},t)-g_{\mu}C\left(\partial\omega_{+}(\mathbf{x},\,\mu),\,t\right)\frac{(x_{\mu}-z_{+\mu})^{2-\alpha_{\mu}}}{\Gamma(3-\alpha_{\mu})}\right\} +\right.\\
\qquad\qquad\qquad\qquad\qquad\left.(1-p_{\mu})\left\{ I_{x_{\mu}-}^{2-\alpha_{\mu}}g_{\mu}C(\mathbf{x},t)-g_{\mu}C\left(\partial\omega_{-}(\mathbf{x},\,\mu),\,t\right)\frac{(z_{-\mu}-x_{\mu})^{2-\alpha_{\mu}}}{\Gamma(3-\alpha_{\mu})}\right\} \right] & \mbox{if}\,\,i\neq0
\end{cases}\label{eq:CoeffA_FADE3DPS}
\end{equation}
instead of (\ref{eq:CoeffA_FADE3D}). We check this equilibrium function
in two dimensions by setting $\ell_{\mu}=0$, $L_{\mu}=1$, $D_{\mu}=1$,
$g_{\mu}=1$ for $\mu=1,\,2$ and noting that $C(\mathbf{x},\,t)=e^{-t}\Pi_{\mu=1}^{2}(x_{\mu}^{\alpha_{\mu}}+a_{\mu})$
satisfies Eq. (\ref{eq:FADE_moreGeneral}) equipped with $\tilde{\boldsymbol{\mathcal{F}}}^{\boldsymbol{\alpha}\boldsymbol{p}\boldsymbol{g}}(C)$
instead of $\boldsymbol{\mathcal{F}}^{\boldsymbol{\alpha}\boldsymbol{p}\boldsymbol{g}}(C)$
provided we also set $\mathbf{u}=\mathbf{0}$, $p_{\mu}=1$ for $\mu=1,\,2$,
and 
\begin{equation}
S_{c}(\mathbf{x},\,t)=-e^{-t}\left[\Pi_{\mu=1}^{2}(x_{\mu}^{\alpha_{\mu}}+a_{\mu})+D_{1}\Gamma(\alpha_{1}+1)(x_{2}^{\alpha_{2}}+a_{2})+D_{2}\Gamma(\alpha_{2}+1)(x_{1}^{\alpha_{1}}+a_{1})\right].\label{eq:SourceTerm_PS}
\end{equation}
With the modified equilibrium distribution function, the LBM of Section
\ref{sec:LBM-FADE3D} equipped with the equilibrium function defined
by (\ref{eq:CoeffA_FADE3DPS}) retrieves the exact solution if we
apply Dirichlet boundary conditions (resp. initial condition) dictated
by $C(\mathbf{x},\,t)=e^{-t}\Pi_{\mu=1}^{2}(x_{\mu}^{\alpha_{\mu}}+a_{\mu})$
(for $a_{\mu}\neq0$) on the boundaries of $\Omega$ (resp. at $t=0$).
A look at the figures \ref{fig:Comparison-LBM-AS_Patie-Simon}(a),(b)
checks this fact. For comparisons of Figs. \ref{fig:Comparison-LBM-AS_Patie-Simon}
(a), (b), we choose fractional parameters $\alpha_{1}=1.7$, $\alpha_{2}=1.2$,
$a_{1}=0.3$, $a_{2}=1.2$ and $p_{1}=p_{2}=1$. The other numerical
parameters for LBM are $\Delta x=0.02$, $\Delta t=2.5\times10^{-5}$,
$N_{x_{1}}=N_{x_{2}}=101$. Comparisons between LBM (symbols) and
analytical solution (solid lines) are presented for two times $t_{1}=0.025$
and $t_{2}=0.425$ along $x_{1}$- and $x_{2}$-profiles respectively
for $x_{2}=1$ and $x_{1}=1$.

\section{\label{sec:Conclusion}Conclusion}

We discussed Lattice Boltzmann schemes that simulate $d$-dimensional
fractional equation (Eq. (\ref{eq:FADE_moreGeneral})) for $d=2$
or $3$. In this conservation equation, the solute flux splits into
classical advective part and dispersive contribution accounting for
super-diffusion. The latter is obtained by applying a regular diffusivity
tensor to a $d$-dimensional vector whose entry of rank $\mu$ is
 the partial derivative w.r.t. $x_{\mu}$ of a linear combination
of two fractional integrals w.r.t. this coordinate.  Moreover, the
LBM splits the concentration into $\mathcal{N}+1$ populations whose
velocities belong to a lattice (here D2Q5 for $d=2$ and D3Q7 for
$d=3$), and uses $\mathcal{N}+1$ equilibrium distributions for these
populations. The above mentioned fractional integrals provide us equilibrium
distributions which cause that the total concentration approximately
solves Eq. (\ref{eq:FADE_moreGeneral}). The complementarity of the
equilibrium function with BGK or MRT collision rules returns accurate
numerical solutions provided the fractional integrals are discretized
with precision.

Eq. (\ref{eq:FADE_moreGeneral}) with Dirichlet boundary conditions
has exact solutions and random walk approximations valid when its
parameters satisfy incompatible conditions. Exact solutions hold for
$\boldsymbol{p}=\boldsymbol{0}$ or $\boldsymbol{1}$ supplemented
by specific space dependent coefficients. Random walk approximations
are more generic. They may be flawed by noise (if we do not use sufficiently
many walkers) but not by numerical diffusion or instability. Comparisons
with random walks and exact solutions revealed that MRT collision
rule approximates the solutions of Eq. (\ref{eq:FADE_moreGeneral})
as accurately as BGK collision rule, even near the boundaries. Yet,
the BGK rule does not adapt to all anisotropic diffusion tensor and
returns solutions that become unstable at moderately large advection
speed. Compared with the ADE, space fractional diffusion equations
are more sensitive to such numerical instabilities. MRT LBM is accurate
in more general conditions, and less unstable because it includes
additional freedom degrees whose values can be chosen so as to damp
perturbations.

The LBM scheme approximates Eq. (\ref{eq:FADE_moreGeneral}) associated
to dispersive flux based on Riemann-Liouville or Caputo derivatives
as well provided we choose appropriate equilibrium function.

Random walks and LBM can be viewed as complementary methods that approximate
the solutions of p.d.es as Eq. (\ref{eq:FADE_moreGeneral}) and check
each other accuracy. The latter depends of their time steps (independent
of each other), which of course influence the computing times which
are different. According to our comparisons (not using parallel computing)
the LBM is four times more rapid in the dimension 3, except when we
can use the independence of the coordinates of $\mathbf{X}^{\Omega}(t)$
(with spatially homogeneous parameters when domain $\Omega$ is a
rectangle). In this case the random walk is ten times more rapid because
it relies upon one-dimensional simulations. In the dimension $d=2$
its computing time is similar to that of the LBM if we do not use
the independence property. Hence, the LBM will be preferred in the
dimension three in general domain or with spatially heterogeneous
parameters, and will become especially useful in tasks that accumulate
many successive simulations. Nevertheless, the LBM needs being assessed
by point-wise comparisons with random walk.

\appendix

\section{\label{sec:asp}Asymptotic analysis of the LBE associated with the
equilibrium function adapted to the fractional ADE}

With the equilibrium function defined by Eqs. (\ref{eq:Feq})-(\ref{eq:CoeffA_FADE3D}),
the LBE approximates the solutions of Eq. (\ref{eq:FADE_moreGeneral})
if $\boldsymbol{\alpha}=\boldsymbol{2}$, within BGK as well as MRT
setting. Allowing $\boldsymbol{\alpha}<\boldsymbol{2}$ only requires
to modify the final steps of the classical Chapman-Enskog expansion.
The reasoning is based on the Taylor expansion of $\mathbb{T}\bigr|\boldsymbol{f}\bigr\rangle$
(defined by Eq. (\ref{eq:Def_Translation})) combined with Eq. (\ref{eq:multiscale})
and the expansion $\bigr|\boldsymbol{f}\bigr\rangle=\sum_{k\geq0}\varepsilon^{k}\bigr|\boldsymbol{f}^{(k)}\bigr\rangle$.
Moreover it assumes bounded derivatives w.r.t $t_{1}$, $t_{2}$ and
$x'_{1}$, ..., $x'_{d}$ for $\bigr|\boldsymbol{f}^{(k)}\bigr\rangle$.
We recall for convenience the method that sequentially collects the
items of each order $\varepsilon^{k}$, and yields a sequence of equations
satisfied by the first moments of the $\bigr|\boldsymbol{f}^{(k)}\bigr\rangle$,
especially $\langle\mathbf{1}|\boldsymbol{f}^{(0)}\rangle=C$. More
specifically, we show that Eqs. (\ref{eq:Diffusion_Coefficient_3D})
and (\ref{eq:M0_FADE_3D})-(\ref{eq:M2_FADE_3D}) imply that $\langle\mathbf{1}|\boldsymbol{f}^{(0)}\rangle$
is solution of Eq. (\ref{eq:FADE_moreGeneral}) within an error of
$O(\varepsilon^{2})$.

Order $\varepsilon^{0}$ writes $0=\mathbf{M}^{-1}\boldsymbol{\Lambda}\mathbf{M}\left[\bigr|\boldsymbol{f}^{eq}\bigr\rangle-\bigr|\boldsymbol{f}^{(0)}\bigr\rangle\right]$
and implies 
\begin{equation}
\bigr|\boldsymbol{f}^{(0)}\bigr\rangle=\bigr|\boldsymbol{f}^{eq}\bigr\rangle\label{eq:fzero}
\end{equation}
because $\mathbf{M}^{-1}\boldsymbol{\Lambda}\mathbf{M}$ is invertible
with BGK or MRT matrices $\mathbf{M}$ and $\boldsymbol{\Lambda}$.
Collisions preserving the total solute mass imply Eq. (\ref{eq:M0_FADE_3D}),
hence 
\begin{equation}
\langle\mathbf{1}\bigr|\boldsymbol{f}^{(k)}\bigr\rangle=0\quad\mathrm{for}\quad k>0.\label{eq:zero}
\end{equation}
If we assume $S_{c}=S\varepsilon^{2}$ for the source rate, orders
$\varepsilon$ and $\varepsilon^{2}$ yield respectively 
\begin{equation}
\Delta t\frac{\partial}{\partial t_{1}}\bigr|\boldsymbol{f}^{(0)}\bigr\rangle+\Delta x\frac{\partial}{\partial x_{\mu}'}\bigr|e_{\mu}\boldsymbol{f}^{(0)}\bigr\rangle=-\mathbf{M}^{-1}\boldsymbol{\Lambda}\mathbf{M}\bigr|\boldsymbol{f}^{(1)}\bigr\rangle\label{eq:44}
\end{equation}
and 
\begin{align}
\Delta t\left[\frac{\partial}{\partial t_{1}}\bigr|\boldsymbol{f}^{(1)}\bigr\rangle+\frac{\partial}{\partial t_{2}}\bigr|\boldsymbol{f}^{(0)}\bigr\rangle\right]+\Delta x\frac{\partial}{\partial x'_{\mu}}\bigr|e_{\mu}\boldsymbol{f}^{(1)}\bigr\rangle+\frac{\Delta t^{2}}{2}\frac{\partial^{2}}{\partial t_{1}^{2}}\bigr|\boldsymbol{f}^{(0)}\bigr\rangle\qquad\nonumber \\
+\Delta t\Delta x\frac{\partial^{2}}{\partial t_{1}\partial x'_{\mu}}\bigr|e_{\mu}\boldsymbol{f}^{(0)}\bigr\rangle+\frac{\Delta x^{2}}{2}\frac{\partial^{2}}{\partial x'_{\mu}\partial x'_{\nu}}\bigr|e_{\mu}e_{\nu}\boldsymbol{f}^{(0)}\bigr\rangle & =-\mathbf{M}^{-1}\boldsymbol{\Lambda}\mathbf{M}\bigr|\boldsymbol{f}^{(2)}\bigr\rangle+\Delta tS\left|\mathbf{w}\right\rangle \label{eq:46}
\end{align}
in which $a_{\mu}b_{\mu}$ and $\partial b_{\mu}/\partial a_{\mu}$
represent $\sum_{\mu=1}^{d}a_{\mu}b_{\mu}$ and $\sum_{\mu=1}^{d}\partial b_{\mu}/\partial a_{\mu}$
respectively.

Eq. (\ref{eq:44}) is equivalent to 
\begin{equation}
-\mathbf{M}\bigr|\boldsymbol{f}^{(1)}\bigr\rangle=\boldsymbol{\Lambda}^{-1}\mathbf{M}\left[\Delta t\frac{\partial}{\partial t_{1}}\bigr|\boldsymbol{f}^{(0)}\bigr\rangle+\Delta x\frac{\partial}{\partial x'_{\mu}}\bigr|e_{\mu}\boldsymbol{f}^{(0)}\bigr\rangle\right].\label{eq:444}
\end{equation}
which determines $\bigr|\boldsymbol{f}^{(1)}\bigr\rangle$ and implies
(projection on $\bigr|\boldsymbol{1}\bigr\rangle$)
\begin{equation}
\Delta t\frac{\partial}{\partial t_{1}}\langle\mathbf{1}|\boldsymbol{f}^{(0)}\rangle+\Delta x\frac{\partial}{\partial x'_{\mu}}\langle e_{\mu}|\boldsymbol{f}^{(0)}\rangle=0\label{eq:A7}
\end{equation}
in view of (\ref{eq:zero}). Summing up $\varepsilon$ times (\ref{eq:A7})
and $\varepsilon^{2}$ times the projection of (\ref{eq:46}) on $\bigr|\boldsymbol{1}\bigr\rangle$
yields 
\begin{align}
\Delta t\left[\varepsilon\frac{\partial}{\partial t_{1}}+\varepsilon^{2}\frac{\partial}{\partial t_{2}}\right]\bigr\langle\boldsymbol{1}|\boldsymbol{f}^{(0)}\bigr\rangle+\Delta x\varepsilon\frac{\partial}{\partial x'_{\mu}}\bigr\langle e_{\mu}|\boldsymbol{f}^{(0)}\bigr\rangle+\Delta x\varepsilon^{2}\frac{\partial}{\partial x'_{\mu}}\bigr\langle e_{\mu}|\boldsymbol{f}^{(1)}\bigr\rangle\qquad\nonumber \\
+\frac{\Delta x\Delta t}{2}\varepsilon^{2}\frac{\partial^{2}}{\partial x'_{\mu}\partial t_{1}}\bigr\langle e_{\mu}|\boldsymbol{f}^{(0)}\bigr\rangle+\frac{\Delta x^{2}}{2}\varepsilon^{2}\frac{\partial^{2}}{\partial x'_{\mu}\partial x'_{\nu}}\bigr\langle e_{\mu}|e_{\nu}\boldsymbol{f}^{(0)}\bigr\rangle & =\Delta t\varepsilon^{2}S\bigr|\mathbf{w}\bigr\rangle\label{eq:477}
\end{align}
in which the two first items are $\Delta t\partial_{t}\bigr\langle\boldsymbol{1}|\boldsymbol{f}^{(0)}\bigr\rangle$
and $\Delta t\boldsymbol{\nabla}\cdot\mathbf{u}C$ (in view of Eqs.
(\ref{eq:multiscale}) and (\ref{eq:M1_FADE_3D})) with an error of
the order of $\varepsilon^{2}$. Then, Eq. (\ref{eq:44}) implies
more or less simple expression for $|\boldsymbol{f}^{(1)}\bigr\rangle$
in BGK or MRT setting.

\subsection{\label{sec:aspBGK}BGK}

The BGK collision operator assumes $\mathbf{M}=\mathbf{I}$ and $\boldsymbol{\Lambda}^{-1}=\lambda\mathbf{I}$,
so that Eq. (\ref{eq:44}) implies 
\begin{equation}
-\bigr|\boldsymbol{f}^{(1)}\bigr\rangle=\lambda\left[\Delta t\frac{\partial}{\partial t_{1}}|\boldsymbol{f}^{(0)}\rangle+\Delta x\frac{\partial}{\partial x'_{\mu}}\bigr|e_{\mu}\boldsymbol{f}^{(0)}\bigr\rangle\right].\label{eq:f1BGK}
\end{equation}
hence 
\begin{equation}
-\bigl\langle e_{\nu}\bigr|\boldsymbol{f}^{(1)}\bigr\rangle=\lambda\left[\Delta t\frac{\partial}{\partial t_{1}}\bigl\langle e_{\nu}\bigr|\boldsymbol{f}^{(0)}\bigr\rangle+\Delta x\frac{\partial}{\partial x'_{\mu}}\bigl\langle e_{\nu}e_{\mu}\bigr|\boldsymbol{f}^{(0)}\bigr\rangle\right].\label{eq:m01BGK}
\end{equation}
Combining the latter with Eqs (\ref{eq:multiscale}) and (\ref{eq:477})
yields 
\begin{equation}
\Delta t\frac{\partial}{\partial t}\bigl\langle\mathbf{1}\bigr|\boldsymbol{f}^{(0)}\bigr\rangle+\Delta x\frac{\partial}{\partial x_{\mu}}\bigl\langle e_{\mu}\bigr|\boldsymbol{f}^{(0)}\bigr\rangle-\left(\lambda-\frac{1}{2}\right)\Delta x\Delta t\frac{\partial^{2}}{\partial t\partial x_{\mu}}\bigl\langle e_{\mu}\bigr|\boldsymbol{f}^{(0)}\bigr\rangle-\left(\lambda-\frac{1}{2}\right)\Delta x^{2}\frac{\partial^{2}}{\partial x_{\mu}\partial x_{\nu}}\bigl\langle e_{\mu}e_{\nu}\bigr|\boldsymbol{f}^{(0)}\bigr\rangle=S_{c}\Delta t.\label{eq:46BGK}
\end{equation}
Neglecting the third term $(\lambda-1/2)\Delta x\Delta t\partial_{tx_{\mu}}^{2}\bigl\langle e_{\mu}\bigr|\boldsymbol{f}^{(0)}\bigr\rangle$
and using Eq. (\ref{eq:M2_FADE_3D}) yields Eq. (\ref{eq:FADE_moreGeneral})
provided $\overline{\overline{\mathbf{D}}}=D\overline{\overline{\mathbf{Id}}}$
if, moreover, Eq. (\ref{eq:Diffusion_Coefficient_3D}) is satisfied.

\subsection{\label{sec:aspMRT}MRT}

With MRT matrices $\mathbf{M}$ and $\boldsymbol{\Lambda}^{-1}$,
Eq. (\ref{eq:444}) implies 
\begin{equation}
-\langle e_{\nu}|\boldsymbol{f}^{(1)}\rangle=\lambda_{\nu\rho}\left[\Delta t\frac{\partial}{\partial t_{1}}\langle e_{\rho}|\boldsymbol{f}^{(0)}\rangle+\Delta x\frac{\partial}{\partial x'_{\mu}}\langle e_{\rho}e_{\mu}|\boldsymbol{f}^{(0)}\rangle\right],\label{eq:m21_MRT}
\end{equation}
very similar to Eq. (\ref{eq:m01BGK}) if we set $\overline{\overline{\boldsymbol{\lambda}}}$
for the $d\times d$ matrix of entries $\lambda_{\mu\nu}$. Recalling
Eqs (\ref{eq:zero}) and (\ref{eq:multiscale}) in Eq. (\ref{eq:477})
yields 
\begin{align}
\Delta t\left[\frac{\partial}{\partial t}\bigl\langle\mathbf{1}\bigr|\boldsymbol{f}^{(0)}\bigr\rangle+\nabla\cdot(C\boldsymbol{u})\right]+\Delta t\Delta x\left(-\frac{\partial}{\partial t}\lambda_{\mu\rho}\frac{\partial}{\partial x_{\rho}}\bigl\langle e_{\rho}\bigr|\boldsymbol{f}^{(0)}\bigr\rangle+\frac{1}{2}\frac{\partial^{2}}{\partial t\partial x_{\mu}}\bigl\langle e_{\rho}\bigr|\boldsymbol{f}^{(0)}\bigr\rangle\right)\qquad\qquad\qquad\label{eq:tott}\\
+\Delta x^{2}\left[\frac{1}{2}\frac{\partial^{2}}{\partial x_{\mu}\partial x_{\nu}}\bigl\langle e_{\mu}e_{\nu}\bigr|\boldsymbol{f}^{(0)}\bigr\rangle-\frac{\partial}{\partial x_{\mu}}\lambda_{\mu\rho}\frac{\partial}{\partial x_{\nu}}\bigl\langle e_{\rho}e_{\nu}\bigr|\boldsymbol{f}^{(0)}\bigr\rangle\right] & =S_{c}\Delta t,
\end{align}
in view of $\bigl\langle e_{\mu}\bigr|\boldsymbol{f}^{(0)}\bigr\rangle=u_{\mu}C\Delta t/\Delta x$
(i.e. (\ref{eq:M1_FADE_3D})). Using Eq. (\ref{eq:M2_FADE_3D}), $\bigl\langle e_{\rho}e_{\nu}\bigr|\boldsymbol{f}^{(0)}\bigr\rangle=\delta_{\rho\nu}[p_{\nu}I_{x_{\nu}+}^{2-\alpha_{\nu}}(g_{\nu}C)+(1-p_{\nu})I_{x_{\nu}-}^{2-\alpha_{\nu}}(g_{\nu}C)]$
and neglecting the second term $\Delta t\Delta x(-\partial_{t}\lambda_{\mu\rho}\partial_{x_{\rho}}\bigl\langle e_{\rho}\bigr|\boldsymbol{f}^{(0)}\bigr\rangle+\partial_{tx_{\mu}}^{2}\bigl\langle e_{\rho}\bigr|\boldsymbol{f}^{(0)}\bigr\rangle/2)$
as above, we recognize Eq. (\ref{eq:FADE_moreGeneral}) if $\overline{\overline{\boldsymbol{\lambda}}}$
satisfies Eq. (\ref{eq:D_ab}) which we copy below for convenience
\begin{equation}
\overline{\mathbf{\overline{D}}}=e^{2}\left(\overline{\overline{\boldsymbol{\lambda}}}-\frac{1}{2}\overline{\overline{\mathbf{Id}}}\right)\frac{\Delta x^{2}}{\Delta t}.\label{eq:condition}
\end{equation}
The matrix $\overline{\overline{\boldsymbol{\lambda}}}$ may depend
on $\mathbf{x}$ to account for spatially non-uniform tensor $\overline{\overline{\mathbf{D}}}$.

\section{\label{sec:AppendixC}Stable process and fractional equation}

\subsection{\label{sec:ApA}Stable Lévy laws}

\begin{figure}
\centering{}\includegraphics[clip,scale=0.3]{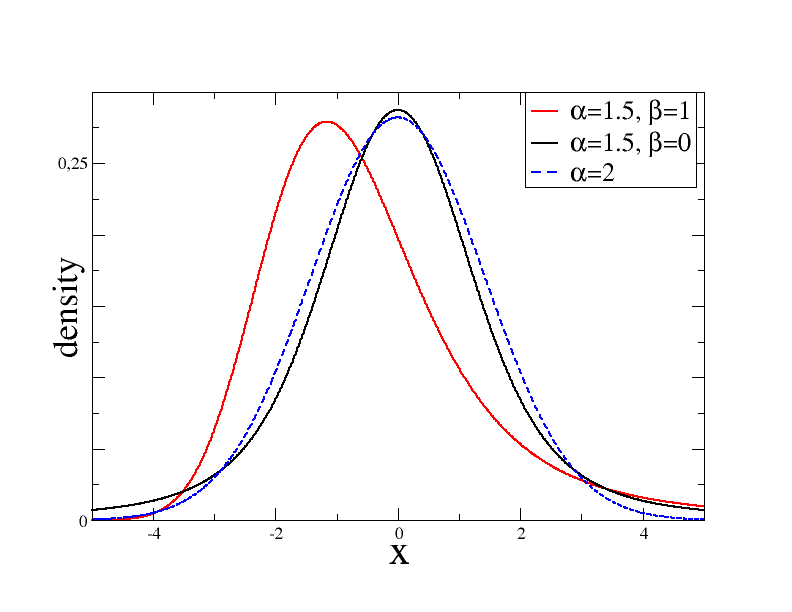}\includegraphics[clip,scale=0.3]{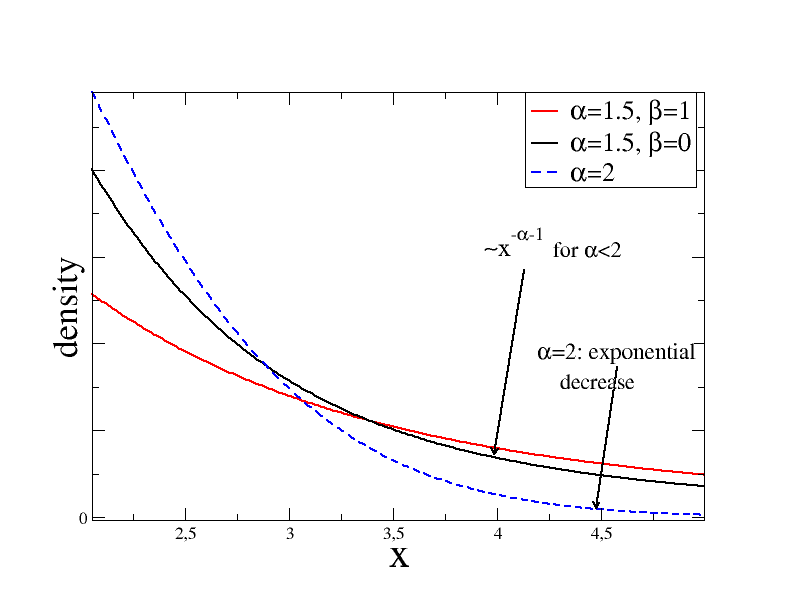}
\protect\caption{\label{fig:densial15}Lévy variable densities $\mathbb{P}^{\alpha,\beta}(x)$.
Represented densities correspond to $\alpha=2$ (dashed blue line)
and $\alpha=1.5$. The latter case is represented for two values of
$\beta$: the solid black line represents the symmetric case $\beta=0$
and the solid red line represents $\beta=1$. Right panel shows a
magnification of the asymptotic behavior of the densities at large
positive values. }
\end{figure}

A look at the characteristic function 
\begin{align}
\langle e^{jkS(\alpha,\,\beta)}\rangle & =e^{-\varphi_{\alpha,\,\beta}}(k)\label{eq:stb}\\
\varphi_{\alpha,\,\beta}(k) & =|k|^{\alpha}\left(1-j\beta\textrm{sign}(k)\tan\frac{\pi\alpha}{2}\right)\label{eq:stbb}
\end{align}
gives an idea of the role played by its two parameters: $\beta$ which
ranges between $-1$ and $+1$ describes the degree of skewness of
$\varphi_{\alpha,\,\beta}$ and the distribution of $S(\alpha,\,-\beta)$
is the mirror image of $S(\alpha,\,\beta)$. We also see that the
influence of $\beta$ vanishes when $\alpha$ approaches $2$, a special
value at which $S(\alpha,\,\beta)$ is standard centered Gaussian.
The stability index $\alpha\in]0,\,2]$ describes the asymptotic decrease
of the density $\mathbb{P}^{\alpha,\beta}(x)$, proportional to $x^{-\alpha-1}$
when $|x|\to\infty$ except for $\alpha=2$ and $x<0$ if $\beta=\pm1$:
decreasing $\alpha$ makes large values more probable and thickens
the tails of the distribution of $S(\alpha,\,\beta)$ illustrated
by Fig. \ref{fig:densial15}. Large positive values of $\beta$ re-inforce
the positive tail without changing the exponent of the asymptotic
behavior, except if $\beta=1$, a special case in which the negative
tail tapers off exponentially. The figure also shows that the most
probable value of $S(\alpha,\,\beta)$ has the sign of $-\beta$ while
Eqs (\ref{eq:stb})-(\ref{eq:stbb}) show that the average is always
zero.

\subsection{\label{sto}Stochastic process and fractional equation in bounded
domain }

That killed process p.d.fs satisfy Eq. (\ref{eq:FADE_moreGeneral})
can be proved for rectangular domain $\Omega$ if the coefficients
do not depend on $\mathbf{x}$. Indeed, reference \citep{Chen-Meerschaert-Nane_JMAA2012} 
implies this result for one-dimensional symmetric process provided
$\mathbf{u}=\mathbf{0}$, moreover assuming smooth initial condition
compactly supported in $\Omega$. We easily extend it to higher dimension
in $\Omega=\Pi_{\mu=1}^{d}]\ell_{\mu},\,L_{\mu}[$ for possibly non-symmetric
process whose projections on the $\mathbf{b}_{\mu}$ are independent,
as here. Indeed, the recent paper \cite{Baeumer_etal_submitted2018} 
states the result for general stable process and initial data provided
$\Omega$ is regular in the sense of Probability (not of numerical
analysis). However, one-dimensional intervals are regular. Moreover
in our case the components of $\mathbf{X}(t)$ are mutually independent,
and this extends to $\mathbf{X}^{\Omega}(t)$ when $\Omega$ is a
rectangle of $\mathbb{R}^{d}$. Consequently, the density of $\mathbf{X}^{\Omega}(t)$
is of the form of $P=\Pi_{\mu=1}^{d}P_{\mu}(x_{\mu},\,t)$ where each
one-dimensional density $P_{\mu}(x_{\mu},\,t)$ satisfies a one-dimensional
version of Eq. (\ref{eq:FADE_moreGeneral}). Hence $P$ satisfies
the $d$-dimensional version. This proves that histograms of samples
of $\mathbf{X}^{\Omega}(t)$ approximate solutions of Eq. (\ref{eq:FADE_moreGeneral3})
satisfying homogeneous Dirichlet conditions at the boundary of $\Omega$.

Validations 1 and 2 in Section \ref{sub:Validations_LBM-BGK} illustrate
this result, predicted by \citep{Du_etal_DCDSB2014}. Other boundary
conditions were proved to achieve the equivalence of processes deduced
from $\mathbf{X}$ and Eq. (\ref{eq:FADE_moreGeneral}) in the case
of spatially homogeneous parameters \citep{Baeumer_etal_JCOAM2018,Baeumer_etal_TAMS2016},
but there also exists boundary constraints whose application to $\mathbf{X}(t)$
severely modifies the equation that rules the p.d.f \citep{Krepysheva_etal_PhysA2006,Cusimano_etal_AMM2017}.

\bibliographystyle{elsarticle-num}
\bibliography{Biblio-CPC-FADE}

\end{document}